\documentclass[%
 reprint,
 amsmath,amssymb,
 aps,
]{revtex4-2}

\usepackage{graphicx}
\usepackage{dcolumn}
\usepackage{bm}
\usepackage{color}
\usepackage{multirow}
\usepackage{array}
\usepackage[dvipsnames]{xcolor}

\begin{document}
\newcommand{\sah}[1]{{\color{teal}#1}}
\newcommand{\jess}[1]{{\color{red}#1}}
\newcommand{\sss}[1]{{\color{pink}#1}}
\newcommand\xleftrightarrow[1]{%
  \mathbin{\ooalign{$\,\xrightarrow{#1}$\cr$\xleftarrow{\hphantom{#1}}\,$}}
}
\preprint{APS/123-QED}

\title{Characterizing Multimode Effects in a Guided Matterwave Gyroscope}

\author{Jessica K. Eastman}
 \email{Jessica.Eastman@anu.edu.au}
 \affiliation{Department of Quantum Science and Technology, Research School of Physics, The Australian National University, Canberra 2601, Australia}

 \author{Ellen Zheng}
  \affiliation{Department of Quantum Science and Technology, Research School of Physics, The Australian National University, Canberra 2601, Australia}
  
\author{Stuart S. Szigeti}
\affiliation{Department of Quantum Science and Technology, Research School of Physics, The Australian National University, Canberra 2601, Australia}

\author{Simon A. Haine}
\email{Simon.Haine@anu.edu.au}

\affiliation{Department of Quantum Science and Technology, Research School of Physics, The Australian National University, Canberra 2601, Australia}

\date{\today}

\begin{abstract}
We theoretically investigate the performance of a compact matterwave vortex gyroscope formed by a two-component Bose-Einstein condensate in a toroidal potential. Unlike conventional atomic gyroscopes that rely on the Sagnac effect, the topological stability of the vortex state yields rotation sensitivity independent of the enclosed area, making the device robust against geometric drifts. Using fully quantum multimode simulations, we quantify two interaction-driven mechanisms that degrade performance: phase diffusion from one-axis-twisting and four-wave mixing from intercomponent scattering. We identify regimes where tuning interaction and trapping parameters produces a trade-off between these effects, and find that reducing the intercomponent scattering length can counterintuitively worsen sensitivity. Finally, we compare the vortex gyroscope to a guided Sagnac interferometer, demonstrating superior scaling, establishing it as a promising candidate for compact precision rotation sensing.
\end{abstract}

\maketitle

\section{Introduction}

Quantum sensors based on atom interferometry have demonstrated state-of-the-art measurements of rotation rate with a long-term stability that outperforms classical alternatives by many orders of magnitude~\cite{Gustavson:1997,Durfee:2006,Dutta:2016,Gautier:2022,Salducci:2024}. Incorporating such low-drift rotation sensing technology into inertial navigation systems could substantially improve the duration over which these systems provide accurate and reliable positioning and navigation without GNSS~\cite{Wright:2022, Wang:2023, Narducci:2022, Gersemann:2025}, unlocking revolutionary capabilities in aerospace, autonomous vehicles, space science, and geophysical exploration~\cite{El-Sheimy:2020}. However, realizing these benefits requires the development of atomic gyroscope technology that is sufficiently compact and resilient for deployment on highly dynamic platforms (e.g. crewed aircraft, uncrewed aerial systems, satellites).

Guided atomic matterwave interferometry is a promising candidate for realizing compact and resilient atomic gyroscopes. Here an optical waveguide is used to confine the atomic matterwaves during the interferometry sequence, potentially enabling compact device geometries and mitigating the impact of platform dynamics on performance. Proof-of-principle laboratory demonstrations of guided atomic Sagnac interferometry have been performed~\cite{Jo:2007,Wu:2007,Burke:2009,Qi:2017,Woffinden:2023} and have recently demonstrated per shot sensitivities of 10$^{-5}$~rad/s~\cite{Moan:2020} and phase stability better than 0.2~rad~\cite{Beydler:2024}. However, in guided configurations, the accumulated phase is proportional to the area enclosed by the matterwave paths. Consequently, the device response is highly sensitive to the details of the trapping geometry, rendering its calibration vulnerable to parameter drifts and undermining one of the key advantages of quantum sensing. This motivates the investigation of alternative guided configurations for rotation measurement.

\begin{figure}
\centering

  \includegraphics[width=.99\linewidth]{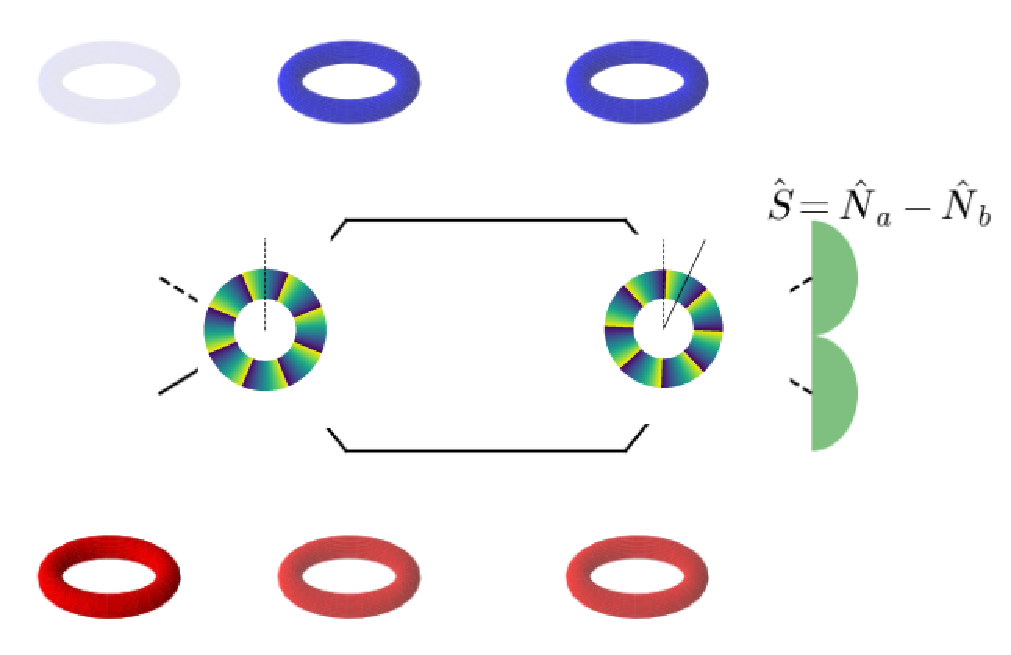}
  
  \label{fig:sub1}

\caption{Schematic for the matterwave vortex gyroscope. The atoms are initiated in the ground state of the toroidal trap. A two-photon Raman transition coherently transfers half the population into a second internal state and a  laser beam with a transverse phase profile imparts Orbital Angular Momentum (OAM) to the atoms. As a result we have a two component BEC where the two components are counter-rotating with angular momentum $\pm l \hbar$ for integer $l$. After some interrogation time, there will be a phase difference between the two components. A second OAM laser pulse is applied and a final readout of population difference between the two components gives a measured rotation rate. }
\label{fig:schematic}
\end{figure}
One promising alternative is the matterwave vortex gyroscope~\cite{Halkyard:2010, Nolan2016, Haine:2016b}, which uses the topological stability imparted by confinement in a toroidal trap \cite{Ryu2007,Ramanathan:2011,Yakimenko:2013,Beattie:2013,Polo:2025}. Recently, the topological stability of similar ring-trap configurations has been proposed for precision sensing of linear accelerations \cite{Borysenko:2025} and vortex transport \cite{Chaika:2026}. Instead of transferring linear momentum from light to atoms to direct them around some closed (guided) path, light with orbital angular momentum is employed to imprint quantized vortices in the superfluid. Since a matterwave vortex manifests as an integer-valued phase winding in the superfluid, the laser pulse imprints a phase winding onto the matterwave. Ideally, this phase winding is topologically stable, and so remains stationary in the inertial frame. Consequently, this phase winding can be compared to the phase winding of the light (which is created in the rotating frame) at later times via quantum interference, yielding a rotation angle, and ultimately a rotation rate. As the phase winding is integer valued, the calibration of this phase response is immune from drifts in the geometry of the confining potential, yielding  superior scale factor stability and reduced environmentally-induced systematic drifts compared to the guided Sagnac configuration.

However, in practice, atom-atom interactions within the atomic superfluid introduce complex multimode quantum many-body behavior that can degrade the vortex gyroscope's performance relative to the ideal operation described above. As we show in detail in this work, two key mechanisms are responsible for this degradation: phase diffusion arising from One-Axis Twisting (OAT)~\cite{Haine:2018, Szigeti:2020, Szigeti:2021} and four-wave mixing (FWM) due to intercomponent scattering~\cite{Haine:2011}. Neither mechanism is captured by conventional mean-field approaches that have been used in many theoretical analyses of guided atomic matterwave interferometry to date~\cite{Halkyard:2010, Helm:2018}. However, detailed modeling that includes both OAT and FWM effects is essential to determine the true performance achievable by the vortex gyroscope. 

Here, we present a comprehensive characterization of the vortex gyroscope's sensitivity to rotation that incorporates first-principles modeling of quantum many-body dynamics induced by atom-atom interactions. This is achieved through detailed numerical simulations via the multimode truncated Wigner (TW) approach~\cite{Steel:1998, Sinatra:2002, Blakie:2008, Polkovnikov:2010}, which captures quantum fluctuations and many-body dynamics beyond the scope of mean-field approaches such as the Gross-Pitaevskii Equation (GPE) \cite{Gross:1961,Pitaevskii:1961}. Our numerical analysis is supported by simple few-mode analytic models, which allow us to isolate and determine the relative importance of OAT and FWM on the sensitivity degradation for different parameter regimes. By doing so, we identify optimal operating regimes and propose mitigation strategies, such as vertical separation of components during interrogation or tuning the interaction strength via Feshbach resonances~\cite{Roberts:1998}. Thus, although these processes are intrinsic to the system, they are somewhat controllable, allowing a trade-off between phase diffusion and FWM through adjustment of cross-scattering parameters.

The paper is structured as follows: In section \ref{section:overview}, we begin by introducing the vortex gyroscope and the many-body physics underpinning its operation. In sections \ref{section:OAT} and \ref{section:FWM}, we then investigate the two dominant interaction-induced degradation mechanisms (phase diffusion via OAT and FWM) using simple few-mode analytical models that qualitatively agree with our multimode numerical TW simulations. In section \ref{section:87Rb}, we apply these insights to ring-trapped Bose-Einstein condensate (BEC) system of $^{87}$Rb atoms with realistic experimental parameters. Finally, in section \ref{section:GS}, we compare the vortex gyroscope with the guided Sagnac interferometer configuration, highlighting the role of atom-atom interactions in both rotation sensing configurations. Although we find that each approach has its merits, our analysis shows that the vortex gyroscope offers superior scaling and robustness against atomic-interaction-induced noise. 

\section{Overview of the matterwave Vortex Gyroscope}
\label{section:overview}
We consider the vortex gyroscope scheme introduced in Ref.~\cite{Halkyard2010} and summarized in Fig.~\ref{fig:schematic}. A BEC of ultracold atoms with two hyperfine states, denoted by $|a\rangle$ and $|b\rangle$, 
is confined in a toroidal potential 
\begin{equation}
    V({\bf r}) = \frac{1}{2} m \omega_r^2 (r_\perp-R)^2 + \frac{1}{2} m\omega_z^2 z^2,
\end{equation}
where $R$ is the radius of the toroid, $r_\perp = \sqrt{x^2 + y^2}$, and $\omega_r$ and $\omega_z$ are trapping frequencies describing the strength of confinement along the $r_\perp$ and $z$ directions, respectively.
The atoms are initiated in the ground state of the ring in $\vert a \rangle$, before a two-photon Raman transition coherently transfers half the population to $\vert b \rangle$, acting as a matterwave beamsplitter. Unlike conventional atom interferometry used to measure linear accelerations, the two lasers used to implement the Raman transition are co-propagating along the z-axis, such that they transfer no linear momentum during the beamsplitting process. Instead, the optical modes carry well-defined orbital angular momentum, such as via Laguerre-Gauss (LG) beams~\cite{Nolan2016, Husband:2026, Husband:2026b}. Specifically, if the two optical modes are LG beams with angular momentum $l_a$ and $l_b$, respectively, then the Raman transition will transfer $(l_a-l_b)\hbar$ to each atom in the beamsplitting process. This results in two counter-rotating components, with angular momentum of the state $\vert b \rangle$ atoms differing by $2l = (l_a-l_b)\hbar$ from the $\vert a \rangle$ atoms.  To simplify the dynamics of the device, we initially transfer angular momentum $+l$ to the initial state before the first beamsplitter, such that the two counter-propagating atomic modes $|a\rangle$ and $|b\rangle$ have angular momentum of equal magnitude and opposite direction after the first beamsplitter. In this case, the relative phase shift after the second beamsplitting pulse is $\phi = 2 l \Omega T$. 

After an interrogation time $T$, a second beamsplitter is implemented via the same set of optical beams. If the sensor has experienced a rotation around the $z$-axis by an angle $\theta = \Omega T$, then this shifts the relative phase of the Raman lasers by $\phi = 2l \theta$, which the second beamsplitter converts to a measurable difference in the populations of internal states $|a\rangle$ and $|b\rangle$ via quantum interference. From this population measurement, the phase (and therefore the rotation rate, $\Omega$) can be inferred. The uncertainty in the estimate of $\Omega$ is given by 
\begin{align}
\Delta \Omega &= \frac{\sqrt{\mathrm{Var}(\hat{S})}}{\vert \partial_\Omega \langle \hat{S}\rangle \vert}, \label{Delta_Omega}
\end{align}
where $\hat{S} = \hat{N}_a - \hat{N}_b$ and $\hat{N}_{a(b)}$ is the number of atoms in component $|a(b)\rangle$. Using standard error propagation, we can express this as
\begin{align}
\Delta \Omega &= \frac{\Delta \phi}{\vert \partial_\Omega \phi\vert} = \frac{\Delta \phi}{2 l T}, 
\end{align}
where $\Delta \phi$ is the uncertainty in the estimate of the phase from the population measurement. Under ideal operation with no interactions, and assuming $N$ initially uncorrelated atoms and perfect overlap between the two modes, we find $\langle \hat{S}\rangle = -N \cos \phi$ and $\Delta \phi = \frac{1}{\sqrt{N}}$, giving
\begin{align}
\Delta \Omega_\mathrm{SQL} &= \frac{1}{2lT \sqrt{N}}.
\end{align} 
However, the non-negligible atomic interactions in a confined BEC can introduce complications such as atomic correlations and imperfect mode-overlap. This can induce complicated many-body dynamics in the trap such that $\Delta \Omega \neq \Delta \Omega_\mathrm{SQL}$. This motivates us to introduce the degradation factor $\xi$, defined by
\begin{align}
\xi &= \frac{\Delta \Omega}{\Delta \Omega_\mathrm{SQL}} = \sqrt{N} \frac{\sqrt{\mathrm{Var}(\hat{S})}}{\vert \partial_\phi \langle \hat{S}\rangle \vert} , 
\end{align}
which is the deviation from ideal non-interacting operation resulting from the many-body dynamics.

\subsection{Many-body Hamiltonian}
The Hamiltonian for the system is
\begin{align}
\label{eq:hammb}
\hat{H} &= \sum_{i = a,b}\hat{H}_i + \hat{H}_\mathrm{int} + \Pi_1(t)\hat{H}_{\mathrm{P}_1} + \Pi_2(t)\hat{H}_{\mathrm{P}_2}.
\end{align}
First,
\begin{align}
    \hat H_{i} &= \int d{\bf r} \hat \Psi_i^\dagger ({\bf r}) \left[\frac{-\hbar^2}{2m} \nabla^2 + V({\bf r})   \right]  \hat \Psi_i ({\bf r}) 
\end{align} 
describes the single-particle dynamics of the individual components $a$ and $b$, and $\hat{\Psi}_{a(b)}$ are the usual bosonic field operators for atoms in state $|a(b)\rangle$, obeying
\begin{equation}
    \left[\hat \Psi_i ({\bf r}) , \hat \Psi_j^\dagger ({\bf r'})  \right] = \delta_{i,j} \delta({\bf r} - {\bf r'}).
 \end{equation}
Second,
\begin{equation}
    \hat H_{\mathrm{int}} = \sum_{i,j}\frac{U_{ij}}{2} \int d{\bf r}  \hat \Psi_i^\dagger ({\bf r})  \hat \Psi_j^\dagger ({\bf r})   \hat \Psi_j ({\bf r})  \hat \Psi_i ({\bf r})  , \label{Hint}
\end{equation}
accounts for the $s$-wave scattering between atoms, where $U_{ij} = 4\pi \hbar^2 a_{ij} /m$ is the interaction strength for contact interactions between atoms in states $\vert i \rangle$ and $\vert j \rangle$, with scattering length $a_{ij}$. Finally, the interactions with the laser pulses that implement that atomic beamsplitters, expressed in cylindrical coordinates, are described by
\begin{equation}
    \hat H_{\mathrm{P}_j} = \frac{\hbar \Omega_R}{2} \int d{\bf r} \left(\Psi_a^\dagger ({\bf r})   \hat \Psi_b ({\bf r})e^{i(2 l \theta + \phi_j)} + \mathrm{h.c.}\right)
\end{equation} 
where $\Omega_R$ is the effective two-photon Rabi frequency, and $\phi_j$ is the relative phase difference of the two lasers used to implement the Raman transition.  As only the phase difference $\phi_2-\phi_1$ can affect observable outcomes, by convention we set $\phi_1 = 0$ and $\phi_2 = \phi$. Here we have assumed a plane wave phase imprinting from the OAM beams. This could be achieved by a Vortex Gaussian beam which has a minimal core and an approximately flat intensity profile \cite{Husband:2026}. The factors $\Pi_{1(2)}(t)$ are top-hat functions in time representing the short duration for which the coupling pulses are applied. In order to implement 50/50 beamsplitters, we set $\Omega_R\int dt \Pi(t)  = \frac{\pi}{2}$. 

The signal $\hat{S} = \hat{N}_a - \hat{N}_b$ and the form of $\hat{H}_{\mathrm{P}_1}$ and $\hat{H}_{\mathrm{P}_2}$,  motivate the introduction of the pseudo spin operators
\begin{subequations}
\begin{align}
\hat{J}_x &= \frac{1}{2} \int d{\bf r} \left(\hat \Psi_a^\dagger \hat \Psi_b e^{i2l\theta} + \hat \Psi_b^\dagger \hat \Psi_a e^{-i2l\theta}\right), \\
\hat{J}_y &= \frac{-i}{2} \int d{\bf r} \left(\hat \Psi_a^\dagger \hat \Psi_b e^{i2l\theta} - \hat \Psi_b^\dagger \hat \Psi_a e^{-i2l\theta}\right), \\
\hat{J}_z &= \frac{1}{2} \int d{\bf r} \left(\hat \Psi_a^\dagger \hat \Psi_a - \hat \Psi_b^\dagger \hat \Psi_b\right),
\end{align}
\end{subequations}
which satisfy the usual angular momentum commutation relations, and hence form an $\mathfrak{su}(2)$ algebra, such that
\begin{align}
\hat{H}_{\mathrm{P}_j} &= \hbar \Omega_R \left(\cos \phi_j \hat{J}_x -\sin \phi_j \hat{J}_y\right) \equiv \hbar \Omega_R\hat{J}_{\phi_j}.
\end{align}

In the limit where $\Pi_{\mathrm{P}_{1(2)}}$ are very narrow in time, the $\hat H_{\mathrm{P}_j}$ terms will dominate the dynamics and we can neglect $\hat H_{a(b)}$ and $\hat H_\mathrm{int}$ for the duration of the pulse. In this limit, the second coupling pulse simply acts as a rotation around $\hat{J}_\phi$ by an angle $\frac{\pi}{2}$, which allows us to express the signal as
\begin{align}
\hat{S} &= 2\hat{J}_z(t_f) = 2\hat{J}_{\phi + \frac{\pi}{2}}(t_0), 
\end{align}
where $t_0$ and $t_f$ are times immediately before and after the final beamsplitter, respectively. Noting that 
\begin{align}
\partial_\phi \hat{J}_{\phi + \frac{\pi}{2}} = -\hat{J}_\phi ,
\end{align}
we can express the degradation factor in the convenient form
\begin{align}
\xi &= \sqrt{N}\frac{\sqrt{\mathrm{Var}(\hat{J}_{\phi+\frac{\pi}{2}}(t_0))}}{\vert \langle \hat{J}_\phi(t_0)\rangle \vert} . \label{eq:degradation}
\end{align}

\subsection{Multimode Dynamics and Quantum Noise} 
Although the GPE provides an adequate mean-field description of many BEC phenomena \cite{Gross:1961, Pitaevskii:1961}, it is incapable of describing the quantum fluctuations needed to quantitatively assess the sensitivity of the vortex gyro. In order to incorporate this effect, we simulate the dynamics of the system using the TW method, which has been previously used to model the beyond mean-field dynamics of quantum gases \cite{Steel:1998, Sinatra:1995, Norrie:2006, Drummond:2017}. Unlike the GPE, it can be used to model non-classical particle correlations \cite{Haine:2014, Haine:2016, Szigeti:2017, Haine:2018, Szigeti:2020} and spontaneous scattering events \cite{Haine:2011}. The derivation of the TW method has been described in detail elsewhere \cite{Drummond:1993, Steel:1998, Blakie:2008, Polkovnikov:2010}. Briefly, the equation of motion for the system's Wigner function can be found from the von Neumann equation by using correspondences between differential operators on the Wigner function and the original quantum operators \cite{Gardiner:2004b}. By truncating third- and higher-order derivatives (the TW approximation), a Fokker-Planck equation (FPE) is obtained. The FPE is then mapped to a set of stochastic partial differential equations for complex fields $\psi_j(\mathbf{r},t)$, which loosely correspond to the original field operators $\hat{\Psi}_j(\mathbf{r}, t)$, with initial conditions stochastically sampled from the appropriate Wigner distribution \cite{Blakie:2008, Olsen:2009}. For the Hamiltonian in Eq. (\ref{eq:hammb}) (in the absence of laser fields), the complex fields obey the partial differential equations
\begin{subequations}
\begin{align}
i \hbar \frac{d  \psi_a}{dt} = \frac{-\hbar^2}{2m}\nabla^2  \psi_a  + V  \psi_a + U_{aa} \vert  \psi_a \vert^2  \psi_a + U_{ab} \vert  \psi_b \vert^2  \psi_a , \\
i \hbar \frac{d  \psi_b}{dt} = \frac{-\hbar^2}{2m}\nabla^2  \psi_b  + V  \psi_b + U_{bb} \vert  \psi_b \vert^2  \psi_b + U_{ab} \vert  \psi_a \vert^2  \psi_b.
\end{align}
\end{subequations}
By averaging over many trajectories with stochastically sampled initial conditions, expectation values of quantities corresponding to symmetrically-ordered operators in the full quantum theory can be obtained via the correspondence $\langle \{ f(\hat{\Psi}^\dag_j, \hat{\Psi}_j)\}_\mathrm{sym}\rangle = \overline{f[\psi_j^*, \psi_j]}$, where `sym' denotes symmetric ordering and the overline denotes the mean over many stochastic trajectories. The initial conditions are sampled stochastically; for a zero temperature BEC wholly occupying state $\vert a \rangle$, the initial condition samples $\psi_a({\bf r},0) = \Psi_0({\bf r}) + \eta_a({\bf r})$, $\psi_b({\bf r},0) = \eta_b({\bf r})$ \cite{Blakie:2008, Olsen:2009}, where $\Psi_0({\bf r})$ is the ground state of the single-component time-independent GPE found by taking the imaginary time evolution in the trapping potential. $\eta_j(\xi)$ are complex Gaussian noises satisfying $\overline{\eta^*_i({\bf r}_n)\eta_j({\bf r}_m)} = \frac{1}{2}\delta_{m,n}\delta_{i,j}/\Delta$, where $\Delta$ is the volume element of the spatial grid. 

In this work, we assume a tight trapping ($\omega_z/2\pi = 120$Hz) to freeze dynamics in $z$ and integrate out the $z$ dimension from our equations. We calculate the effective 2D interaction strength using a Gaussian ansatz and using the variational principle, this is given by
\begin{equation}
U^{2D}_{ij} = \frac{1}{R_z}\sqrt{\frac{1}{2\pi}}U_{ij},
\end{equation}
where the length scale in the $z$ direction, $R_z$, is found by numerically minimizing the energy using the Gaussian ansatz (details can be found in the appendix).
One could also choose to make a Thomas-Fermi ansatz in order to obtain an effective 2D model, corresponding to a strongly-interacting regime. For the purposes of this work, where the introduction of interactions becomes an undesirable consequence, it is preferable to consider a weakly interacting regime, and thus we take a Gaussian ansatz instead. 

For numerical efficiency in the multimode TW simulations, a special ring basis is employed \cite{Mehdi:2021,Prikhodko:2021}. Details of the parameters used in the simulations can be found in the appendix. 

The atomic interactions result in two distinct mechanisms of degradation. To see this, we expand the field operators into the angular momentum basis
\begin{subequations}
\begin{align}
 \hat \Psi_a ({\bf r}) &\approx \left(\sum_k^\infty e^{ik\theta} \hat a_k \right) \varphi_a({ r}), \\
 \hat \Psi_b ({\bf r}) &\approx \left(\sum_m^\infty e^{im\theta} \hat b_k \right) \varphi_b({r}),
\end{align}
\end{subequations}
where $\varphi_i(r)$ are the normalized ground states of the confining potentials for each component, and $\hat a_k$ is the annihilation operator for the angular momentum mode $k$, where the mode operators obey the usual commutation relations:
\begin{equation}
    \left[\hat a_k,\hat a_j^\dagger\right] = \delta_{i,j} ,\left[\hat a_k,\hat a_j\right] = 0, \left[\hat a_k^\dagger,\hat a_j^\dagger\right] = 0 . 
\end{equation}
Substituting this into $\hat{H}_\mathrm{int}$ (Eq. (\ref{Hint})), and assuming an initial state with only the target modes ($\hat{a}_l$, and $\hat{b}_{-l}$) occupied, the only terms that provide non-trivial dynamics on this initial state are
\begin{align}
\hat{H}_\mathrm{int} &= \hat{H}_\mathrm{OAT} + \hat{H}_\mathrm{FWM}.
\end{align}
Here
\begin{align}
\hat{H}_\mathrm{OAT} &= \hbar \chi(t) \hat{j}_z^2 \label{Hoat}
\end{align}
is the well known One-Axis Twisting (OAT) Hamiltonian \cite{Kitagawa1993} with
\begin{align}
\hat{j}_z &= \frac{1}{2}(\hat{a}^\dag_l \hat{a}_l - \hat{b}^\dag_{-l}\hat{b}_{-l}) \, ,
\end{align}
$\chi = \chi_{aa} + \chi_{bb} - 2\chi_{ab}$, and
\begin{equation}
    \chi_{ij} (t) = \tfrac{U_{ij}}{2\hbar}  \int d{r} \vert \phi_i ({\bf r},t)\vert^2 \vert \phi_j ({\bf r},t)\vert^2. 
\end{equation}
This Hamiltonian results in the relative population fluctuations coupling into relative phase fluctuations.  Similarly, assuming an undepleted pump approximation such that the target modes are assumed to be undepleted by the dynamics, 
\begin{align}
\hat{H}_\mathrm{FWM} &= \chi_{ab}(t) \sum_{\kappa}(\hat a_{\kappa}^\dagger \hat b_{-\kappa}^\dagger\hat a_{l}\hat b_{-l}+h.c.) , 
\end{align} 
 is the well known four-wave mixing (FWM) Hamiltonian, which results in atoms from the two target modes colliding and scattering into unoccupied modes. Setting the scattering lengths to be equal ($\chi_{aa} = \chi_{bb} = \chi_{ab}$) sets $\hat{H}_\mathrm{OAT}$ to zero, and as a result turns off phase diffusion and leaves only FWM. However, turning off the cross-scattering ($\chi_{ab}= 0$) effectively removes the FWM and leaves only the phase diffusion dynamics which occurs entirely in the SU(2) phase space for the two target modes. Physically, the latter could be achieved by spatially separating the two components in $z$. 
We will investigate the two extreme cases in isolation, looking at how OAT and FWM dynamics separately contribute to the degradation of the sensitivity. We will also investigate experimentally-derived scattering lengths for $^{87}$Rb atoms, which lie in between these two extreme regimes. 

\begin{figure}
\centering
  \includegraphics[width=.99\linewidth]{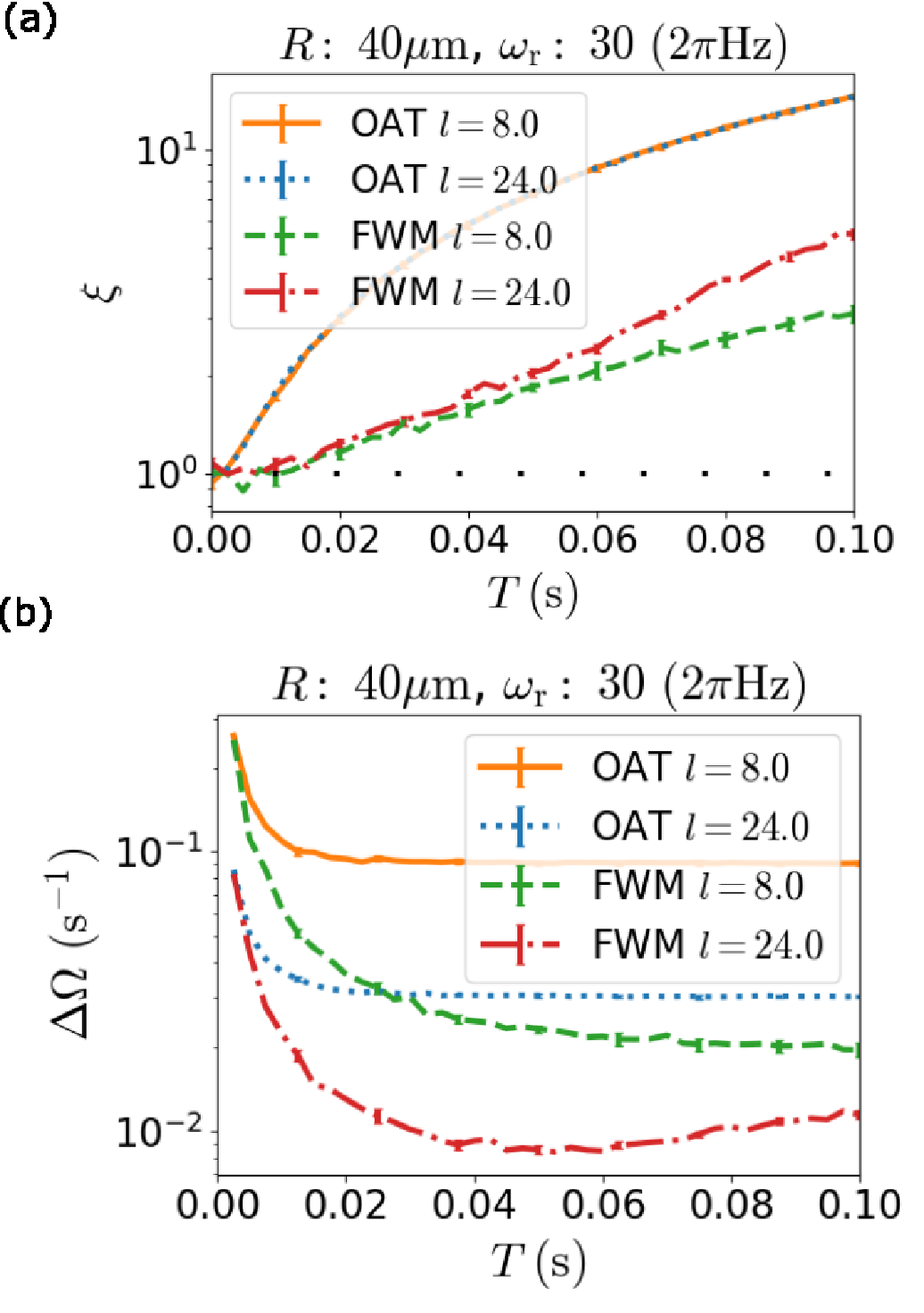} 
 \label{fig:sub1}
\caption{(a) Degradation factor $\xi$ and (b) rotation sensitivity $\Delta \Omega$ as a function of the interrogation time $T$ in the regime dominated by OAT dynamics ($l=8$ orange line, $l=24$ blue dots) and the FWM regime ($l=8$ green dash, $l=24$ red dash-dots) for trapping parameters \{$R=40 \mu $m, $\omega_r/2\pi = 30 $Hz\}.  The ideal non-interacting case gives $\xi=1$ for all time. Error bars (plotted every 4 points) give the standard error in the mean over 960 trajectories.  }
\label{fig:initbadness}
\end{figure}

We begin by examining the multimode TW simulations for both cases where the dynamics are dominated by OAT and FWM individually. Figure~\ref{fig:initbadness}(a) shows the degradation factor for the case in which OAT dynamics dominate (${U_{aa} = U_{bb}, U_{ab} = 0}$) and the case in which FWM dynamics dominate (${U_{aa} = U_{bb} = U_{ab}}$), for two different charges ($l = 8, 24$) and a single set of trapping parameters (ring radius $R = 40,\mu\text{m}$, radial trapping frequency $\omega_r/2\pi = 30,\text{Hz}$). These initial observations reveal a clear distinction between OAT and FWM at short times, with significant FWM-induced degradation emerging only at later times. For OAT, $\xi$ is independent of the charge, whereas a pronounced charge dependence is evident in the FWM case, particularly at longer times. Figure~\ref{fig:initbadness}(b) shows the corresponding rotation sensitivity for the same parameters, illustrating how this degradation impacts performance. These results suggest that retaining cross-scattering between the two components may be advantageous. To gain deeper insight into the underlying physics, we first analyze a few-mode model before turning to a full many-body characterization.

\begin{figure*}
\centering
  \includegraphics[width=.75\linewidth]{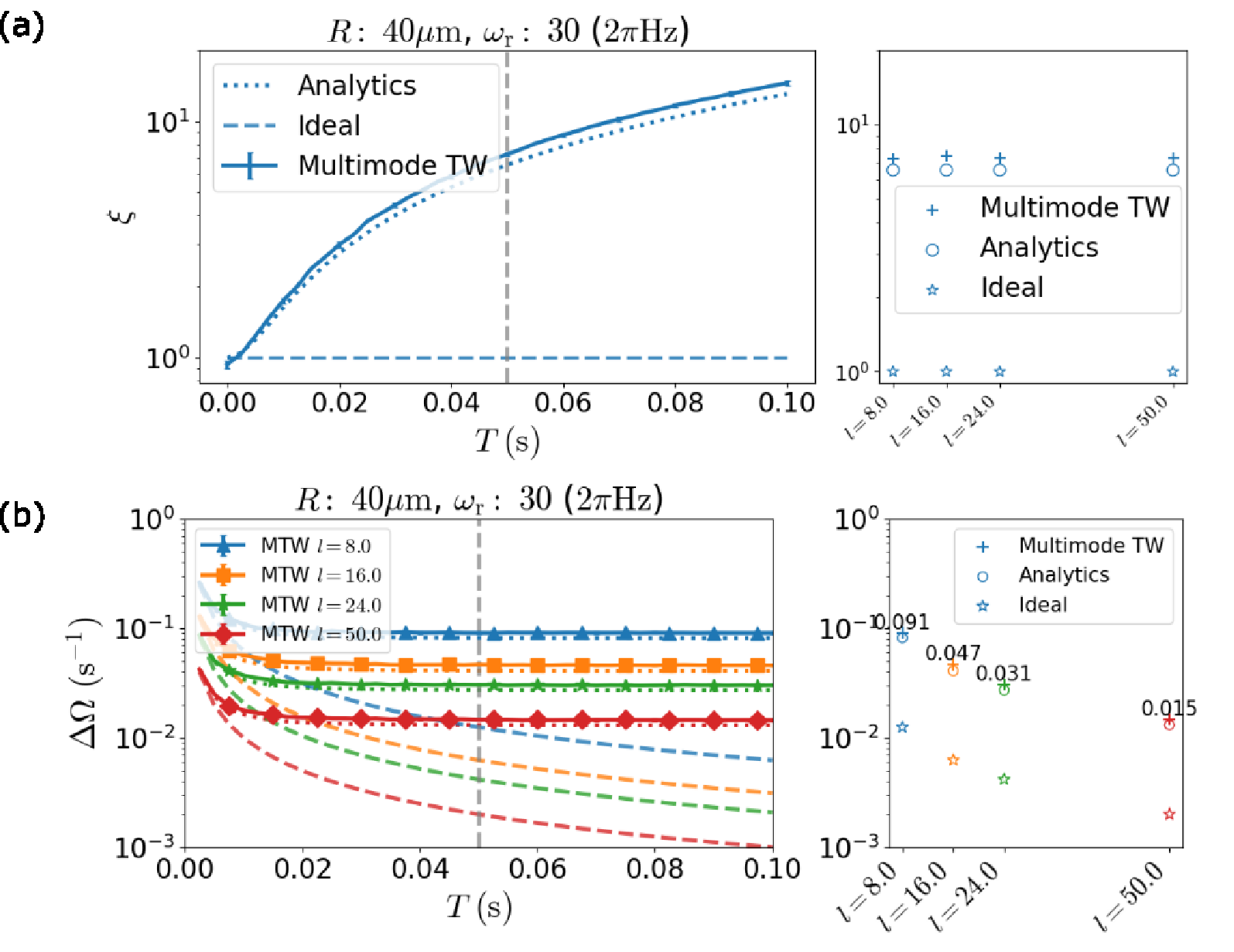}
  \label{fig:sub1}
\caption{(a) Rotation sensitivity as a function of interrogation time $T$ (left) and the corresponding asymptotic limit of sensitivity at $T=0.05$s (right) for OAT for ($l=8$, $16$, $24$ and $50$) and trapping parameters \{$R=40 \mu $m, $\omega_r/2\pi = 30$ Hz\}. The non-interacting ideal case is also plotted for each (dashed lines). (b) Corresponding degradation factor $\xi$ (which is kick independent). The error bar plot corresponds to the multimode TW results with error given by standard error in the mean from 960 sampled trajectories. The two-mode analytics is given for each charge by the dotted lines. Both the multimode TW result and the two-mode analytics are plotted on the right plot. 
}
\label{fig:OATresultmain}
\end{figure*}
\section{Phase diffusion regime}
\label{section:OAT}
To understand the process of phase diffusion that occurs, we will look at the extreme case where we turn off the cross scattering and consider equal scattering lengths for the two components ($\chi_{aa}=\chi_{bb}$, $\chi_{ab}=0$), eliminating any dynamics caused by FWM. 
\subsection{Two-mode model}
The dynamics of phase diffusion is well characterized by Eq. (\ref{Hoat}). The exact solution of this Hamiltonian has previously been studied extensively \cite{Kitagawa1993,Haine:2014, Haine:2018}.
The solution is given by
\begin{subequations}
    \begin{equation}
        \hat a(t) =  e^{i\chi(\hat j_z + \frac{1}{4})t}\hat a(0),
    \end{equation}
    \begin{equation}
        \hat b(t) =  e^{-i\chi(\hat j_z-\frac{1}{4})t}\hat b(0),
    \end{equation}
\end{subequations}
where we assume $\chi$ is time independent (i.e. the modes overlap at all times).
In the two-mode model we are interested in the terms $\mathrm{Var}(j_y(t))$, and $\langle \hat{j}_x(t) \rangle$. To evaluate these terms, we expand the exponential operator, accounting for quantum fluctuations.  Assuming an initial coherent spin state, i.e. $\langle \hat j_x(0) \rangle = \tfrac{N}{2}$, $\langle \hat j_y(0) \rangle = \langle \hat j_z(0) \rangle = 0$, with variances $\mathrm{Var}(j_x) = 0$, $\mathrm{Var}(j_y) = \mathrm{Var}(j_z)=\tfrac{N}{4}$ and zero covariance, the $\chi$ dependence of the equations will only enter into the variance of $\hat j_y$ and the expectation of $\hat j_x$ in the second order expansion with respect to $\chi t$. We therefore expand to second order in $\chi t$ to obtain 
\begin{equation}
\begin{split}
    \mathrm{Var}(j_y(t)) &= \mathrm{Var}(j_y(0)) + (2\chi t)^2 \mathrm{Var}(j_z(0))\langle \hat j_x(0)^2\rangle\\
    &= \frac{N}{4}\left(1+(\chi t)^2N^2\right),
    \end{split}
\end{equation}
where $\langle \hat{j}_x(0)^2\rangle =N^2/4$, and
\begin{equation}
\begin{split}
    \langle \hat j_x(t)\rangle   &= (1-2(\chi t)^2 \langle \hat j_z(0)^2\rangle) \langle \hat j_x(0)\rangle + 2 \chi t \langle \hat j_z(0) \rangle \langle \hat j_y(0)\rangle\\
    &= \frac{N}{2} - (\chi t)^2 \frac{N^2}{4}.
    \end{split}
\end{equation}
The resulting analytical expression for the degradation factor is given by
\begin{equation}
    \xi = \sqrt{\frac{1+(Nt\chi)^2}{(1 - \frac{1}{2}N(\chi t)^2 )^2}}.
    \label{eq:oatdeg}
\end{equation}
Using a Gaussian ansatz (now for the radial modes $\phi_i(r)$) and assuming no time-dependence of the ground-state mode functions, we find an effective $0$D interaction strength
\begin{equation}
    \chi_{ij} = \frac{U_{ij}}{8 \hbar \pi^2 R R_\perp R_z},
\end{equation}
where $R_\perp$ is the length scale in the radial direction. We find this by numerical minimization of the energy. Details can be found in Appendix section \ref{appendix:Gauss}. This degradation factor is shown in Fig. \ref{fig:OATresultmain}.
\subsection{Multimode results}
\begin{figure}
\centering
  \includegraphics[width=.99\linewidth]{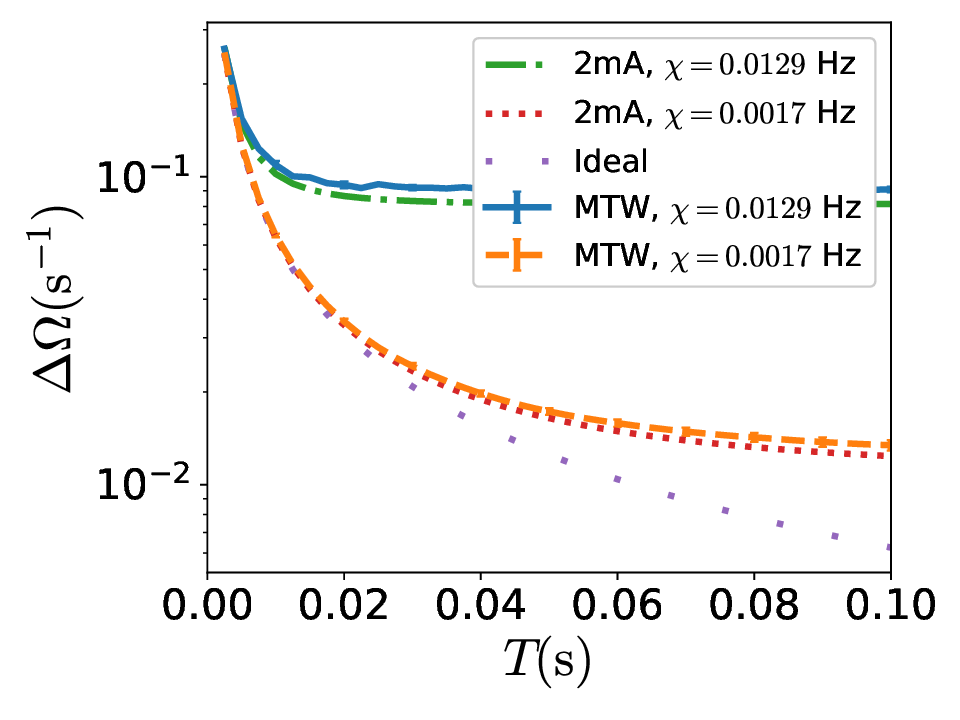}
  \label{fig:sub1}
\caption{Rotation sensitivity for the OAT $l=8$ case for Multimode TW (MTW) and two-mode analytics (2mA) for the two sets of parameters $R=40 \mu $m, $\omega_r/2\pi = 30$ Hz, $\chi= 0.0129$ (MTW - blue solid line, 2mA - green dot-dashed) and $R=200 \mu $m, $\omega_r/2\pi = 10$ Hz, $\chi= 0.0017$ (MTW - orange dashed, 2mA - red dotted). The ideal, non-interacting case is also plotted (purple dotted with larger spacing).
}
\label{fig:sensIdealtwomode}
\end{figure}
We now compare the two-mode analytics ($2$mA) with the full Multimode TW (MTW) results in the regime dominated by OAT dynamics. We will use $a_{aa}=100.4a_0$ for the scattering lengths in this case, which corresponds to the hyperfine state $\vert F=1,m_f=0 \rangle \equiv \vert a\rangle$ in $^{87}$Rb atoms \cite{Mertes2007}.  Figure \ref{fig:OATresultmain} (a) shows the rotation sensitivity both as a function of interrogation time $T$ (left) and at a time $T = 50$ms (right) by which point the sensitivity has reached an asymptotic limit and cannot be improved by longer interrogation. We show several different charges ($l=8$, $16$, $24$ and $50$) and show both the MTW result and the $2$mA result. The trapping parameters we use are a radius of $R=40\mu$m, and a radial trapping frequency of $\omega_r/2\pi=30$Hz. This results in an effective interaction strength $\chi = 0.013$Hz. We also show in this figure the non-interacting case ($\xi = 1$) as dashed lines in comparison for each charge. (b) Shows the corresponding degradation factor which is charge independent (charge dependence in the sensitivity is simply the linear scaling in $\Delta \Omega = \xi/\sqrt{N} /2lT$). 

As can be seen from the figure, the multimode and two-mode results are in good agreement for the OAT dynamics, demonstrating that in this case, a simple two-mode ansatz does very well at describing the resulting dynamics. The degradation is well-described by the analytic expression given in Eq. (\ref{eq:oatdeg}), where no explicit dependence on charge is present.  In the regime where 
\begin{equation}
\frac{1}{N} \ll \chi t \ll \frac{1}{\sqrt{N}},
\end{equation}
the degradation reduces to $\xi \to N \chi t$, giving a sensitivity $\Delta \Omega \to \sqrt{N} \chi/2l$. For $\chi = 0.0129$Hz and $10,000$ atoms, this occurs for an interrogation time of $0.008$s$ \ll T \ll 0.78$s. Thus, the growth of the degradation factor, after some initial time, exactly cancels out the improvement that can be made to the sensitivity by increasing the interrogation time. Notably, increasing the number of atoms reduces sensitivity instead of enhancing precision. In this regime, the sensitivity asymptotes to $\Delta \Omega = 0.08,0.04,0.0269,0.0129$ s$^{-1}$ for charge $l=8,16,24,50$, respectively.

The radius and trapping frequency, while not explicit in the degradation factor, will implicitly change the effective interaction strength $\chi_{ij}$.
One would expect the OAT dynamics to converge to the ideal case when the radius is increased and the trapping frequency is decreased, resulting in a more dilute condensate. The question is: where does this occur? With trapping frequency fixed, a radius of $R \approx 350 \mu$m will ensure $\xi <2$. Alternatively, fixing the radius, trapping frequency $\omega_r/2\pi \approx 1$Hz will also ensure minimal degradation. 

More realistically, a trap radius of $R=200 \mu$m, and a radial trapping frequency of $\omega_r/2\pi = 10$Hz will give a desired minimal degradation from phase diffusion. The effective interaction strength in this case is $\chi=0.0017$Hz. The rotation sensitivity for this case can be found in Fig. \ref{fig:sensIdealtwomode}. This optimization of the parameters indeed results in an enhancement of sensitivity for the OAT case, negating the limitation of quantum phase diffusion for a longer duration. However, in the OAT regime, the phase diffusion will always limit the interrogation time compared with the ideal non-interacting case.

\begin{figure*}
\centering
 \includegraphics[width=.7\linewidth]{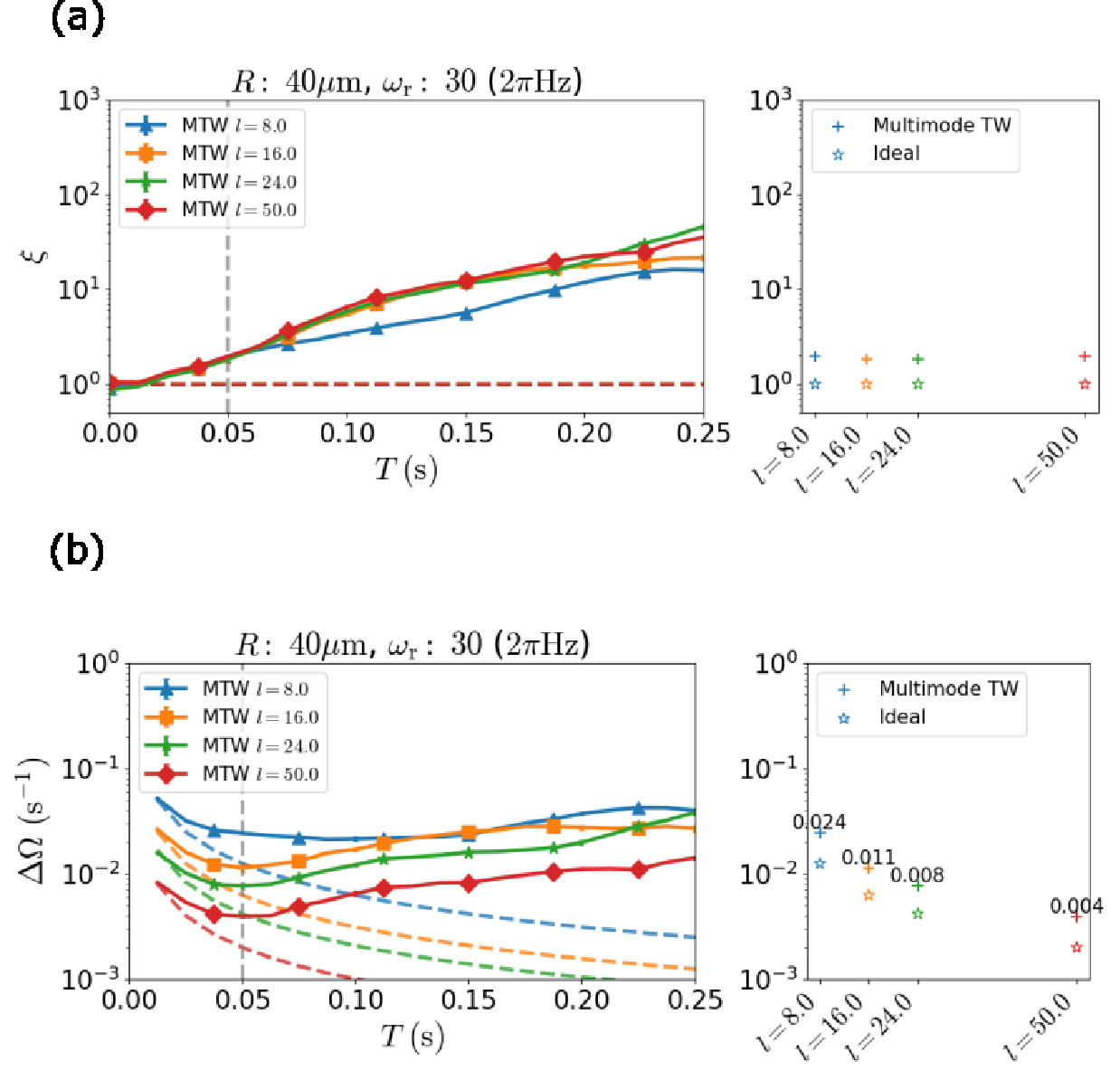}
 \label{fig:sub1}
\caption{(a) Rotation sensitivity $\Delta \Omega$ as a function of interrogation time $T$ (left) and the corresponding minimum at $T=0.05s$ (right) for Four Wave Mixing for ($l=8$, $16$, $24$ and $50$) and the set of trapping parameters ($R=40 \mu$m, $\omega_r/2\pi = 30$ Hz). (b) Corresponding degradation factors. The non-interacting ideal case is also plotted for each (dashed lines). The error bar plot corresponds to the multimode TW results with error given by standard error in the mean from 960 sampled trajectories. Plotted on a log scale for $\Delta \Omega$. 
}
\label{fig:FWMresultmain}
\end{figure*}

\section{Scattering regime}
\label{section:FWM}
To understand the process of FWM that occurs due to the cross scattering between the components, we will look at the extreme case where we have all equal scattering lengths ($\chi_{aa}=\chi_{bb}=\chi_{ab}$), eliminating any dynamics caused by phase diffusion. We will use $a_{aa}=100.4a_0$ for all scattering lengths in this case, which corresponds to the hyperfine state $\vert F=1,m_f=0 \rangle \equiv \vert a\rangle$ in $^{87}$Rb atoms \cite{Mertes2007}. In fig. \ref{fig:FWMresultmain}, we show the sensitivity in the FWM regime for different charges ($l=8,16,24,50$) as a function of interrogation time $T$ for the set of trapping parameters ($R=40 \mu$m, $\omega_r/2\pi = 30$ Hz). We show the corresponding minimum sensitivity achieved to the right, as compared with the standard quantum limit for non interacting atoms. 

To begin to understand these results we will look at a simple few-mode model to analytically investigate the process.

\subsection{Four-mode model}
In order to accurately model the case where we have cross scattering leading to four wave mixing, we will need to consider both the target modes and the modes that they scatter into. Our target modes are $\hat a_l$ and $\hat b_{-l}$ that ideally form the two counter-rotating interferometer arms. Conservation of momentum means that our scatter modes for the two components will come in pairs of equal and opposite angular momentum $\pm \kappa$. Thus we can just consider four modes at a time, the two target modes and the two scatter modes and scan over all possible scatter mode pairs to gain a full picture of the behavior.
In so doing we will neglect the scattering from the scatter modes into other scatter modes (essentially decoupling the scattering modes from each other). This neglect will be valid in the short time limit in which the interferometer would ideally operate.  Neglecting this will enable us to develop an analytical understanding of the underlying processes.

The approximation to the field operators in a four-mode model is given by
\begin{subequations}
\begin{align}
    \hat \Psi_a ({\bf r}) &\approx\frac{1}{2\sqrt{ \pi}} \left( e^{il\theta} \hat a_l +   e^{i\kappa\theta} \hat a_\kappa\right) \phi_a({\bf r}), \\
    \hat \Psi_b ({\bf r}) &\approx \frac{1}{2\sqrt{ \pi}}\left( e^{-il\theta} \hat b_{-l} +   e^{-i\kappa\theta} \hat b_{-\kappa}\right) \phi_b({\bf r}),
\end{align}
\end{subequations}
where $\pm l$ are our target modes and $\pm \kappa$ are the scatter modes (note that we will account for all possible $\kappa$ in the angular momentum mode spectrum). The Heisenberg equations of motion for the target modes in the four-mode model are given by
\begin{subequations}
\begin{equation}
    i \frac{d  \hat a_{l}}{dt} = \left[ \omega_l + \tilde \omega_{l}\right]\hat a_l + 2 \chi_{ab}\hat b^\dagger_{-l} \hat b_{-\kappa} \hat a_{\kappa},
\end{equation}
and
\begin{equation}
    i \frac{d  \hat b_{-l}}{dt} = \left[ \omega_{-l} + \tilde \omega_{-l}\right]\hat b_{-l} + 2 \chi_{ab}\hat a^\dagger_{l} \hat b_{-\kappa} \hat a_{\kappa},
\end{equation}
\end{subequations}
and similarly for the scatter modes. The rotation term from the linear Hamiltonian is given by
$\omega_l = \hbar l^2/(2mR^2)$, and the rotation from the interaction Hamiltonian is given by 
\begin{equation}
    \tilde \omega_{l} = 2 \chi_{aa} \left(  \hat N_{l} + 2\hat N_{\kappa}  \right) + 2 \chi_{ab}\left(\hat N_{{-l}}  +  \hat N_{{-\kappa}} \right),
\end{equation}
where $N_\kappa = \hat a^\dagger_\kappa\hat a_\kappa$, $N_{-\kappa} = \hat b^\dagger_{-\kappa}\hat b_{-\kappa}$. 
Note that we have ignored any radial effects here by assuming perfect spatial overlap of modes.

\subsubsection{Undepleted pump approximation}
Typically the solution to the four-wave mixing employs an undepleted pump approximation, where the target modes are assumed to be undepleted by the dynamics \cite{Haine:2011}. This is only valid for short times because the target modes are finite in size and will be depleted over time, seeing an effect from the interaction with the scatter modes. We will employ the undepleted pump approximation here in order to produce an analytical solution and compare with a numerical simulation of truncated Wigner equations for the four-mode model. We additionally account for the phase evolution resulting from the kinetic term and the self interaction term in the equations of motion for the target modes, producing the following approximation
\begin{subequations}
\begin{align}
    \hat a_{l} (t) &\rightarrow (\langle \hat N_l(0) \rangle)^{1/2} e^{-i(\omega_l + \tilde \omega_l) t}, \\
    \hat b_{-l} (t) &\rightarrow (\langle \hat N_{-l}(0) \rangle)^{1/2} e^{-i(\omega_{-l}+\tilde \omega_{-l}) t} ,
\end{align}
\end{subequations}
where $\langle \hat N_l(0) \rangle$ is the initial number of atoms in the $l$ mode (pump mode). We make an additional assumption that because we assume an undepleted pump, the imbalance of $ \hat N_{a_l}  +  2\hat N_{a_\kappa} $ in the equation of motion is assumed to stay constant (ie. $\tilde \omega_{l}$ is a constant). 

Using the above approximation and moving to a rotating frame $\tilde a_{\kappa}(t) = \exp(+i (\omega_{l} +\tilde \omega_{l}) t) \hat a_{\kappa} (t)$ and similarly for $\tilde b_{-\kappa}(t)$, the equations of motion for the scatter modes come in the pair
\begin{subequations}
\begin{align}
     \frac{d \tilde a_{\kappa}}{dt} &= -i \left[\Delta \lambda \tilde a_{\kappa}+\tilde\chi \tilde b_{-\kappa}^\dagger \right], \\
     \frac{d \tilde b^\dagger_{-\kappa}}{dt} &= i \left[\Delta \lambda \tilde b^\dagger_{-\kappa}+\tilde\chi \tilde a_{\kappa} \right],
\end{align}
\end{subequations}
where we define $\tilde \chi = 2 \chi_{ab} (\langle N_{l}(0) \rangle)^{1/2} (\langle N_{-l}(0) \rangle)^{1/2}$, and $\Delta \lambda = \Delta \omega + \Delta \tilde\omega$, and we define the detunings to be $ \Delta \tilde\omega = \tilde \omega_\kappa - \tilde \omega_l$, and $\Delta \omega = \omega_\kappa - \omega_l$. The general solution to this is given by
\begin{subequations}
\begin{align}
    \tilde a_\kappa (t) &= d(t) \tilde a_\kappa(0) + e(t) \tilde b^\dagger_{-\kappa}(0), \\
    \tilde b_{-\kappa} (t) &= f(t) \tilde b_{-\kappa}(0) + g(t) \tilde a^\dagger_{\kappa}(0),
\end{align}
\end{subequations}
where the time-dependent coefficients are given by 
\begin{subequations}
\begin{align}
     d(t)=f(t) &= \cosh\left(\frac{\sqrt{\nu}t}{2}\right) - \frac{2i\Delta \lambda}{\sqrt{\nu}}\sinh\left(\frac{\sqrt{\nu}t}{2}\right)  , \\
     e(t)=g(t) &= - \frac{2i\tilde \chi}{\sqrt{\nu}}\sinh\left(\frac{\sqrt{\nu}t}{2}\right) ,
\end{align}
\end{subequations}
and the parameter $\nu =4\Delta \lambda ^2 - 4\tilde \chi^2$, rather importantly determines the regime of the solution as we will discuss.
 
\subsubsection{Analysis of the scatter mode population dynamics (the regimes of $\nu$)}
\begin{figure}
\centering
  \includegraphics[width=.99\linewidth]{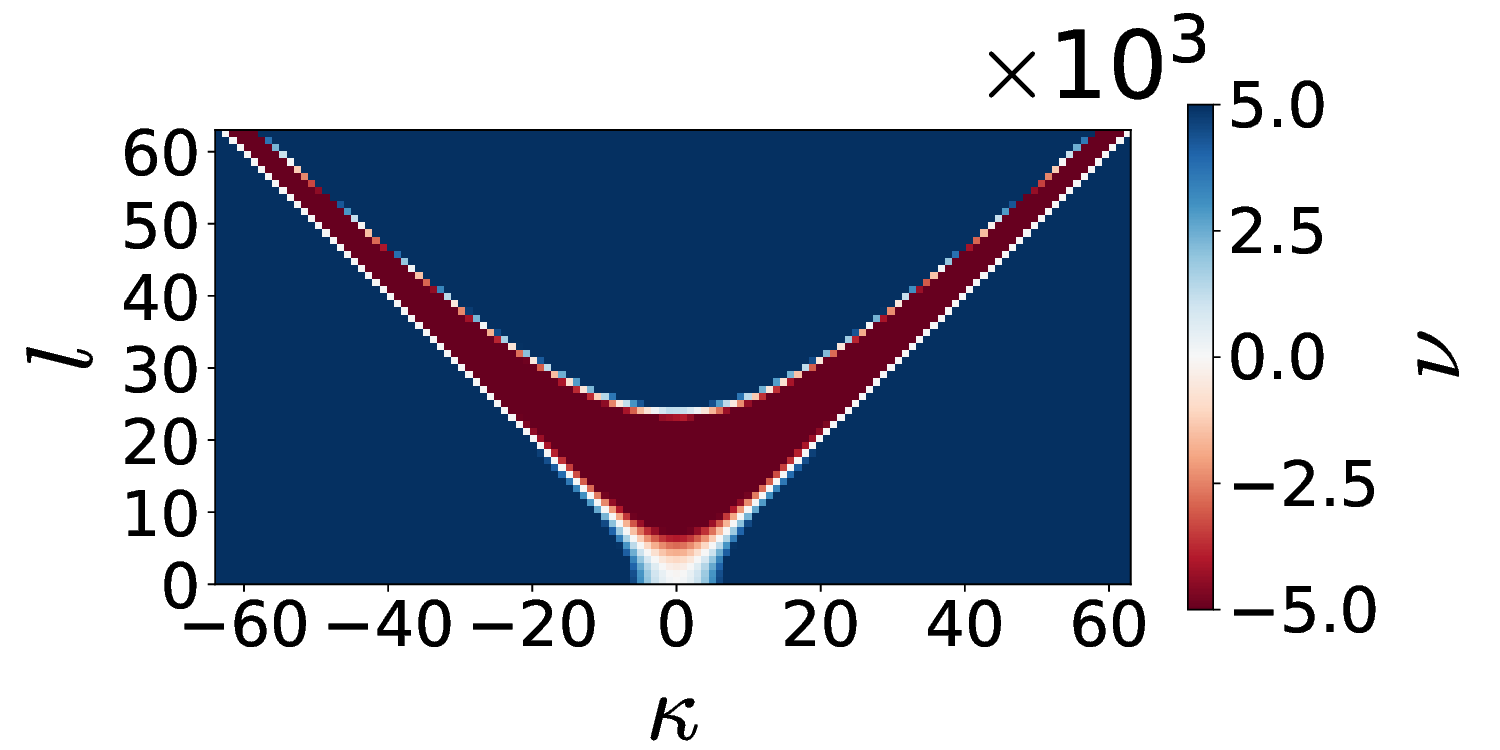}
  
  \label{fig:sub1}
\caption{The detuning parameter $\nu$ which determines the solution regime for the mode population of the four-mode analytics. The red region shows when $\sqrt{\nu}$ is purely imaginary, while the blue region shows when it is real. We see an increased population growth within the red region and a damped out growth in the blue region. }
\label{fig:nutransition1}
\end{figure}
There are three regimes of importance for this solution that depend on this detuning parameter (and primarily is determined by the parameter $\Delta \omega$), that is when $\sqrt{\nu}$ is real or imaginary. The population growth of the scatter modes is either sinusoidal (real regime),  exponential (imaginary regime) or zero. 
In the imaginary regime where $\Delta \lambda^2 < \tilde \chi^2$, the population of the scatter mode for component $a$ is given by
\begin{equation}
N_\kappa(t) =\left(\frac{4\tilde \chi^2}{\eta^2} \right)\sinh^2\left(\frac{\eta t}{2}\right),
\end{equation}
and similarly for component $b$, where $\eta =\sqrt{-\nu}$. In the real regime where $\Delta \lambda^2 > \tilde \chi^2$, the population is
\begin{equation}
N_\kappa(t) = \left(\frac{4\tilde \chi^2}{\nu} \right)\sin^2\left(\frac{\sqrt{\nu} t}{2}\right).
\end{equation}
The two regimes meet when $\sqrt{\nu}=0$. Taking the limit $\nu \rightarrow 0$ from above and below, we find the population here is $N_\kappa(t) = (\tilde \chi^2)t^2/4$. Depending on this parameter, the population growth of the scatter modes will either be exponential, sinusoidal or quadratic in time.

We can gain a rough comprehension of where these regimes reside for our choice of scatter modes by assuming that $\langle \hat N_{a_l} \rangle = N/2$, and $\langle \hat N_{a_\kappa} \rangle = 0$, such that
\begin{equation}
    \nu = \left(\Delta \omega +  \chi_{aa}N \right)^2 - \left( \chi_{ab} N\right)^2,
\end{equation}
which for the case $\chi_{aa} = \chi_{ab}$, will be negative when $\Delta \omega <0$, and therefore when $\kappa < l$. Figure~\ref{fig:nutransition1} illustrates this, showing the transition for parameter $\nu$ as we scan over target modes and scatter modes. The boundary between the two regimes will occur when $\kappa = -l$.

\begin{figure}
\centering
  \includegraphics[width=.90\linewidth]{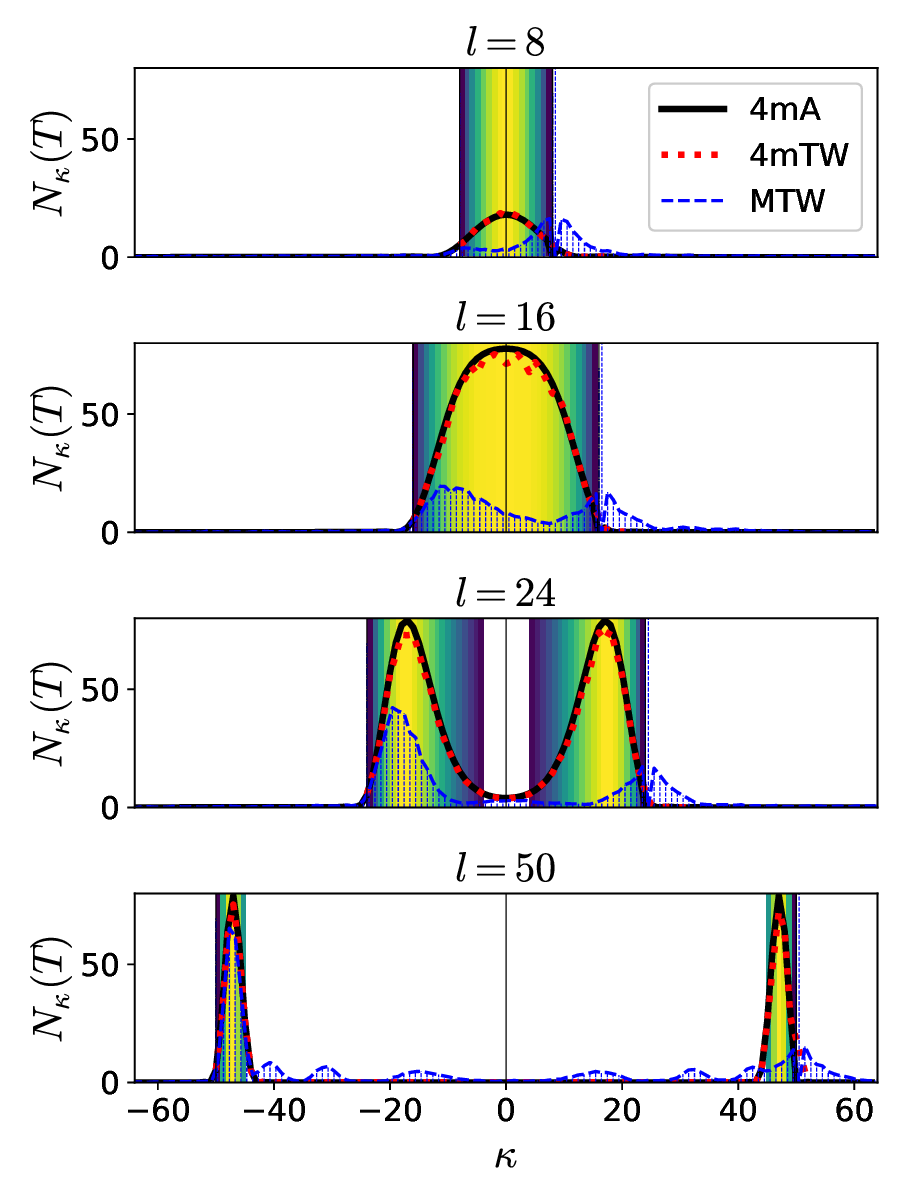}
  \label{fig:sub1}
\caption{Angular momentum population for the component $\vert a\rangle$ scatter modes at an interrogation time $T=50$ms for the multimode TW simulation (blue dashed line, shaded beneath), the numerical four-mode TW (red dotted line), and the four-mode analytics (black solid line) for $l=8,16,24,50$ and trapping parameters \{$R=40 \mu$m, $\omega_r/2\pi = 30$Hz\}. The vertical lines mark the $\pm l$ modes. The shaded gradient region indicates where the critical parameter $\nu(t)<0$ in the numerical four-mode model at this interrogation time, and gives the normalized magnitude of the parameter (max is lightest, min is darkest). Component $\vert b\rangle$ gives the mirrored version of this. 
}
\label{fig:fwmpop}
\end{figure}

The scatter modes which have an angular momentum number smaller in magnitude than the target modes will see enhanced population growth. Scatter modes with angular momentum number larger in magnitude than the target modes will be suppressed from growing. We see this in Fig.~\ref{fig:fwmpop} which shows the population growth of the scatter modes after a short time ($T=50$ms) for a fixed choice of four different charges ($l=8,16,24,50$). Note how the peak population growth is not the mode $\kappa=-l$, as one might naively expect, but is offset from this. The resonant mode gives quadratic growth with time. In the plot, the red full line indicates the target mode, whereas the blue dashed lines indicate the boundary for $\nu =0$. The shaded gradient region indicates the imaginary regime for $\nu$ where we see the largest growth of the scatter modes.

\subsubsection{Degradation in the four-mode model}
We now investigate how this scattering causes degradation in the four-mode model. According to Eq. (\ref{eq:degradation}), we will need the expectation value, variance and covariance of the pseudo spin operators $\hat J_x$ and $\hat J_y$. 

For the four-mode model we insert our ansatz for the field operators into the multimode definition of the pseudo spin operators and we make the gaussian ansatz for the condensate (for the radial mode).

Due to conservation of angular momentum, the expectation and variance of the pseudo spin operators simplify down to the following neat expressions once we integrate over the spatial dimensions (assuming perfect overlap of the components):
\begin{align}
\langle \hat{J}_x \rangle = \langle \hat{j}_x\rangle, \quad \langle \hat{J}_y \rangle = \langle \hat{j}_y\rangle ,
\end{align}
with
\begin{subequations}
\begin{align}
     \hat{j}_x\  &= \frac{1}{2} (\hat a_l^\dagger \hat b_{-l} +\hat b_{-l}^\dagger \hat a_l), \\
      \hat{j}_y &= \frac{-i}{2} (\hat a_l^\dagger \hat b_{-l} -\hat b_{-l}^\dagger \hat a_l).
\end{align}
\end{subequations}
\begin{figure}
\centering
  \includegraphics[width=.90\linewidth]{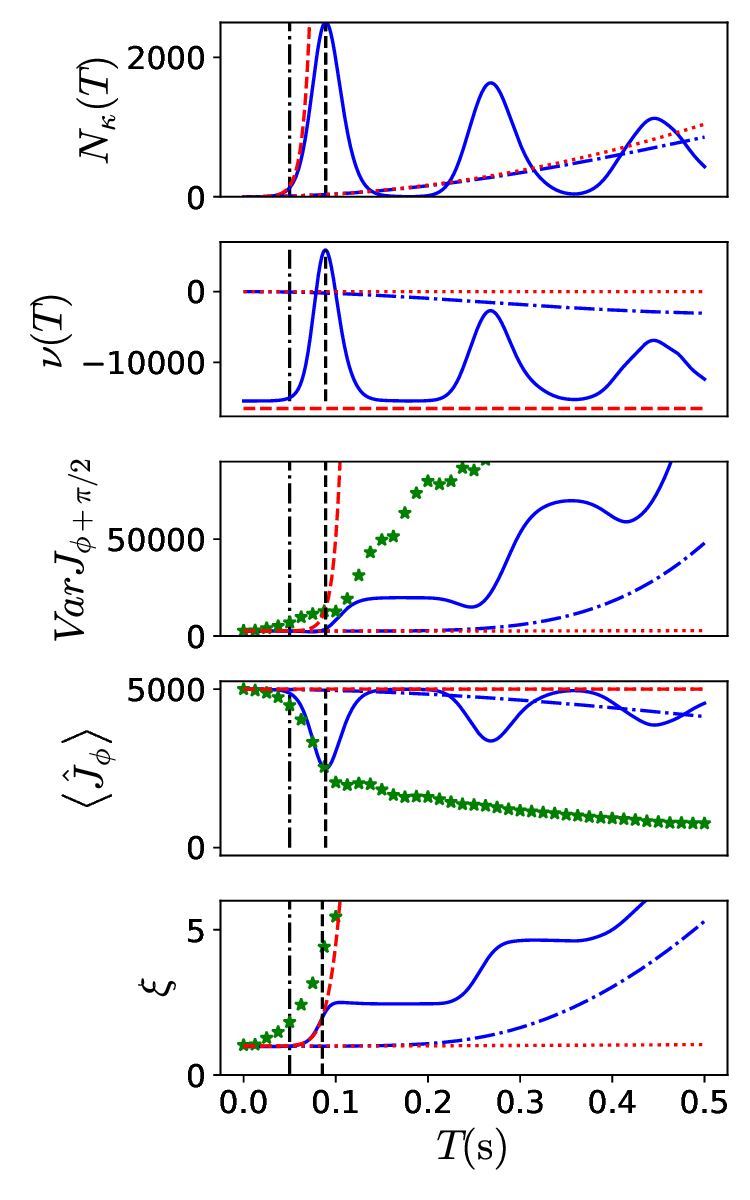}
  \label{fig:sub1}
\caption{Dynamics in the four-mode model (for $l=16$) over the interrogation time for the scatter mode population growth $N_\kappa(T)$, the solution parameter $\nu(T)$, variance $\mathrm{Var}( J_{\phi+\pi/2}(T))$, expectation value $\langle \hat J_\phi (T)\rangle$, and the degradation factor $\xi(T)$. The blue solid (dash-dotted) line gives the fastest initially growing $\kappa = -l+5$ (on resonance $\kappa = -l$) scatter mode for the numerical four-mode model, while the red dashed (dotted) lines give the same for the analytical solution. The MTW data is given by the green star points for comparison. The dash-dotted vertical black line gives the optimal time based on the MTW data, whereas the dashed black line gives the turning point of the population dynamics for $\kappa = -l+5$ in the numerical four-mode model. Trapping parameters are \{$R=40 \mu$m, $\omega_r/2\pi = 30$Hz\}.}
\label{fig:numericalfourmodeanalysis}
\end{figure}
Similarly
\begin{equation}
    \langle \hat J_z \rangle = \frac{1}{2} \langle (\hat a_l^\dagger\hat a_l + \hat a_\kappa^\dagger\hat a_\kappa - \hat b_{-l}^\dagger \hat b_{-l} - \hat b_{-\kappa}^\dagger \hat b_{-\kappa})\rangle.
\end{equation}
Likewise the variance is simply given by
\begin{subequations}
\begin{align}
 \mathrm{Var}(\hat{J}_x) &= \mathrm{Var}(\hat{j}_x) + \frac{1}{4}\langle \hat N_\kappa \rangle + \frac{1}{4}\langle \hat N_{-\kappa} \rangle \\
 \mathrm{Var}(\hat{J}_y) &= \mathrm{Var}(\hat{j}_y) +\frac{1}{4} \langle \hat N_\kappa \rangle + \frac{1}{4}\langle \hat N_{-\kappa} \rangle \\
 \mathrm{CoVar}(\hat{J}_x\hat{J}_y) &= \mathrm{CoVar}(\hat{j}_x\hat{j}_y) \, .
\end{align}
\end{subequations}
The additional terms $\frac{1}{4}\langle \hat N_\kappa \rangle + \frac{1}{4}\langle \hat N_{-\kappa} \rangle$ simply ensure the conservation of atom number, since they complement the terms $\frac{1}{4}\langle \hat N_l \rangle + \frac{1}{4}\langle \hat N_{-l} \rangle$ that appear in $\mathrm{Var}(\hat j_x)$ and $\mathrm{Var}(\hat j_y)$ when we normally order the operators. The variance will be the same for the full multimode system, with the additional terms being instead given by the sum of the number of atoms in all of the other modes $\sum_s( \frac{1}{4}\langle \hat N_{a,s} \rangle + \frac{1}{4}\langle \hat N_{b,-s} \rangle)$.
Together these terms are a constant of the evolution due to the atom number conservation. Therefore the evolution of the pseudo spin operators $\hat J_y$, $\hat J_x$ will entirely depend on the effect the scatter modes have on the mode operators for the target modes $\hat a_l(t)$ and $\hat b_{-l}(t)$.
As a result, the effect will not be captured at all by the undepleted pump approximation, where we assume the target modes are unaffected by the scatter modes and where number conservation is not accounted for. This means that the undepleted pump solution will see the variance grow in the regime with negative solution parameter $\nu<0$, directly proportional to the growth of the scatter modes. But this is merely an artifact and does nothing more than signify the breakdown of the model. The degradation caused by the four-wave mixing occurs later than this time and to understand why it occurs, we must consider the numerical four-mode model. 

\subsubsection{Comparison with numerical four-mode TW}
We now compare these results to the numerical four-mode problem solving the TWEs (see equations in the appendix section \ref{appendix:4mTWE}). In this case, we are accounting for the nonlinear coupling between the target modes and scatter modes, and as a result we have conservation of atom number (albeit only between the four modes, neglecting all other modes). We once again consider the individual cases for all possible target/scatter mode pairs. 

We show the mode population for the numerical four-mode model at $T=50$ms in Fig. \ref{fig:fwmpop}. 
At a short interrogation time $T=50$ms, the numerical four-mode model is in agreement with the analytical expression obtained from the undepleted pump approximation, giving the same population distribution in the scatter modes. 
However, the nonlinear effects kick in after this time, resulting in different dynamical behavior. Figure \ref{fig:numericalfourmodeanalysis} shows the dynamics in the four-mode model (for $l=16$) over the interrogation time for the scatter mode population growth $N_\kappa$, the solution parameter $\nu$, variance $\mathrm{Var}( J_{\phi+\pi/2})$, expectation value $\langle \hat J_\phi \rangle$, and the degradation factor $\xi$. This plot shows both the numerical results (blue solid and dash-dotted lines), and the undepleted pump approximation (red dashed and dotted lines) for both the resonant mode ($\kappa = -l$), and the fastest initially growing mode (in this case $\kappa = -l+5$). The plot also shows the MTW for comparison with the green starred plot.
For the numerical case we estimate the $\nu$ parameter by assuming $\tilde \chi \approx 2 \chi_{ab} \sqrt{N_l(T)N_{-l}(T)}$. We see in the numerical version that (like the analytical version) the solution parameter $\nu(T)$ is still vital for determining the rate of population growth of the scatter modes, however this parameter is now time dependent and in fact changes sign for some of the scatter modes with the greatest growth. This occurs before the turning point in the scatter mode growth (see the dashed black line in the figure). The nonlinear nature of the equations works to stabilize the population growth as a result of the conservation of atom number. For comparison, the dot-dash black line gives the optimum sensitivity achieved in the full multimode simulations. After this point the degradation that results from the four-mode scattering effects start to kick in.
 Prior to this point the degradation is essentially non-existent, and the four-mode numerics follows the non-interacting ideal case.

\begin{figure}
\centering
  \includegraphics[width=.80\linewidth]{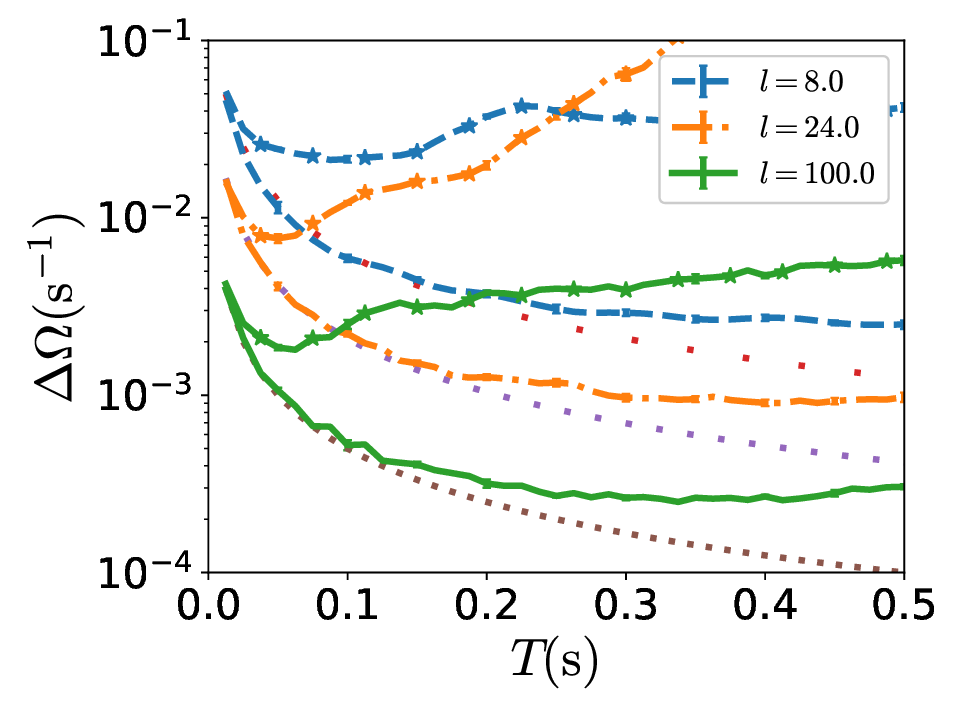}
  \label{fig:sub1}
\caption{Rotation sensitivity for the FWM $l=8,24,100$ case for multimode TW (MTW) for trap parameters $R=200 \mu $m, $\omega_r/2\pi = 10$ Hz (unmarked lines), compared with the parameters $R=40 \mu$m, $\omega_r/2\pi = 30$ Hz (starred lines). The ideal, non-interacting case is also plotted (dotted lines) for each.
}
\label{fig:sensIdealfwmlimit}
\end{figure}

\begin{figure*}
\centering
  \includegraphics[width=.7\linewidth]{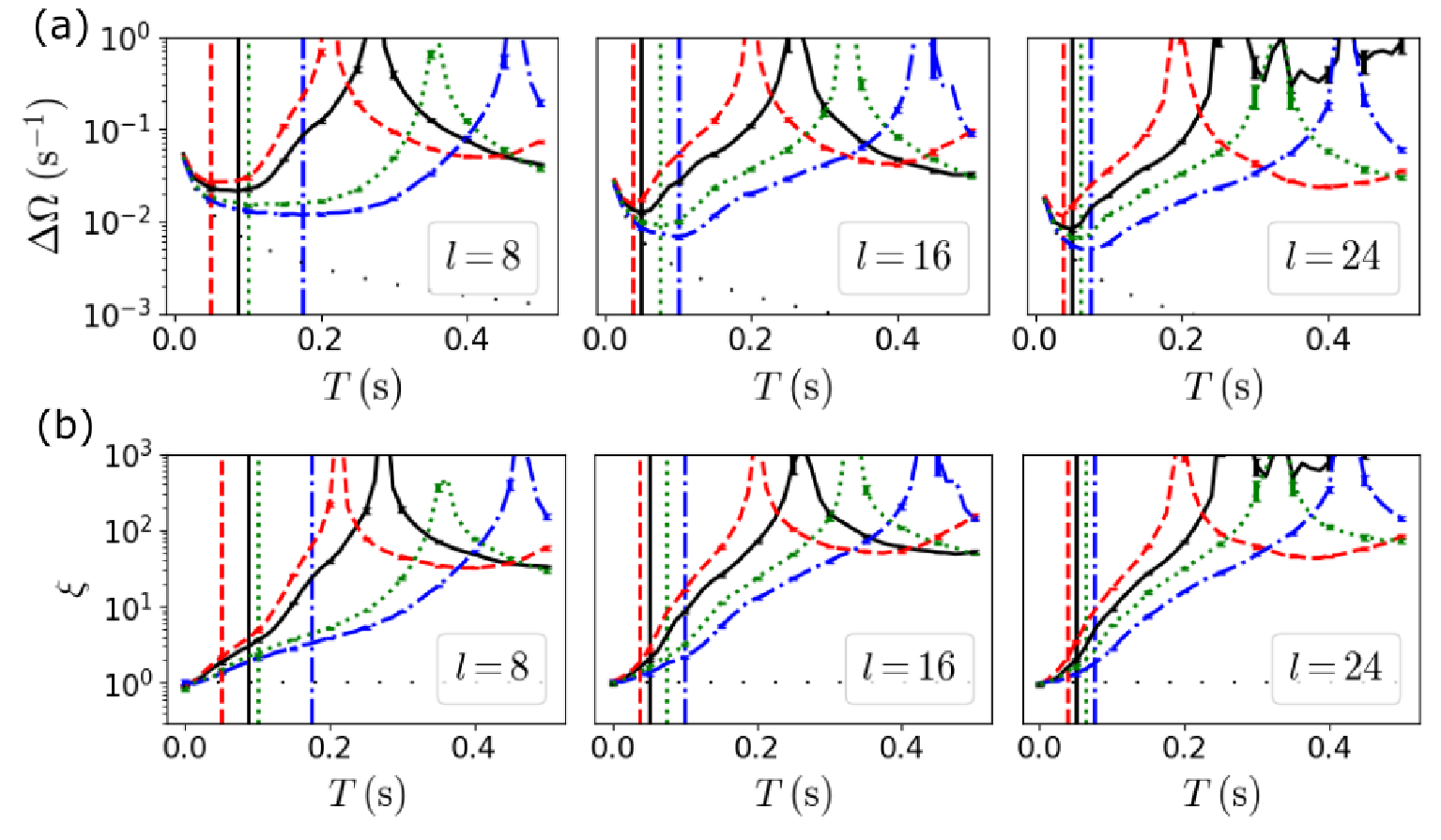}
  \label{fig:sub1}
\caption{(a) Sensitivity and (b) degradation factor vs interrogation time (multimode TW $^{87}$Rb atoms: $a_{aa} = 100.4a_0$, $a_{bb}= 95.0a_0$, $a_{ab} = 97.66a_0$) for different charges (left to right plots) and different trapping parameters: \{$R=40 \mu$ m, $\omega_r/2\pi = 30$ Hz\} (Solid black line), \{$R=40 \mu$ m, $\omega_r/2\pi = 50$ Hz)\} (red dashed line) ,  \{$R=70 \mu$ m, $\omega_r/2\pi = 30$ Hz\} (blue dash dotted line), and  \{$R=70 \mu$ m, $\omega_r/2\pi = 50$ Hz\} (green dotted line). 
Error bars (plotted every few data points) give the standard error in the mean ensemble averaged over 480  trajectories. Vertical lines give the time of minimum sensitivity. The horizontal line gives the ideal limit $\xi=1$.
}
\label{fig:badnessALLtrapparams}
\end{figure*}
 
A key result of the four-mode model is that it analytically gives the three different solution regimes for the scatter modes. Importantly, it demonstrates how the modes with the largest growth in the exponential growth rate region contribute the most to the degradation of the sensitivity. We now investigate the full multimode simulations to examine how this model breaks down.

\subsection{Multimode TW results (FWM)}
We now compare the analytical and numerical results obtained from the four-mode model with those from the full MTW simulations, and assess how well they agree. As we did with the four-mode model, we will look at the population of the scatter modes in the full multimode numerical simulations. Figure \ref{fig:fwmpop} shows the population of the angular momentum modes for the full MTW simulations for several charges ($l=8,16,24,50$) at an interrogation time ($T=50$ms). 
The gradient shaded region gives the space in which the critical parameter $\nu(t)$ is negative for the four-mode TWEs with the gradient defined as the normalized magnitude of the parameter (max is lightest, min is darkest). This corresponds to exponential growth of the scatter modes and the magnitude of that growth. 

The distribution of scatter modes in proximity to the target mode does not reflect the simple analytics, revealing that the dynamics is dominated at small times by radial coupling to the nonlinear self-interacting terms. This is evidenced by similar behavior present in the OAT case (see appendix section \ref{appendix:OATMode} for more details). Further away from the target mode, in proximity to the resonant mode, we do see behavior that mimics the simple analytics. This similarity increases as we go to higher charge.

The formation of smaller intermediate bumps in the MTW scatter mode population can be explained by the coupling of the scatter modes to each other (a feature we neglected in the four-mode model). This coupling results in new regions of exponential growth from the already populated scatter modes. This effect only really comes into play for a larger charge, and longer evolution time. This does mean we will not witness the revival seen in the numerical four-mode model, but instead something more akin to the depletion expected from the undepleted pump approximation (at a slower rate).


The scattering regime demonstrates degradation that kicks in on long time scales but limits phase diffusion due to the balanced scattering lengths, leading to an improved sensitivity over the OAT case initially.
We can optimize the trapping parameters in the same way as done for the OAT case above. Figure \ref{fig:sensIdealfwmlimit} demonstrates the improved sensitivity for the FWM case when the parameter set $R=200 \mu $m, $\omega_r/2\pi = 10$ Hz is used instead. As we saw in the OAT case, the widening of the trap removes the initial degradation caused by the radial coupling to the self-interaction terms. This optimization leads to an incredible improvement, pushing the sensitivity in the FWM case to follow the ideal limit for a large interrogation time. This demonstrates the superiority of operating in the scattering regime over the phase diffusion regime. We can conclude that the population imbalance-energy coupling is the true cause of limitation in the rotation sensor. Effectively removing it and operating in an atomically-diffuse regime can lead to almost ideal behavior for long interrogation times.

\section{Results for $^{87}\rm{Rb}$ ($\chi_{aa} \neq \chi_{bb} \neq \chi_{ab}$), Phase diffusion and scattering combined }
\label{section:87Rb}
Having characterized the degradation of each effect in isolation, we now move on to exploring the resulting degradation in a set up of $^{87}$Rb atoms that would be the perfect candidate for an experimental realization of the scheme.
In this section we use the experimentally-measured scattering lengths $a_{aa} = 100.4a_0$, $a_{bb}= 95.0a_0$, $a_{ab} = 97.66a_0$ for the hyperfine states $\vert F=1,m_f=0 \rangle \equiv \vert a\rangle$ and $\vert F=2,m_f=0 \rangle \equiv \vert b\rangle$ in a BEC of $^{87}$Rb atoms \cite{Mertes2007}. In this case we will have both phase diffusion and four-wave mixing mechanisms at play within the evolution of the quantum system.

\begin{figure}
\centering
  \includegraphics[width=.99\linewidth]{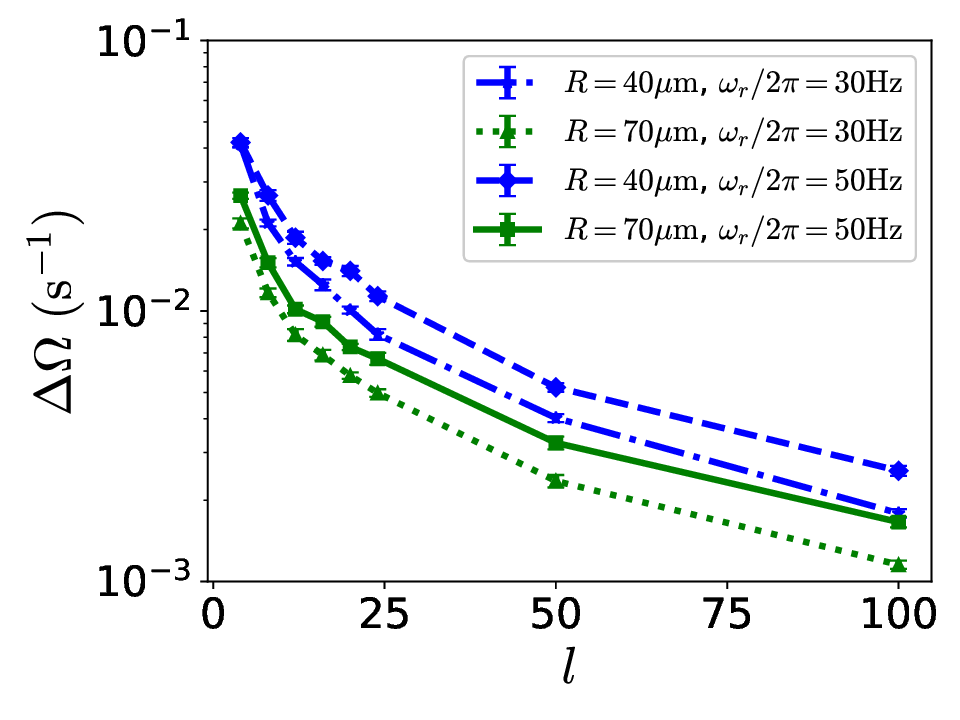}
  
  \label{fig:sub1}

\caption{Minimum sensitivity for $^{87}$Rb atoms ($a_{aa} = 100.4a_0$, $a_{bb}= 95.0a_0$, $a_{ab} = 97.66a_0$) for two different ring radii and radial trapping frequency. 
These results are constructed from 480 trajectories and error bars are standard error in mean. 
}
\label{fig:allresult}
\end{figure}
\begin{figure*}
\centering
  \includegraphics[width=.90\linewidth]{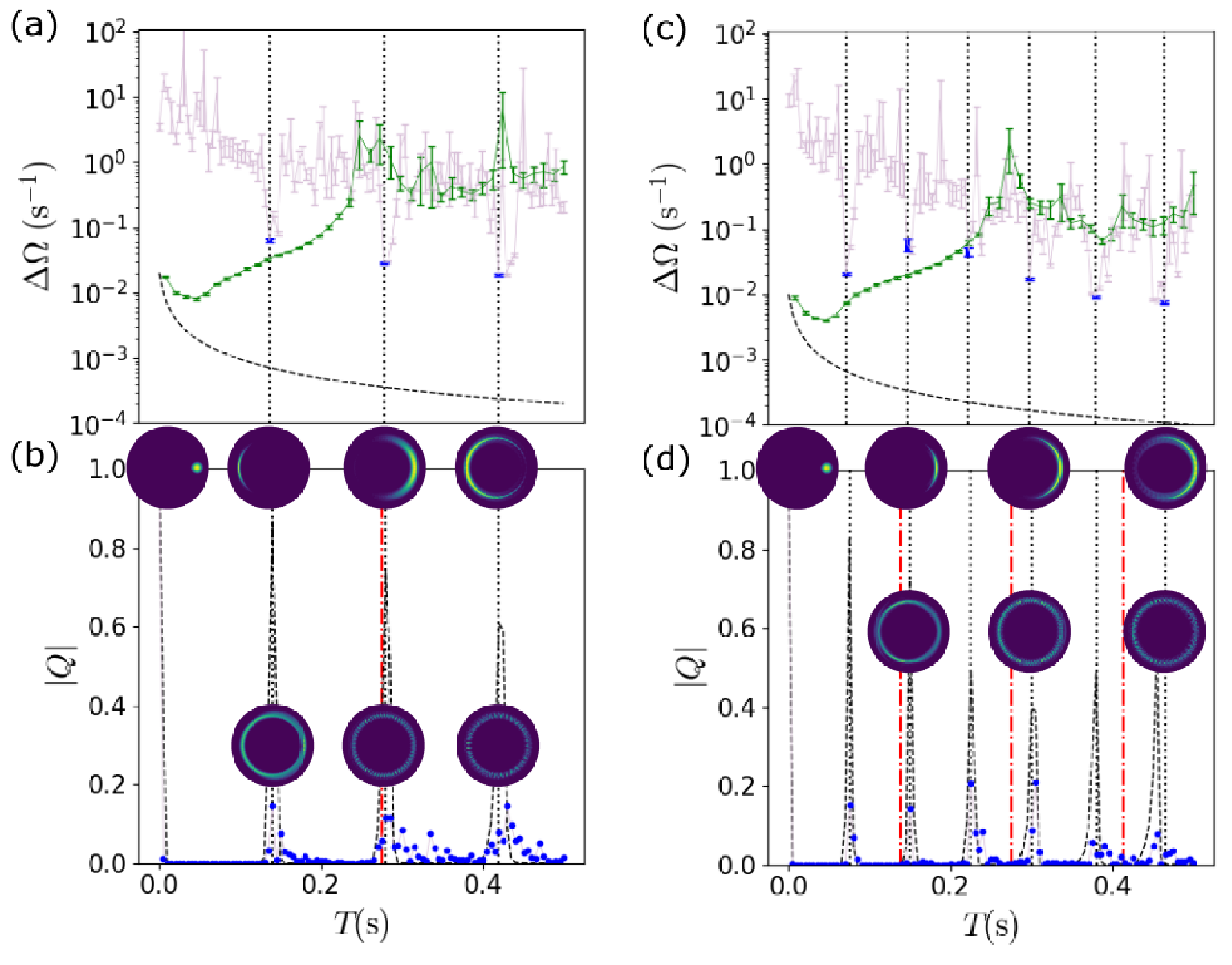}
  \label{fig:sub1}
\caption{(a) Sensitivity for the guided Sagnac interferometer ($a_{aa} = 100.4a_0$, $a_{bb}= 95.0a_0$, $a_{ab} = 97.66a_0$) as function of interrogation time (up to $0.5$s) for $l=50$ for a ring radius of $R=40\mu$m and radial trapping frequency of $\omega_r/2\pi=30  $Hz. The black dashed line marks the ideal non-interacting sensitivity. The highlighted blue points mark the point where the components are overlapping. The green error bar plot shows the VG for comparison for the same $l$. (b) The corresponding relative overlap of the two components $\vert Q\vert$ in the non-interacting GPE simulation (black dashed) and interacting GPE (blue dots). The red dot-dashed line marks the analytical period of orbit for the center of the wavepacket. The top line of density plots shows the non interacting GPE evolution at $t=0,0.14,0.28,0.42$s, the bottom line shows the density for the interacting GPE. In this case the atoms have completed one and a half full orbits of the ring trap, resulting in three points of overlap. These are the points at which we would wish to stop the interrogation and act the second beamsplitter. (c) and (d) show the same thing for $l=100$, with non-interacting GPE snap shots at $t=0,0.15,0.30,0.47$s }
\label{fig:Bromtimeplot100}
\end{figure*}

In Fig. \ref{fig:badnessALLtrapparams} we show the evolution of the degradation for three different charges ($l=8,16,24$) and four sets of trapping parameters ($R=40,70 \mu$m, $\omega_r/2\pi = 30,50$ Hz). The initial growth resulting from the radial coupling can be mitigated somewhat with a larger trap; even for a tighter trap this is found to be better. The vertical lines show the corresponding time at which a minimum sensitivity is achieved. The scattering that can be described by the four-mode model kicks in around the point after we reach the minimum sensitivity. And this can also be alleviated by a larger/ weaker trap.

In Fig.~\ref{fig:allresult}, we show the optimal sensitivities achieved for these trapping parameters as a function of the charge. Though we have both FWM and OAT dynamics present, the FWM dominates the dynamics, due to the nearly identical scattering lengths significantly reducing the magnitude of the effective OAT Hamiltonian.

The scattering lengths can experimentally be altered by tuning the Feshbach resonance \cite{Hanna2010}. If this is done it is important to keep in mind the regime we are aiming for. The goal is to minimize the effective interaction strength $\chi(t)$, to ensure optimal operation of the device. We could think of a worst case scenario, that is, where $a_{aa} = a_{bb} = 0$, and $a_{ab} \neq 0$. In that case, we would have the worst of both worlds, with scattering degrading the sensor in the long time, and phase diffusion limiting the sensor in the short time. Avoiding this limit is recommended. As evidence suggests, the best practice would be to minimize all scattering lengths $a_{aa},a_{bb},a_{ab}$, while also keeping them similar in magnitude. Or alternatively to minimize the effective interaction strengths via the other parameters at play, ie. weakening the radial trap, increasing the ring radius.

\section{Comparison of the Vortex gyroscope to the guided Sagnac Interferometer}
\label{section:GS}

We now compare the matterwave vortex gyroscope (VG) to a competing approach, namely, the guided Sagnac (GS) interferometer. 
The GS interferometer comprises two localized wavepackets that are counter-propagating around the toroidal trap with $\pm l \hbar$ angular momentum. The second beamsplitter pulse is accomplished when the two wavepackets are overlapping once again. Numerically we simulate a GS interferometer with the TW simulations just as we do for the VG, however in this case we begin by finding the ground state in an harmonic trap located and centered within the ring trap radial width. This state is then allowed to evolve within the ring trap. We split into two counter-propagating components which orbit the ring.
We will use the scattering lengths of $^{87}$Rb atoms just as we did for the VG in the previous section.

Unlike the VG, the components of the GS do not overlap spatially for the entire interrogation period. Therefore, the interferometer will only achieve a sensitive measure of rotation when the magnitude of the mode overlap 
\begin{equation}
Q  = \langle\int d{\bf r} \hat \Psi^\dagger_a  \hat\Psi_b e^{2il\theta}\rangle,
\end{equation}
is maximum, which occurs when the atoms have completed a half- or full-orbit of the ring and we must take the interrogation time to be at these times. 
We can express the magnitude of the mode overlap in terms of the pseudo spin operators
\begin{equation}
\vert Q \vert = \sqrt{\langle\hat J_x\rangle^2 + \langle\hat J_y\rangle^2}.
\end{equation}
The period for a classical particle in a ring is 
\begin{equation}
T_\mathrm{period} = \frac{2\pi R^2 m}{\hbar l}.
\end{equation}
To obtain several half orbits in the allotted simulation time of $500$ms (for fair comparison with the VG), a charge of $l=20,24,50,100$ is used. A consequence of this overlap constraint in time is that increasing the ring radius to a more diffuse regime will change the period time and will not trivially improve sensitivity.

In Fig. \ref{fig:Bromtimeplot100}, we show the sensitivity (in the interacting TW sims) over the interrogation time for both $l=50,100$ for the trapping parameters $R = 40\mu$m, $\omega_r/2\pi = 30$Hz in the top plot, matched with the overlap of the two components in the interacting and non-interacting GPE. The minimum sensitivity is reached at the points corresponding to the overlap of the two components in the GPE sims. 
There are several reasons why the GS would be a less desirable choice for the rotation sensor. Capturing the exact interrogation time where the two components are overlapping would be finicky and prone to error. Moreover, because the components are not in the ground state of the ring trap, there is undesirable breathing of the components. Unlike the VG, the GS is composed of a broad spread of angular momentum components. Additionally, the phase response of the VG is immune to drifts in the radius of the toroid. This is not the case for the GS, as both time of maximum overlap, and phase accumulated per circuit, depend on this parameter \cite{Haine:2016b}.

\begin{figure*}
\centering
  \includegraphics[width=.80\linewidth]{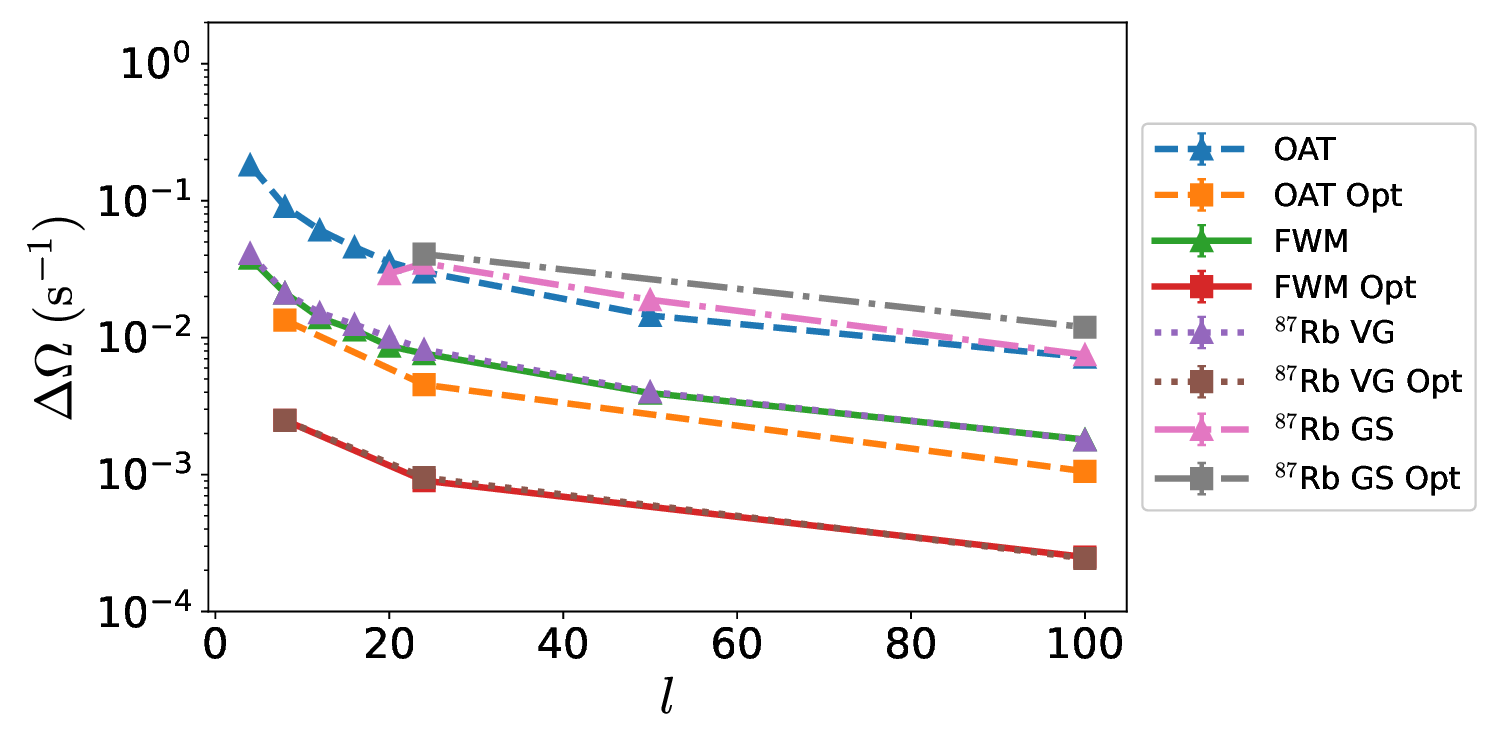}
  \label{fig:sub1}
\caption{Comparison of the optimal sensitivity for the Vortex gyroscope with the Guided Sagnac interferometer ($^{87}$Rb atoms $a_{aa} = 100.4a_0$, $a_{bb}= 95.0a_0$, $a_{ab} = 97.66a_0$), phase diffusion (OAT) ($a_{aa} = 100.4a_0$, $a_{bb}= 100.4a_0$, $a_{ab} = 0$), and scattering (FWM) ($a_{aa} = 100.4a_0$, $a_{bb}= 100.4a_0$, $a_{ab} = 100.4a_0$). Both the Vortex Gyroscope and the Guided Sagnac results are averaged over 480 numerical trajectories. All simulations performed in the ring trap basis. For the set of trapping parameters ($R=40 \mu$m, $\omega_r/2\pi = 30$Hz) and ($R=200 \mu$m, $\omega_r/2\pi = 10$Hz).}
\label{fig:Bromsensopt}
\end{figure*}

In Fig. \ref{fig:Bromsensopt}, we plot the optimal sensitivity achieved in the GS and the VG for a cloud of $^{87}$Rb atoms, compared with the FWM and OAT regimes. The GS sees a regime in which OAT dynamics dominate, whereas the VG sees a regime where the FWM dynamics dominates due to the nearly identical scattering lengths in $^{87}$Rb atoms. It is clear that the VG performs better against atom-atom interactions as a result of the cross scattering that occurs between the two components. Operating in a regime where cross scattering is present, and atomic density is diffuse ($R=200 \mu$m and $\omega_r/2\pi=10$Hz), will lead to optimal performance, pushing operation to nearly ideal behavior for interrogation times up to $150$ms.

\section{Conclusion and Outlook}
The vortex gyroscope has been proposed as a compact rotation sensor, harnessing the topological stability of ultracold atoms in a toroidal trap and the transverse phase profile of OAM carrying laser pulses in order to achieve high rotation sensitivity. Atomic interactions within the device will limit the performance, driving it away from ideal operation.
In this paper we have characterized the degradation to rotation sensitivity in a vortex gyroscope, caused by atom-atom interactions in a two component BEC.
The mechanisms by which the interactions cause degradation to the sensitivity are four-wave mixing and phase diffusion. These two processes have been isolated and studied, respectively, before a realistic experimental setting of $^{87}$Rb atoms was characterized. Evidence suggests that the phase diffusion is a process which limits sensitivity for very short interrogation times, whereas the four-wave mixing leads to degradations that happen on a longer timescale. 

This paper has compared the results for the vortex gyroscope with another proposed device, the guided Sagnac interferometer. As is known, this device does not have the stability to fluctuations that the VG has. We have shown in this paper that the guided Sagnac is also dominated by phase diffusion, whereas the VG is dominated by four-wave mixing, resulting in a more ideal rotation sensor when interactions must be accounted for. Optimizing the trapping parameters is essential for ensuring near ideal behavior of the device.

The best practices will be to operate the device in a regime where four-wave mixing processes dominate, where lower sensitivity can be achieved in a shorter interrogation time. The degradation in this regime can be mitigated almost entirely by increasing the radius of the ring, weakening the trapping potential in the radial direction and making the system more diffuse. It is essential however to ensure that cross scattering is maintained by operating with the two components of the BEC spatially overlapping at all times.
If one wishes to optimize the interactions, by tuning the Feshbach resonance for instance, one should ensure that the scattering lengths all remain on par with each other as they are lowered. Asymmetry in the scattering lengths, particularly the cross scattering compared with the intra-component scattering lengths will render an increased degradation as a result of the phase diffusion.

Simple few-mode analytics have provided essential insight into the operation of the device in the differing regimes. Phase diffusion from OAT can be accurately described by a simple two-mode model, revealing the key parameter for optimization is the effective interaction strength $\chi(t)$. Reducing this is key to optimizing performance in this regime.
Scattering from FWM, does not follow a simple four-mode model with an undepleted pump approximation, due to initial degradation from radial coupling to the self interactions. This is seen identically in the OAT case. However the simple few-mode ansatz in this case does analytically demonstrate the regions of scatter mode population growth once the FWM starts to dominate. 
In our characterization of scattering resulting from four wave mixing, we have shown that several mechanisms are at play. A four-mode model that accounts for interaction between the two target modes and a scatter mode pair downplays the degradation seen in the full multimode case. The model assumes no radial wobble and no cross talk between the scattering modes. Results from the multimode simulations demonstrate the necessity to incorporate these factors into building a reliable model of the device. 
Analytics obtained from assuming an undepleted pump approximation allow us to analytically describe the initial growth of the scatter modes, with results agreeing with the multimode simulations, however it does not allow for a valid expression of the degradation that occurs initially. The four-mode numerical results indicate that this device would follow the standard quantum limit for longer than observed in the multimode results. The coupling in of the radial modes and the coupling between the scattering modes result in the initial degradation in the four wave mixing regime.

Moreover, the results presented here assume a Gaussian ansatz for the $z$ direction, and the four-mode model assumes a Gaussian with perfect overlap for the radial direction. This is a best case scenario for the interaction strength, and something closer to a Thomas Fermi distribution will only lead to worse outcomes. For a full characterization of the degradation, a full 3D simulation is suggested as future work.
As interactions will act to degrade the device precision, operating in a regime of diffuse atoms will provide the best performance. Regardless of the regime, operating without phase diffusion is optimal.
Although the toroidal trap is heralded for its topological stability and persistent currents, the multimode effects will limit the precision of rotation sensors within such a device. Recent work has proposed using superintegrability in a four-well model in a ring configuration for enhanced rotation sensing \cite{Ymai:2026}.

\section{Acknowledgments}
The authors acknowledge Kaiwen Zhu, Zain Mehdi and Joe Hope for stimulating theoretical discussions and Ryan Husband, Ryan Thomas, Sam Legge, John Debs, John Close, Cass Sackett, Mike Larson and Eric Imhof for stimulating experimental discussions.
We acknowledge the Ngunnawal and Ngambri peoples as the traditional custodians of the land on which this research was conducted.
This research was undertaken with the assistance of resources from the National Computational Infrastructure (NCI Australia), an NCRIS enabled capability supported by the Australian Government.
This research was funded by the Australian Research council project No. LP190100621 and the Australian government Department of Industry, Science, and Resources via the Australia-India Strategic Research Fund (AIRXIV000025). S.A.H acknowledges support through an Australian Research Council Future Fellowship, Grant No. FT210100809. S.S.S was supported by an Australian Research Council Discovery Early Career Researcher Award, Project No. DE200100495.

\bibliography{bib}

\begin{thebibliography}{58}%
\makeatletter
\providecommand \@ifxundefined [1]{%
 \@ifx{#1\undefined}
}%
\providecommand \@ifnum [1]{%
 \ifnum #1\expandafter \@firstoftwo
 \else \expandafter \@secondoftwo
 \fi
}%
\providecommand \@ifx [1]{%
 \ifx #1\expandafter \@firstoftwo
 \else \expandafter \@secondoftwo
 \fi
}%
\providecommand \natexlab [1]{#1}%
\providecommand \enquote  [1]{``#1''}%
\providecommand \bibnamefont  [1]{#1}%
\providecommand \bibfnamefont [1]{#1}%
\providecommand \citenamefont [1]{#1}%
\providecommand \href@noop [0]{\@secondoftwo}%
\providecommand \href [0]{\begingroup \@sanitize@url \@href}%
\providecommand \@href[1]{\@@startlink{#1}\@@href}%
\providecommand \@@href[1]{\endgroup#1\@@endlink}%
\providecommand \@sanitize@url [0]{\catcode `\\12\catcode `\$12\catcode `\&12\catcode `\#12\catcode `\^12\catcode `\_12\catcode `\%12\relax}%
\providecommand \@@startlink[1]{}%
\providecommand \@@endlink[0]{}%
\providecommand \url  [0]{\begingroup\@sanitize@url \@url }%
\providecommand \@url [1]{\endgroup\@href {#1}{\urlprefix }}%
\providecommand \urlprefix  [0]{URL }%
\providecommand \Eprint [0]{\href }%
\providecommand \doibase [0]{https://doi.org/}%
\providecommand \selectlanguage [0]{\@gobble}%
\providecommand \bibinfo  [0]{\@secondoftwo}%
\providecommand \bibfield  [0]{\@secondoftwo}%
\providecommand \translation [1]{[#1]}%
\providecommand \BibitemOpen [0]{}%
\providecommand \bibitemStop [0]{}%
\providecommand \bibitemNoStop [0]{.\EOS\space}%
\providecommand \EOS [0]{\spacefactor3000\relax}%
\providecommand \BibitemShut  [1]{\csname bibitem#1\endcsname}%
\let\auto@bib@innerbib\@empty
\bibitem [{\citenamefont {Gustavson}\ \emph {et~al.}(1997)\citenamefont {Gustavson}, \citenamefont {Bouyer},\ and\ \citenamefont {Kasevich}}]{Gustavson:1997}%
  \BibitemOpen
  \bibfield  {author} {\bibinfo {author} {\bibfnamefont {T.~L.}\ \bibnamefont {Gustavson}}, \bibinfo {author} {\bibfnamefont {P.}~\bibnamefont {Bouyer}},\ and\ \bibinfo {author} {\bibfnamefont {M.~A.}\ \bibnamefont {Kasevich}},\ }\bibfield  {title} {\bibinfo {title} {Precision rotation measurements with an atom interferometer gyroscope},\ }\href {https://doi.org/10.1103/PhysRevLett.78.2046} {\bibfield  {journal} {\bibinfo  {journal} {Phys. Rev. Lett.}\ }\textbf {\bibinfo {volume} {78}},\ \bibinfo {pages} {2046} (\bibinfo {year} {1997})}\BibitemShut {NoStop}%
\bibitem [{\citenamefont {Durfee}\ \emph {et~al.}(2006)\citenamefont {Durfee}, \citenamefont {Shaham},\ and\ \citenamefont {Kasevich}}]{Durfee:2006}%
  \BibitemOpen
  \bibfield  {author} {\bibinfo {author} {\bibfnamefont {D.~S.}\ \bibnamefont {Durfee}}, \bibinfo {author} {\bibfnamefont {Y.~K.}\ \bibnamefont {Shaham}},\ and\ \bibinfo {author} {\bibfnamefont {M.~A.}\ \bibnamefont {Kasevich}},\ }\bibfield  {title} {\bibinfo {title} {Long-term stability of an area-reversible atom-interferometer sagnac gyroscope},\ }\href@noop {} {\bibfield  {journal} {\bibinfo  {journal} {Phys. Rev. Lett.}\ }\textbf {\bibinfo {volume} {97}},\ \bibinfo {pages} {240801} (\bibinfo {year} {2006})}\BibitemShut {NoStop}%
\bibitem [{\citenamefont {Dutta}\ \emph {et~al.}(2016)\citenamefont {Dutta}, \citenamefont {Savoie}, \citenamefont {Fang}, \citenamefont {Venon}, \citenamefont {Garrido~Alzar}, \citenamefont {Geiger},\ and\ \citenamefont {Landragin}}]{Dutta:2016}%
  \BibitemOpen
  \bibfield  {author} {\bibinfo {author} {\bibfnamefont {I.}~\bibnamefont {Dutta}}, \bibinfo {author} {\bibfnamefont {D.}~\bibnamefont {Savoie}}, \bibinfo {author} {\bibfnamefont {B.}~\bibnamefont {Fang}}, \bibinfo {author} {\bibfnamefont {B.}~\bibnamefont {Venon}}, \bibinfo {author} {\bibfnamefont {C.~L.}\ \bibnamefont {Garrido~Alzar}}, \bibinfo {author} {\bibfnamefont {R.}~\bibnamefont {Geiger}},\ and\ \bibinfo {author} {\bibfnamefont {A.}~\bibnamefont {Landragin}},\ }\bibfield  {title} {\bibinfo {title} {Continuous cold-atom inertial sensor with $1\text{ }\text{ }\mathrm{nrad}/\mathrm{sec}$ rotation stability},\ }\href@noop {} {\bibfield  {journal} {\bibinfo  {journal} {Phys. Rev. Lett.}\ }\textbf {\bibinfo {volume} {116}},\ \bibinfo {pages} {183003} (\bibinfo {year} {2016})}\BibitemShut {NoStop}%
\bibitem [{\citenamefont {Gautier}\ \emph {et~al.}(2022)\citenamefont {Gautier}, \citenamefont {Guessoum}, \citenamefont {Sidorenkov}, \citenamefont {Bouton}, \citenamefont {Landragin},\ and\ \citenamefont {Geiger}}]{Gautier:2022}%
  \BibitemOpen
  \bibfield  {author} {\bibinfo {author} {\bibfnamefont {R.}~\bibnamefont {Gautier}}, \bibinfo {author} {\bibfnamefont {M.}~\bibnamefont {Guessoum}}, \bibinfo {author} {\bibfnamefont {L.~A.}\ \bibnamefont {Sidorenkov}}, \bibinfo {author} {\bibfnamefont {Q.}~\bibnamefont {Bouton}}, \bibinfo {author} {\bibfnamefont {A.}~\bibnamefont {Landragin}},\ and\ \bibinfo {author} {\bibfnamefont {R.}~\bibnamefont {Geiger}},\ }\bibfield  {title} {\bibinfo {title} {Accurate measurement of the sagnac effect for matter waves},\ }\href {https://doi.org/10.1126/sciadv.abn8009} {\bibfield  {journal} {\bibinfo  {journal} {Science Advances}\ }\textbf {\bibinfo {volume} {8}},\ \bibinfo {pages} {eabn8009} (\bibinfo {year} {2022})}\BibitemShut {NoStop}%
\bibitem [{\citenamefont {Salducci}\ \emph {et~al.}(2024)\citenamefont {Salducci}, \citenamefont {Bidel}, \citenamefont {Cadoret}, \citenamefont {Darmon}, \citenamefont {Zahzam}, \citenamefont {Bonnin}, \citenamefont {Schwartz}, \citenamefont {Blanchard},\ and\ \citenamefont {Bresson}}]{Salducci:2024}%
  \BibitemOpen
  \bibfield  {author} {\bibinfo {author} {\bibfnamefont {C.}~\bibnamefont {Salducci}}, \bibinfo {author} {\bibfnamefont {Y.}~\bibnamefont {Bidel}}, \bibinfo {author} {\bibfnamefont {M.}~\bibnamefont {Cadoret}}, \bibinfo {author} {\bibfnamefont {S.}~\bibnamefont {Darmon}}, \bibinfo {author} {\bibfnamefont {N.}~\bibnamefont {Zahzam}}, \bibinfo {author} {\bibfnamefont {A.}~\bibnamefont {Bonnin}}, \bibinfo {author} {\bibfnamefont {S.}~\bibnamefont {Schwartz}}, \bibinfo {author} {\bibfnamefont {C.}~\bibnamefont {Blanchard}},\ and\ \bibinfo {author} {\bibfnamefont {A.}~\bibnamefont {Bresson}},\ }\bibfield  {title} {\bibinfo {title} {Quantum sensing of acceleration and rotation by interfering magnetically launched atoms},\ }\href@noop {} {\bibfield  {journal} {\bibinfo  {journal} {Science Advances}\ }\textbf {\bibinfo {volume} {10}},\ \bibinfo {pages} {eadq4498} (\bibinfo {year} {2024})}\BibitemShut {NoStop}%
\bibitem [{\citenamefont {Wright}\ \emph {et~al.}(2022)\citenamefont {Wright}, \citenamefont {Anastassiou}, \citenamefont {Mishra}, \citenamefont {Davies}, \citenamefont {Phillips}, \citenamefont {Maskell},\ and\ \citenamefont {Ralph}}]{Wright:2022}%
  \BibitemOpen
  \bibfield  {author} {\bibinfo {author} {\bibfnamefont {M.~J.}\ \bibnamefont {Wright}}, \bibinfo {author} {\bibfnamefont {L.}~\bibnamefont {Anastassiou}}, \bibinfo {author} {\bibfnamefont {C.}~\bibnamefont {Mishra}}, \bibinfo {author} {\bibfnamefont {J.~M.}\ \bibnamefont {Davies}}, \bibinfo {author} {\bibfnamefont {A.~M.}\ \bibnamefont {Phillips}}, \bibinfo {author} {\bibfnamefont {S.}~\bibnamefont {Maskell}},\ and\ \bibinfo {author} {\bibfnamefont {J.~F.}\ \bibnamefont {Ralph}},\ }\bibfield  {title} {\bibinfo {title} {Cold atom inertial sensors for navigation applications},\ }\href {https://doi.org/10.3389/fphy.2022.994459} {\bibfield  {journal} {\bibinfo  {journal} {Frontiers in Physics}\ }\textbf {\bibinfo {volume} {10}},\ \bibinfo {pages} {994459} (\bibinfo {year} {2022})}\BibitemShut {NoStop}%
\bibitem [{\citenamefont {Wang}\ \emph {et~al.}(2023)\citenamefont {Wang}, \citenamefont {Kealy}, \citenamefont {Gilliam}, \citenamefont {Haine}, \citenamefont {Close}, \citenamefont {Moran}, \citenamefont {Talbot}, \citenamefont {Williams}, \citenamefont {Hardman}, \citenamefont {Freier}, \citenamefont {Wigley}, \citenamefont {White}, \citenamefont {Szigeti},\ and\ \citenamefont {Legge}}]{Wang:2023}%
  \BibitemOpen
  \bibfield  {author} {\bibinfo {author} {\bibfnamefont {X.}~\bibnamefont {Wang}}, \bibinfo {author} {\bibfnamefont {A.}~\bibnamefont {Kealy}}, \bibinfo {author} {\bibfnamefont {C.}~\bibnamefont {Gilliam}}, \bibinfo {author} {\bibfnamefont {S.}~\bibnamefont {Haine}}, \bibinfo {author} {\bibfnamefont {J.}~\bibnamefont {Close}}, \bibinfo {author} {\bibfnamefont {B.}~\bibnamefont {Moran}}, \bibinfo {author} {\bibfnamefont {K.}~\bibnamefont {Talbot}}, \bibinfo {author} {\bibfnamefont {S.}~\bibnamefont {Williams}}, \bibinfo {author} {\bibfnamefont {K.}~\bibnamefont {Hardman}}, \bibinfo {author} {\bibfnamefont {C.}~\bibnamefont {Freier}}, \bibinfo {author} {\bibfnamefont {P.}~\bibnamefont {Wigley}}, \bibinfo {author} {\bibfnamefont {A.}~\bibnamefont {White}}, \bibinfo {author} {\bibfnamefont {S.}~\bibnamefont {Szigeti}},\ and\ \bibinfo {author} {\bibfnamefont {S.}~\bibnamefont {Legge}},\ }\bibfield  {title} {\bibinfo {title} {Improving measurement performance via fusion of classical and quantum accelerometers},\
  }\href {https://www.cambridge.org/core/journals/journal-of-navigation/article/improving-measurement-performance-via-fusion-of-classical-and-quantum-accelerometers/42D24E88A7F588960C60583DAA390469} {\bibfield  {journal} {\bibinfo  {journal} {The Journal of Navigation}\ }\textbf {\bibinfo {volume} {76}},\ \bibinfo {pages} {91} (\bibinfo {year} {2023})}\BibitemShut {NoStop}%
\bibitem [{\citenamefont {Narducci}\ \emph {et~al.}(2022)\citenamefont {Narducci}, \citenamefont {Black},\ and\ \citenamefont {Burke}}]{Narducci:2022}%
  \BibitemOpen
  \bibfield  {author} {\bibinfo {author} {\bibfnamefont {F.~A.}\ \bibnamefont {Narducci}}, \bibinfo {author} {\bibfnamefont {A.~T.}\ \bibnamefont {Black}},\ and\ \bibinfo {author} {\bibfnamefont {J.~H.}\ \bibnamefont {Burke}},\ }\bibfield  {title} {\bibinfo {title} {Advances toward fieldable atom interferometers},\ }\href {https://doi.org/10.1080/23746149.2021.1946426} {\bibfield  {journal} {\bibinfo  {journal} {Advances in Physics: X}\ }\textbf {\bibinfo {volume} {7}},\ \bibinfo {pages} {1946426} (\bibinfo {year} {2022})}\BibitemShut {NoStop}%
\bibitem [{\citenamefont {Gersemann}\ \emph {et~al.}(2025)\citenamefont {Gersemann}, \citenamefont {Rajagopalan}, \citenamefont {Abidi}, \citenamefont {Barbey}, \citenamefont {Sabu}, \citenamefont {Chen}, \citenamefont {Weddig}, \citenamefont {Tennstedt}, \citenamefont {Petring}, \citenamefont {Droese}, \citenamefont {Kassner}, \citenamefont {K{\"u}nzler}, \citenamefont {Keinert}, \citenamefont {Xiao}, \citenamefont {Dencker}, \citenamefont {Wurz}, \citenamefont {L{\"o}wer}, \citenamefont {von Hin{\"u}ber}, \citenamefont {Schlippert}, \citenamefont {Rasel}, \citenamefont {Sch{\"o}n},\ and\ \citenamefont {Abend}}]{Gersemann:2025}%
  \BibitemOpen
  \bibfield  {author} {\bibinfo {author} {\bibfnamefont {M.}~\bibnamefont {Gersemann}}, \bibinfo {author} {\bibfnamefont {A.}~\bibnamefont {Rajagopalan}}, \bibinfo {author} {\bibfnamefont {M.}~\bibnamefont {Abidi}}, \bibinfo {author} {\bibfnamefont {P.}~\bibnamefont {Barbey}}, \bibinfo {author} {\bibfnamefont {A.}~\bibnamefont {Sabu}}, \bibinfo {author} {\bibfnamefont {X.}~\bibnamefont {Chen}}, \bibinfo {author} {\bibfnamefont {N.~B.}\ \bibnamefont {Weddig}}, \bibinfo {author} {\bibfnamefont {B.}~\bibnamefont {Tennstedt}}, \bibinfo {author} {\bibfnamefont {J.}~\bibnamefont {Petring}}, \bibinfo {author} {\bibfnamefont {N.}~\bibnamefont {Droese}}, \bibinfo {author} {\bibfnamefont {A.}~\bibnamefont {Kassner}}, \bibinfo {author} {\bibfnamefont {C.}~\bibnamefont {K{\"u}nzler}}, \bibinfo {author} {\bibfnamefont {L.}~\bibnamefont {Keinert}}, \bibinfo {author} {\bibfnamefont {X.}~\bibnamefont {Xiao}}, \bibinfo {author} {\bibfnamefont {F.}~\bibnamefont {Dencker}}, \bibinfo {author} {\bibfnamefont {M.~C.}\ \bibnamefont
  {Wurz}}, \bibinfo {author} {\bibfnamefont {A.}~\bibnamefont {L{\"o}wer}}, \bibinfo {author} {\bibfnamefont {E.}~\bibnamefont {von Hin{\"u}ber}}, \bibinfo {author} {\bibfnamefont {D.}~\bibnamefont {Schlippert}}, \bibinfo {author} {\bibfnamefont {E.~M.}\ \bibnamefont {Rasel}}, \bibinfo {author} {\bibfnamefont {S.}~\bibnamefont {Sch{\"o}n}},\ and\ \bibinfo {author} {\bibfnamefont {S.}~\bibnamefont {Abend}},\ }\bibfield  {title} {\bibinfo {title} {Developments for quantum inertial navigation systems employing bose--einstein condensates},\ }\href@noop {} {\bibfield  {journal} {\bibinfo  {journal} {Applied Physics Reviews}\ }\textbf {\bibinfo {volume} {12}},\ \bibinfo {pages} {031306} (\bibinfo {year} {2025})}\BibitemShut {NoStop}%
\bibitem [{\citenamefont {El-Sheimy}\ and\ \citenamefont {Youssef}(2020)}]{El-Sheimy:2020}%
  \BibitemOpen
  \bibfield  {author} {\bibinfo {author} {\bibfnamefont {N.}~\bibnamefont {El-Sheimy}}\ and\ \bibinfo {author} {\bibfnamefont {A.}~\bibnamefont {Youssef}},\ }\bibfield  {title} {\bibinfo {title} {Inertial sensors technologies for navigation applications: state of the art and future trends},\ }\href@noop {} {\bibfield  {journal} {\bibinfo  {journal} {Satellite Navigation}\ }\textbf {\bibinfo {volume} {1}},\ \bibinfo {pages} {2} (\bibinfo {year} {2020})}\BibitemShut {NoStop}%
\bibitem [{\citenamefont {Jo}\ \emph {et~al.}(2007)\citenamefont {Jo}, \citenamefont {Shin}, \citenamefont {Will}, \citenamefont {Pasquini}, \citenamefont {Saba}, \citenamefont {Ketterle}, \citenamefont {Pritchard}, \citenamefont {Vengalattore},\ and\ \citenamefont {Prentiss}}]{Jo:2007}%
  \BibitemOpen
  \bibfield  {author} {\bibinfo {author} {\bibfnamefont {G.-B.}\ \bibnamefont {Jo}}, \bibinfo {author} {\bibfnamefont {Y.}~\bibnamefont {Shin}}, \bibinfo {author} {\bibfnamefont {S.}~\bibnamefont {Will}}, \bibinfo {author} {\bibfnamefont {T.~A.}\ \bibnamefont {Pasquini}}, \bibinfo {author} {\bibfnamefont {M.}~\bibnamefont {Saba}}, \bibinfo {author} {\bibfnamefont {W.}~\bibnamefont {Ketterle}}, \bibinfo {author} {\bibfnamefont {D.~E.}\ \bibnamefont {Pritchard}}, \bibinfo {author} {\bibfnamefont {M.}~\bibnamefont {Vengalattore}},\ and\ \bibinfo {author} {\bibfnamefont {M.}~\bibnamefont {Prentiss}},\ }\bibfield  {title} {\bibinfo {title} {Long phase coherence time and number squeezing of two bose-einstein condensates on an atom chip},\ }\href {https://doi.org/10.1103/PhysRevLett.98.030407} {\bibfield  {journal} {\bibinfo  {journal} {Phys. Rev. Lett.}\ }\textbf {\bibinfo {volume} {98}},\ \bibinfo {pages} {030407} (\bibinfo {year} {2007})}\BibitemShut {NoStop}%
\bibitem [{\citenamefont {Wu}\ \emph {et~al.}(2007)\citenamefont {Wu}, \citenamefont {Su},\ and\ \citenamefont {Prentiss}}]{Wu:2007}%
  \BibitemOpen
  \bibfield  {author} {\bibinfo {author} {\bibfnamefont {S.}~\bibnamefont {Wu}}, \bibinfo {author} {\bibfnamefont {E.}~\bibnamefont {Su}},\ and\ \bibinfo {author} {\bibfnamefont {M.}~\bibnamefont {Prentiss}},\ }\bibfield  {title} {\bibinfo {title} {Demonstration of an area-enclosing guided-atom interferometer for rotation sensing},\ }\href {https://doi.org/10.1103/PhysRevLett.99.173201} {\bibfield  {journal} {\bibinfo  {journal} {Phys. Rev. Lett.}\ }\textbf {\bibinfo {volume} {99}},\ \bibinfo {pages} {173201} (\bibinfo {year} {2007})}\BibitemShut {NoStop}%
\bibitem [{\citenamefont {Burke}\ and\ \citenamefont {Sackett}(2009)}]{Burke:2009}%
  \BibitemOpen
  \bibfield  {author} {\bibinfo {author} {\bibfnamefont {J.~H.~T.}\ \bibnamefont {Burke}}\ and\ \bibinfo {author} {\bibfnamefont {C.~A.}\ \bibnamefont {Sackett}},\ }\bibfield  {title} {\bibinfo {title} {Scalable bose-einstein-condensate sagnac interferometer in a linear trap},\ }\href@noop {} {\bibfield  {journal} {\bibinfo  {journal} {Phys. Rev. A}\ }\textbf {\bibinfo {volume} {80}},\ \bibinfo {pages} {061603} (\bibinfo {year} {2009})}\BibitemShut {NoStop}%
\bibitem [{\citenamefont {Qi}\ \emph {et~al.}(2017)\citenamefont {Qi}, \citenamefont {Hu}, \citenamefont {Valenzuela}, \citenamefont {Zhang}, \citenamefont {Zhai}, \citenamefont {Quan}, \citenamefont {Waltham},\ and\ \citenamefont {Fang}}]{Qi:2017}%
  \BibitemOpen
  \bibfield  {author} {\bibinfo {author} {\bibfnamefont {L.}~\bibnamefont {Qi}}, \bibinfo {author} {\bibfnamefont {Z.}~\bibnamefont {Hu}}, \bibinfo {author} {\bibfnamefont {T.}~\bibnamefont {Valenzuela}}, \bibinfo {author} {\bibfnamefont {Y.}~\bibnamefont {Zhang}}, \bibinfo {author} {\bibfnamefont {Y.}~\bibnamefont {Zhai}}, \bibinfo {author} {\bibfnamefont {W.}~\bibnamefont {Quan}}, \bibinfo {author} {\bibfnamefont {N.}~\bibnamefont {Waltham}},\ and\ \bibinfo {author} {\bibfnamefont {J.}~\bibnamefont {Fang}},\ }\bibfield  {title} {\bibinfo {title} {Magnetically guided cesium interferometer for inertial sensing},\ }\href {https://doi.org/10.1063/1.4980066} {\bibfield  {journal} {\bibinfo  {journal} {Applied Physics Letters}\ }\textbf {\bibinfo {volume} {110}},\ \bibinfo {pages} {153502} (\bibinfo {year} {2017})}\BibitemShut {NoStop}%
\bibitem [{\citenamefont {Woffinden}\ \emph {et~al.}(2023)\citenamefont {Woffinden}, \citenamefont {Groszek}, \citenamefont {Gauthier}, \citenamefont {Mommers}, \citenamefont {Bromley}, \citenamefont {Haine}, \citenamefont {Rubinsztein-Dunlop}, \citenamefont {Davis}, \citenamefont {Neely},\ and\ \citenamefont {Baker}}]{Woffinden:2023}%
  \BibitemOpen
  \bibfield  {author} {\bibinfo {author} {\bibfnamefont {C.~W.}\ \bibnamefont {Woffinden}}, \bibinfo {author} {\bibfnamefont {A.~J.}\ \bibnamefont {Groszek}}, \bibinfo {author} {\bibfnamefont {G.}~\bibnamefont {Gauthier}}, \bibinfo {author} {\bibfnamefont {B.~J.}\ \bibnamefont {Mommers}}, \bibinfo {author} {\bibfnamefont {M.~W.~J.}\ \bibnamefont {Bromley}}, \bibinfo {author} {\bibfnamefont {S.~A.}\ \bibnamefont {Haine}}, \bibinfo {author} {\bibfnamefont {H.}~\bibnamefont {Rubinsztein-Dunlop}}, \bibinfo {author} {\bibfnamefont {M.~J.}\ \bibnamefont {Davis}}, \bibinfo {author} {\bibfnamefont {T.~W.}\ \bibnamefont {Neely}},\ and\ \bibinfo {author} {\bibfnamefont {M.}~\bibnamefont {Baker}},\ }\bibfield  {title} {\bibinfo {title} {{Viability of rotation sensing using phonon interferometry in Bose-Einstein condensates}},\ }\href {https://doi.org/10.21468/SciPostPhys.15.4.128} {\bibfield  {journal} {\bibinfo  {journal} {SciPost Phys.}\ }\textbf {\bibinfo {volume} {15}},\ \bibinfo {pages} {128} (\bibinfo {year}
  {2023})}\BibitemShut {NoStop}%
\bibitem [{\citenamefont {Moan}\ \emph {et~al.}(2020)\citenamefont {Moan}, \citenamefont {Horne}, \citenamefont {Arpornthip}, \citenamefont {Luo}, \citenamefont {Fallon}, \citenamefont {Berl},\ and\ \citenamefont {Sackett}}]{Moan:2020}%
  \BibitemOpen
  \bibfield  {author} {\bibinfo {author} {\bibfnamefont {E.~R.}\ \bibnamefont {Moan}}, \bibinfo {author} {\bibfnamefont {R.~A.}\ \bibnamefont {Horne}}, \bibinfo {author} {\bibfnamefont {T.}~\bibnamefont {Arpornthip}}, \bibinfo {author} {\bibfnamefont {Z.}~\bibnamefont {Luo}}, \bibinfo {author} {\bibfnamefont {A.~J.}\ \bibnamefont {Fallon}}, \bibinfo {author} {\bibfnamefont {S.~J.}\ \bibnamefont {Berl}},\ and\ \bibinfo {author} {\bibfnamefont {C.~A.}\ \bibnamefont {Sackett}},\ }\bibfield  {title} {\bibinfo {title} {Quantum rotation sensing with dual sagnac interferometers in an atom-optical waveguide},\ }\href {https://doi.org/10.1103/PhysRevLett.124.120403} {\bibfield  {journal} {\bibinfo  {journal} {Phys. Rev. Lett.}\ }\textbf {\bibinfo {volume} {124}},\ \bibinfo {pages} {120403} (\bibinfo {year} {2020})}\BibitemShut {NoStop}%
\bibitem [{\citenamefont {Beydler}\ \emph {et~al.}(2024)\citenamefont {Beydler}, \citenamefont {Moan}, \citenamefont {Luo}, \citenamefont {Chu},\ and\ \citenamefont {Sackett}}]{Beydler:2024}%
  \BibitemOpen
  \bibfield  {author} {\bibinfo {author} {\bibfnamefont {M.~M.}\ \bibnamefont {Beydler}}, \bibinfo {author} {\bibfnamefont {E.~R.}\ \bibnamefont {Moan}}, \bibinfo {author} {\bibfnamefont {Z.}~\bibnamefont {Luo}}, \bibinfo {author} {\bibfnamefont {Z.}~\bibnamefont {Chu}},\ and\ \bibinfo {author} {\bibfnamefont {C.~A.}\ \bibnamefont {Sackett}},\ }\bibfield  {title} {\bibinfo {title} {Guided-wave sagnac atom interferometer with large area and multiple orbits},\ }\href@noop {} {\bibfield  {journal} {\bibinfo  {journal} {AVS Quantum Science}\ }\textbf {\bibinfo {volume} {6}},\ \bibinfo {pages} {014401} (\bibinfo {year} {2024})}\BibitemShut {NoStop}%
\bibitem [{\citenamefont {Halkyard}\ \emph {et~al.}(2010{\natexlab{a}})\citenamefont {Halkyard}, \citenamefont {Jones},\ and\ \citenamefont {Gardiner}}]{Halkyard:2010}%
  \BibitemOpen
  \bibfield  {author} {\bibinfo {author} {\bibfnamefont {P.~L.}\ \bibnamefont {Halkyard}}, \bibinfo {author} {\bibfnamefont {M.~P.~A.}\ \bibnamefont {Jones}},\ and\ \bibinfo {author} {\bibfnamefont {S.~A.}\ \bibnamefont {Gardiner}},\ }\bibfield  {title} {\bibinfo {title} {Rotational response of two-component {Bose-Einstein} condensates in ring traps},\ }\href {https://doi.org/10.1103/PhysRevA.81.061602} {\bibfield  {journal} {\bibinfo  {journal} {Phys. Rev. A}\ }\textbf {\bibinfo {volume} {81}},\ \bibinfo {pages} {061602} (\bibinfo {year} {2010}{\natexlab{a}})}\BibitemShut {NoStop}%
\bibitem [{\citenamefont {Nolan}\ \emph {et~al.}(2016)\citenamefont {Nolan}, \citenamefont {Sabbatini}, \citenamefont {Bromley}, \citenamefont {Davis},\ and\ \citenamefont {Haine}}]{Nolan2016}%
  \BibitemOpen
  \bibfield  {author} {\bibinfo {author} {\bibfnamefont {S.~P.}\ \bibnamefont {Nolan}}, \bibinfo {author} {\bibfnamefont {J.}~\bibnamefont {Sabbatini}}, \bibinfo {author} {\bibfnamefont {M.~W.~J.}\ \bibnamefont {Bromley}}, \bibinfo {author} {\bibfnamefont {M.~J.}\ \bibnamefont {Davis}},\ and\ \bibinfo {author} {\bibfnamefont {S.~A.}\ \bibnamefont {Haine}},\ }\bibfield  {title} {\bibinfo {title} {Quantum enhanced measurement of rotations with a spin-1 bose-einstein condensate in a ring trap},\ }\href {https://doi.org/10.1103/PhysRevA.93.023616} {\bibfield  {journal} {\bibinfo  {journal} {Phys. Rev. A}\ }\textbf {\bibinfo {volume} {93}},\ \bibinfo {pages} {023616} (\bibinfo {year} {2016})}\BibitemShut {NoStop}%
\bibitem [{\citenamefont {Haine}(2016)}]{Haine:2016b}%
  \BibitemOpen
  \bibfield  {author} {\bibinfo {author} {\bibfnamefont {S.~A.}\ \bibnamefont {Haine}},\ }\bibfield  {title} {\bibinfo {title} {Mean-field dynamics and fisher information in matter wave interferometry},\ }\href {https://doi.org/10.1103/PhysRevLett.116.230404} {\bibfield  {journal} {\bibinfo  {journal} {Phys. Rev. Lett.}\ }\textbf {\bibinfo {volume} {116}},\ \bibinfo {pages} {230404} (\bibinfo {year} {2016})}\BibitemShut {NoStop}%
\bibitem [{\citenamefont {Ryu}\ \emph {et~al.}(2007)\citenamefont {Ryu}, \citenamefont {Andersen}, \citenamefont {Clad\'e}, \citenamefont {Natarajan}, \citenamefont {Helmerson},\ and\ \citenamefont {Phillips}}]{Ryu2007}%
  \BibitemOpen
  \bibfield  {author} {\bibinfo {author} {\bibfnamefont {C.}~\bibnamefont {Ryu}}, \bibinfo {author} {\bibfnamefont {M.~F.}\ \bibnamefont {Andersen}}, \bibinfo {author} {\bibfnamefont {P.}~\bibnamefont {Clad\'e}}, \bibinfo {author} {\bibfnamefont {V.}~\bibnamefont {Natarajan}}, \bibinfo {author} {\bibfnamefont {K.}~\bibnamefont {Helmerson}},\ and\ \bibinfo {author} {\bibfnamefont {W.~D.}\ \bibnamefont {Phillips}},\ }\bibfield  {title} {\bibinfo {title} {Observation of persistent flow of a bose-einstein condensate in a toroidal trap},\ }\href {https://doi.org/10.1103/PhysRevLett.99.260401} {\bibfield  {journal} {\bibinfo  {journal} {Phys. Rev. Lett.}\ }\textbf {\bibinfo {volume} {99}},\ \bibinfo {pages} {260401} (\bibinfo {year} {2007})}\BibitemShut {NoStop}%
\bibitem [{\citenamefont {Ramanathan}\ \emph {et~al.}(2011)\citenamefont {Ramanathan}, \citenamefont {Wright}, \citenamefont {Muniz}, \citenamefont {Zelan}, \citenamefont {Hill}, \citenamefont {Lobb}, \citenamefont {Helmerson}, \citenamefont {Phillips},\ and\ \citenamefont {Campbell}}]{Ramanathan:2011}%
  \BibitemOpen
  \bibfield  {author} {\bibinfo {author} {\bibfnamefont {A.}~\bibnamefont {Ramanathan}}, \bibinfo {author} {\bibfnamefont {K.~C.}\ \bibnamefont {Wright}}, \bibinfo {author} {\bibfnamefont {S.~R.}\ \bibnamefont {Muniz}}, \bibinfo {author} {\bibfnamefont {M.}~\bibnamefont {Zelan}}, \bibinfo {author} {\bibfnamefont {W.~T.}\ \bibnamefont {Hill}}, \bibinfo {author} {\bibfnamefont {C.~J.}\ \bibnamefont {Lobb}}, \bibinfo {author} {\bibfnamefont {K.}~\bibnamefont {Helmerson}}, \bibinfo {author} {\bibfnamefont {W.~D.}\ \bibnamefont {Phillips}},\ and\ \bibinfo {author} {\bibfnamefont {G.~K.}\ \bibnamefont {Campbell}},\ }\bibfield  {title} {\bibinfo {title} {Superflow in a toroidal bose-einstein condensate: An atom circuit with a tunable weak link},\ }\href {https://doi.org/10.1103/PhysRevLett.106.130401} {\bibfield  {journal} {\bibinfo  {journal} {Phys. Rev. Lett.}\ }\textbf {\bibinfo {volume} {106}},\ \bibinfo {pages} {130401} (\bibinfo {year} {2011})}\BibitemShut {NoStop}%
\bibitem [{\citenamefont {Yakimenko}\ \emph {et~al.}(2013)\citenamefont {Yakimenko}, \citenamefont {Isaieva}, \citenamefont {Vilchinskii},\ and\ \citenamefont {Weyrauch}}]{Yakimenko:2013}%
  \BibitemOpen
  \bibfield  {author} {\bibinfo {author} {\bibfnamefont {A.~I.}\ \bibnamefont {Yakimenko}}, \bibinfo {author} {\bibfnamefont {K.~O.}\ \bibnamefont {Isaieva}}, \bibinfo {author} {\bibfnamefont {S.~I.}\ \bibnamefont {Vilchinskii}},\ and\ \bibinfo {author} {\bibfnamefont {M.}~\bibnamefont {Weyrauch}},\ }\bibfield  {title} {\bibinfo {title} {Stability of persistent currents in spinor bose-einstein condensates},\ }\href {https://doi.org/10.1103/PhysRevA.88.051602} {\bibfield  {journal} {\bibinfo  {journal} {Phys. Rev. A}\ }\textbf {\bibinfo {volume} {88}},\ \bibinfo {pages} {051602(R)} (\bibinfo {year} {2013})}\BibitemShut {NoStop}%
\bibitem [{\citenamefont {Beattie}\ \emph {et~al.}(2013)\citenamefont {Beattie}, \citenamefont {Moulder}, \citenamefont {Fletcher},\ and\ \citenamefont {Hadzibabic}}]{Beattie:2013}%
  \BibitemOpen
  \bibfield  {author} {\bibinfo {author} {\bibfnamefont {S.}~\bibnamefont {Beattie}}, \bibinfo {author} {\bibfnamefont {S.}~\bibnamefont {Moulder}}, \bibinfo {author} {\bibfnamefont {R.~J.}\ \bibnamefont {Fletcher}},\ and\ \bibinfo {author} {\bibfnamefont {Z.}~\bibnamefont {Hadzibabic}},\ }\bibfield  {title} {\bibinfo {title} {Persistent currents in spinor condensates},\ }\href {https://doi.org/10.1103/PhysRevLett.110.025301} {\bibfield  {journal} {\bibinfo  {journal} {Phys. Rev. Lett.}\ }\textbf {\bibinfo {volume} {110}},\ \bibinfo {pages} {025301} (\bibinfo {year} {2013})}\BibitemShut {NoStop}%
\bibitem [{\citenamefont {Polo}\ \emph {et~al.}(2025)\citenamefont {Polo}, \citenamefont {Chetcuti}, \citenamefont {Haug}, \citenamefont {Minguzzi}, \citenamefont {Wright},\ and\ \citenamefont {Amico}}]{Polo:2025}%
  \BibitemOpen
  \bibfield  {author} {\bibinfo {author} {\bibfnamefont {J.}~\bibnamefont {Polo}}, \bibinfo {author} {\bibfnamefont {W.}~\bibnamefont {Chetcuti}}, \bibinfo {author} {\bibfnamefont {T.}~\bibnamefont {Haug}}, \bibinfo {author} {\bibfnamefont {A.}~\bibnamefont {Minguzzi}}, \bibinfo {author} {\bibfnamefont {K.}~\bibnamefont {Wright}},\ and\ \bibinfo {author} {\bibfnamefont {L.}~\bibnamefont {Amico}},\ }\bibfield  {title} {\bibinfo {title} {Persistent currents in ultracold gases},\ }\href {https://doi.org/https://doi.org/10.1016/j.physrep.2025.06.003} {\bibfield  {journal} {\bibinfo  {journal} {Physics Reports}\ }\textbf {\bibinfo {volume} {1137}},\ \bibinfo {pages} {1} (\bibinfo {year} {2025})},\ \bibinfo {note} {persistent currents in ultracold gases}\BibitemShut {NoStop}%
\bibitem [{\citenamefont {Borysenko}\ \emph {et~al.}(2025)\citenamefont {Borysenko}, \citenamefont {Bazhan}, \citenamefont {Prykhodko}, \citenamefont {Pfeiffer}, \citenamefont {Lind}, \citenamefont {Birkl},\ and\ \citenamefont {Yakimenko}}]{Borysenko:2025}%
  \BibitemOpen
  \bibfield  {author} {\bibinfo {author} {\bibfnamefont {Y.}~\bibnamefont {Borysenko}}, \bibinfo {author} {\bibfnamefont {N.}~\bibnamefont {Bazhan}}, \bibinfo {author} {\bibfnamefont {O.}~\bibnamefont {Prykhodko}}, \bibinfo {author} {\bibfnamefont {D.}~\bibnamefont {Pfeiffer}}, \bibinfo {author} {\bibfnamefont {L.}~\bibnamefont {Lind}}, \bibinfo {author} {\bibfnamefont {G.}~\bibnamefont {Birkl}},\ and\ \bibinfo {author} {\bibfnamefont {A.}~\bibnamefont {Yakimenko}},\ }\bibfield  {title} {\bibinfo {title} {Acceleration-driven dynamics of josephson vortices in coplanar superfluid rings},\ }\href {https://doi.org/10.1103/PhysRevA.111.043308} {\bibfield  {journal} {\bibinfo  {journal} {Phys. Rev. A}\ }\textbf {\bibinfo {volume} {111}},\ \bibinfo {pages} {043308} (\bibinfo {year} {2025})}\BibitemShut {NoStop}%
\bibitem [{\citenamefont {Chaika}\ \emph {et~al.}(2026)\citenamefont {Chaika}, \citenamefont {Oliinyk}, \citenamefont {Yatsuta}, \citenamefont {Edwards}, \citenamefont {Proukakis}, \citenamefont {Bland},\ and\ \citenamefont {Yakimenko}}]{Chaika:2026}%
  \BibitemOpen
  \bibfield  {author} {\bibinfo {author} {\bibfnamefont {A.}~\bibnamefont {Chaika}}, \bibinfo {author} {\bibfnamefont {A.~O.}\ \bibnamefont {Oliinyk}}, \bibinfo {author} {\bibfnamefont {I.~V.}\ \bibnamefont {Yatsuta}}, \bibinfo {author} {\bibfnamefont {M.}~\bibnamefont {Edwards}}, \bibinfo {author} {\bibfnamefont {N.~P.}\ \bibnamefont {Proukakis}}, \bibinfo {author} {\bibfnamefont {T.}~\bibnamefont {Bland}},\ and\ \bibinfo {author} {\bibfnamefont {A.~I.}\ \bibnamefont {Yakimenko}},\ }\bibfield  {title} {\bibinfo {title} {Controlled acoustic-driven vortex transport in coupled superfluid rings},\ }\href {https://doi.org/10.1103/sfjj-s62c} {\bibfield  {journal} {\bibinfo  {journal} {Phys. Rev. A}\ }\textbf {\bibinfo {volume} {113}},\ \bibinfo {pages} {053305} (\bibinfo {year} {2026})}\BibitemShut {NoStop}%
\bibitem [{\citenamefont {Haine}(2018)}]{Haine:2018}%
  \BibitemOpen
  \bibfield  {author} {\bibinfo {author} {\bibfnamefont {S.~A.}\ \bibnamefont {Haine}},\ }\bibfield  {title} {\bibinfo {title} {Quantum noise in bright soliton matterwave interferometry},\ }\href {http://stacks.iop.org/1367-2630/20/i=3/a=033009} {\bibfield  {journal} {\bibinfo  {journal} {New Journal of Physics}\ }\textbf {\bibinfo {volume} {20}},\ \bibinfo {pages} {033009} (\bibinfo {year} {2018})}\BibitemShut {NoStop}%
\bibitem [{\citenamefont {Szigeti}\ \emph {et~al.}(2020)\citenamefont {Szigeti}, \citenamefont {Nolan}, \citenamefont {Close},\ and\ \citenamefont {Haine}}]{Szigeti:2020}%
  \BibitemOpen
  \bibfield  {author} {\bibinfo {author} {\bibfnamefont {S.~S.}\ \bibnamefont {Szigeti}}, \bibinfo {author} {\bibfnamefont {S.~P.}\ \bibnamefont {Nolan}}, \bibinfo {author} {\bibfnamefont {J.~D.}\ \bibnamefont {Close}},\ and\ \bibinfo {author} {\bibfnamefont {S.~A.}\ \bibnamefont {Haine}},\ }\bibfield  {title} {\bibinfo {title} {High-precision quantum-enhanced gravimetry with a bose-einstein condensate},\ }\href {https://doi.org/10.1103/PhysRevLett.125.100402} {\bibfield  {journal} {\bibinfo  {journal} {Phys. Rev. Lett.}\ }\textbf {\bibinfo {volume} {125}},\ \bibinfo {pages} {100402} (\bibinfo {year} {2020})}\BibitemShut {NoStop}%
\bibitem [{\citenamefont {Szigeti}\ \emph {et~al.}(2021)\citenamefont {Szigeti}, \citenamefont {Hosten},\ and\ \citenamefont {Haine}}]{Szigeti:2021}%
  \BibitemOpen
  \bibfield  {author} {\bibinfo {author} {\bibfnamefont {S.~S.}\ \bibnamefont {Szigeti}}, \bibinfo {author} {\bibfnamefont {O.}~\bibnamefont {Hosten}},\ and\ \bibinfo {author} {\bibfnamefont {S.~A.}\ \bibnamefont {Haine}},\ }\bibfield  {title} {\bibinfo {title} {Improving cold-atom sensors with quantum entanglement: Prospects and challenges},\ }\href@noop {} {\bibfield  {journal} {\bibinfo  {journal} {Applied Physics Letters}\ }\textbf {\bibinfo {volume} {118}},\ \bibinfo {pages} {140501} (\bibinfo {year} {2021})}\BibitemShut {NoStop}%
\bibitem [{\citenamefont {Haine}\ and\ \citenamefont {Ferris}(2011)}]{Haine:2011}%
  \BibitemOpen
  \bibfield  {author} {\bibinfo {author} {\bibfnamefont {S.~A.}\ \bibnamefont {Haine}}\ and\ \bibinfo {author} {\bibfnamefont {A.~J.}\ \bibnamefont {Ferris}},\ }\bibfield  {title} {\bibinfo {title} {Surpassing the standard quantum limit in an atom interferometer with four-mode entanglement produced from four-wave mixing},\ }\href {https://doi.org/10.1103/PhysRevA.84.043624} {\bibfield  {journal} {\bibinfo  {journal} {Phys. Rev. A}\ }\textbf {\bibinfo {volume} {84}},\ \bibinfo {pages} {043624} (\bibinfo {year} {2011})}\BibitemShut {NoStop}%
\bibitem [{\citenamefont {Helm}\ \emph {et~al.}(2018)\citenamefont {Helm}, \citenamefont {Billam}, \citenamefont {Rakonjac}, \citenamefont {Cornish},\ and\ \citenamefont {Gardiner}}]{Helm:2018}%
  \BibitemOpen
  \bibfield  {author} {\bibinfo {author} {\bibfnamefont {J.~L.}\ \bibnamefont {Helm}}, \bibinfo {author} {\bibfnamefont {T.~P.}\ \bibnamefont {Billam}}, \bibinfo {author} {\bibfnamefont {A.}~\bibnamefont {Rakonjac}}, \bibinfo {author} {\bibfnamefont {S.~L.}\ \bibnamefont {Cornish}},\ and\ \bibinfo {author} {\bibfnamefont {S.~A.}\ \bibnamefont {Gardiner}},\ }\bibfield  {title} {\bibinfo {title} {Spin-orbit-coupled interferometry with ring-trapped bose-einstein condensates},\ }\href@noop {} {\bibfield  {journal} {\bibinfo  {journal} {Phys. Rev. Lett.}\ }\textbf {\bibinfo {volume} {120}},\ \bibinfo {pages} {063201} (\bibinfo {year} {2018})}\BibitemShut {NoStop}%
\bibitem [{\citenamefont {Steel}\ \emph {et~al.}(1998)\citenamefont {Steel}, \citenamefont {Olsen}, \citenamefont {Plimak}, \citenamefont {Drummond}, \citenamefont {Tan}, \citenamefont {Collett}, \citenamefont {Walls},\ and\ \citenamefont {Graham}}]{Steel:1998}%
  \BibitemOpen
  \bibfield  {author} {\bibinfo {author} {\bibfnamefont {M.~J.}\ \bibnamefont {Steel}}, \bibinfo {author} {\bibfnamefont {M.~K.}\ \bibnamefont {Olsen}}, \bibinfo {author} {\bibfnamefont {L.~I.}\ \bibnamefont {Plimak}}, \bibinfo {author} {\bibfnamefont {P.~D.}\ \bibnamefont {Drummond}}, \bibinfo {author} {\bibfnamefont {S.~M.}\ \bibnamefont {Tan}}, \bibinfo {author} {\bibfnamefont {M.~J.}\ \bibnamefont {Collett}}, \bibinfo {author} {\bibfnamefont {D.~F.}\ \bibnamefont {Walls}},\ and\ \bibinfo {author} {\bibfnamefont {R.}~\bibnamefont {Graham}},\ }\bibfield  {title} {\bibinfo {title} {Dynamical quantum noise in trapped {Bose-Einstein} condensates},\ }\href {https://doi.org/10.1103/PhysRevA.58.4824} {\bibfield  {journal} {\bibinfo  {journal} {Phys. Rev. A}\ }\textbf {\bibinfo {volume} {58}},\ \bibinfo {pages} {4824} (\bibinfo {year} {1998})}\BibitemShut {NoStop}%
\bibitem [{\citenamefont {Sinatra}\ \emph {et~al.}(2002)\citenamefont {Sinatra}, \citenamefont {Lobo},\ and\ \citenamefont {Castin}}]{Sinatra:2002}%
  \BibitemOpen
  \bibfield  {author} {\bibinfo {author} {\bibfnamefont {A.}~\bibnamefont {Sinatra}}, \bibinfo {author} {\bibfnamefont {C.}~\bibnamefont {Lobo}},\ and\ \bibinfo {author} {\bibfnamefont {Y.}~\bibnamefont {Castin}},\ }\bibfield  {title} {\bibinfo {title} {The truncated wigner method for bose-condensed gases: limits of validity and applications1},\ }\href {https://doi.org/10.1088/0953-4075/35/17/301} {\bibfield  {journal} {\bibinfo  {journal} {Journal of Physics B: Atomic, Molecular and Optical Physics}\ }\textbf {\bibinfo {volume} {35}},\ \bibinfo {pages} {3599} (\bibinfo {year} {2002})}\BibitemShut {NoStop}%
\bibitem [{\citenamefont {Blakie}\ \emph {et~al.}(2008)\citenamefont {Blakie}, \citenamefont {Bradley}, \citenamefont {Davis}, \citenamefont {Ballagh},\ and\ \citenamefont {Gardiner}}]{Blakie:2008}%
  \BibitemOpen
  \bibfield  {author} {\bibinfo {author} {\bibfnamefont {P.~B.}\ \bibnamefont {Blakie}}, \bibinfo {author} {\bibfnamefont {A.~S.}\ \bibnamefont {Bradley}}, \bibinfo {author} {\bibfnamefont {M.~J.}\ \bibnamefont {Davis}}, \bibinfo {author} {\bibfnamefont {R.~J.}\ \bibnamefont {Ballagh}},\ and\ \bibinfo {author} {\bibfnamefont {C.~W.}\ \bibnamefont {Gardiner}},\ }\bibfield  {title} {\bibinfo {title} {Dynamics and statistical mechanics of ultra-cold {Bose} gases using c-field techniques},\ }\bibfield  {booktitle} {\emph {\bibinfo {booktitle} {Advances in Physics}},\ }\href {https://doi.org/10.1080/00018730802564254} {\bibfield  {journal} {\bibinfo  {journal} {Advances in Physics}\ }\textbf {\bibinfo {volume} {57}},\ \bibinfo {pages} {363} (\bibinfo {year} {2008})}\BibitemShut {NoStop}%
\bibitem [{\citenamefont {Polkovnikov}(2010)}]{Polkovnikov:2010}%
  \BibitemOpen
  \bibfield  {author} {\bibinfo {author} {\bibfnamefont {A.}~\bibnamefont {Polkovnikov}},\ }\bibfield  {title} {\bibinfo {title} {Phase space representation of quantum dynamics},\ }\href {https://doi.org/https://doi.org/10.1016/j.aop.2010.02.006} {\bibfield  {journal} {\bibinfo  {journal} {Annals of Physics}\ }\textbf {\bibinfo {volume} {325}},\ \bibinfo {pages} {1790} (\bibinfo {year} {2010})}\BibitemShut {NoStop}%
\bibitem [{\citenamefont {Gross}(1961)}]{Gross:1961}%
  \BibitemOpen
  \bibfield  {author} {\bibinfo {author} {\bibfnamefont {E.~P.}\ \bibnamefont {Gross}},\ }\bibfield  {title} {\bibinfo {title} {Structure of a quantized vortex in boson systems},\ }\href {https://doi.org/10.1007/BF02731494} {\bibfield  {journal} {\bibinfo  {journal} {Il Nuovo Cimento (1955-1965)}\ }\textbf {\bibinfo {volume} {20}},\ \bibinfo {pages} {454} (\bibinfo {year} {1961})}\BibitemShut {NoStop}%
\bibitem [{\citenamefont {Pitaevskii}(1961)}]{Pitaevskii:1961}%
  \BibitemOpen
  \bibfield  {author} {\bibinfo {author} {\bibfnamefont {L.~P.}\ \bibnamefont {Pitaevskii}},\ }\bibfield  {title} {\bibinfo {title} {Vortex lines in an imperfect bose gas},\ }\href@noop {} {\bibfield  {journal} {\bibinfo  {journal} {Sov. Phys. JETP}\ }\textbf {\bibinfo {volume} {13}},\ \bibinfo {pages} {451} (\bibinfo {year} {1961})}\BibitemShut {NoStop}%
\bibitem [{\citenamefont {Roberts}\ \emph {et~al.}(1998)\citenamefont {Roberts}, \citenamefont {Claussen}, \citenamefont {Burke}, \citenamefont {Greene}, \citenamefont {Cornell},\ and\ \citenamefont {Wieman}}]{Roberts:1998}%
  \BibitemOpen
  \bibfield  {author} {\bibinfo {author} {\bibfnamefont {J.~L.}\ \bibnamefont {Roberts}}, \bibinfo {author} {\bibfnamefont {N.~R.}\ \bibnamefont {Claussen}}, \bibinfo {author} {\bibfnamefont {J.~P.}\ \bibnamefont {Burke}}, \bibinfo {author} {\bibfnamefont {C.~H.}\ \bibnamefont {Greene}}, \bibinfo {author} {\bibfnamefont {E.~A.}\ \bibnamefont {Cornell}},\ and\ \bibinfo {author} {\bibfnamefont {C.~E.}\ \bibnamefont {Wieman}},\ }\bibfield  {title} {\bibinfo {title} {Resonant magnetic field control of elastic scattering in cold $^{85}rb$},\ }\href@noop {} {\bibfield  {journal} {\bibinfo  {journal} {Phys. Rev. Lett.}\ }\textbf {\bibinfo {volume} {81}},\ \bibinfo {pages} {5109} (\bibinfo {year} {1998})}\BibitemShut {NoStop}%
\bibitem [{\citenamefont {Halkyard}\ \emph {et~al.}(2010{\natexlab{b}})\citenamefont {Halkyard}, \citenamefont {Jones},\ and\ \citenamefont {Gardiner}}]{Halkyard2010}%
  \BibitemOpen
  \bibfield  {author} {\bibinfo {author} {\bibfnamefont {P.~L.}\ \bibnamefont {Halkyard}}, \bibinfo {author} {\bibfnamefont {M.~P.~A.}\ \bibnamefont {Jones}},\ and\ \bibinfo {author} {\bibfnamefont {S.~A.}\ \bibnamefont {Gardiner}},\ }\bibfield  {title} {\bibinfo {title} {Rotational response of two-component bose-einstein condensates in ring traps},\ }\href {https://doi.org/10.1103/PhysRevA.81.061602} {\bibfield  {journal} {\bibinfo  {journal} {Phys. Rev. A}\ }\textbf {\bibinfo {volume} {81}},\ \bibinfo {pages} {061602} (\bibinfo {year} {2010}{\natexlab{b}})}\BibitemShut {NoStop}%
\bibitem [{\citenamefont {Husband}\ \emph {et~al.}(2026{\natexlab{a}})\citenamefont {Husband}, \citenamefont {Eastman}, \citenamefont {Haine}, \citenamefont {Eagle}, \citenamefont {Close}, \citenamefont {Thomas},\ and\ \citenamefont {Legge}}]{Husband:2026}%
  \BibitemOpen
  \bibfield  {author} {\bibinfo {author} {\bibfnamefont {R.}~\bibnamefont {Husband}}, \bibinfo {author} {\bibfnamefont {J.}~\bibnamefont {Eastman}}, \bibinfo {author} {\bibfnamefont {S.~A.}\ \bibnamefont {Haine}}, \bibinfo {author} {\bibfnamefont {R.~H.}\ \bibnamefont {Eagle}}, \bibinfo {author} {\bibfnamefont {J.~D.}\ \bibnamefont {Close}}, \bibinfo {author} {\bibfnamefont {R.~J.}\ \bibnamefont {Thomas}},\ and\ \bibinfo {author} {\bibfnamefont {S.}~\bibnamefont {Legge}},\ }\bibfield  {title} {\bibinfo {title} {Propagation of gaussian vortex beams from aperture-limited optics},\ }\href {https://doi.org/10.1088/2040-8986/ae681e} {\bibfield  {journal} {\bibinfo  {journal} {Journal of Optics}\ }\textbf {\bibinfo {volume} {28}},\ \bibinfo {pages} {055603} (\bibinfo {year} {2026}{\natexlab{a}})}\BibitemShut {NoStop}%
\bibitem [{\citenamefont {Husband}\ \emph {et~al.}(2026{\natexlab{b}})\citenamefont {Husband}, \citenamefont {Thomas}, \citenamefont {Ben-Aïcha}, \citenamefont {Eagle}, \citenamefont {Eastman}, \citenamefont {Debs}, \citenamefont {Everitt}, \citenamefont {Larsen}, \citenamefont {Imhof}, \citenamefont {Sackett}, \citenamefont {Close}, \citenamefont {Haine},\ and\ \citenamefont {Legge}}]{Husband:2026b}%
  \BibitemOpen
  \bibfield  {author} {\bibinfo {author} {\bibfnamefont {R.}~\bibnamefont {Husband}}, \bibinfo {author} {\bibfnamefont {R.~J.}\ \bibnamefont {Thomas}}, \bibinfo {author} {\bibfnamefont {Y.}~\bibnamefont {Ben-Aïcha}}, \bibinfo {author} {\bibfnamefont {R.~H.}\ \bibnamefont {Eagle}}, \bibinfo {author} {\bibfnamefont {J.}~\bibnamefont {Eastman}}, \bibinfo {author} {\bibfnamefont {J.~E.}\ \bibnamefont {Debs}}, \bibinfo {author} {\bibfnamefont {P.~J.}\ \bibnamefont {Everitt}}, \bibinfo {author} {\bibfnamefont {M.}~\bibnamefont {Larsen}}, \bibinfo {author} {\bibfnamefont {E.}~\bibnamefont {Imhof}}, \bibinfo {author} {\bibfnamefont {C.~A.}\ \bibnamefont {Sackett}}, \bibinfo {author} {\bibfnamefont {J.~D.}\ \bibnamefont {Close}}, \bibinfo {author} {\bibfnamefont {S.~A.}\ \bibnamefont {Haine}},\ and\ \bibinfo {author} {\bibfnamefont {S.}~\bibnamefont {Legge}},\ }\href {https://arxiv.org/abs/2606.04430} {\bibinfo {title} {Atom interferometry with transverse optical modes}} (\bibinfo {year} {2026}{\natexlab{b}}),\
  \Eprint {https://arxiv.org/abs/2606.04430} {arXiv:2606.04430 [physics.atom-ph]} \BibitemShut {NoStop}%
\bibitem [{\citenamefont {Sinatra}\ \emph {et~al.}(1995)\citenamefont {Sinatra}, \citenamefont {Castelli}, \citenamefont {Lugiato}, \citenamefont {Grangier},\ and\ \citenamefont {Poizat}}]{Sinatra:1995}%
  \BibitemOpen
  \bibfield  {author} {\bibinfo {author} {\bibfnamefont {A.}~\bibnamefont {Sinatra}}, \bibinfo {author} {\bibfnamefont {F.}~\bibnamefont {Castelli}}, \bibinfo {author} {\bibfnamefont {L.~A.}\ \bibnamefont {Lugiato}}, \bibinfo {author} {\bibfnamefont {P.}~\bibnamefont {Grangier}},\ and\ \bibinfo {author} {\bibfnamefont {J.~P.}\ \bibnamefont {Poizat}},\ }\bibfield  {title} {\bibinfo {title} {Effective two-level model versus three-level model},\ }\href {http://stacks.iop.org/1355-5111/7/i=3/a=016} {\bibfield  {journal} {\bibinfo  {journal} {Quantum and Semiclassical Optics: Journal of the European Optical Society Part B}\ }\textbf {\bibinfo {volume} {7}},\ \bibinfo {pages} {405} (\bibinfo {year} {1995})}\BibitemShut {NoStop}%
\bibitem [{\citenamefont {Norrie}\ \emph {et~al.}(2006)\citenamefont {Norrie}, \citenamefont {Ballagh},\ and\ \citenamefont {Gardiner}}]{Norrie:2006}%
  \BibitemOpen
  \bibfield  {author} {\bibinfo {author} {\bibfnamefont {A.~A.}\ \bibnamefont {Norrie}}, \bibinfo {author} {\bibfnamefont {R.~J.}\ \bibnamefont {Ballagh}},\ and\ \bibinfo {author} {\bibfnamefont {C.~W.}\ \bibnamefont {Gardiner}},\ }\bibfield  {title} {\bibinfo {title} {Quantum turbulence and correlations in {Bose-Einstein} condensate collisions},\ }\href {https://doi.org/10.1103/PhysRevA.73.043617} {\bibfield  {journal} {\bibinfo  {journal} {Phys. Rev. A}\ }\textbf {\bibinfo {volume} {73}},\ \bibinfo {pages} {043617} (\bibinfo {year} {2006})}\BibitemShut {NoStop}%
\bibitem [{\citenamefont {Drummond}\ and\ \citenamefont {Opanchuk}(2017)}]{Drummond:2017}%
  \BibitemOpen
  \bibfield  {author} {\bibinfo {author} {\bibfnamefont {P.~D.}\ \bibnamefont {Drummond}}\ and\ \bibinfo {author} {\bibfnamefont {B.}~\bibnamefont {Opanchuk}},\ }\bibfield  {title} {\bibinfo {title} {Truncated {Wigner} dynamics and conservation laws},\ }\href {https://doi.org/10.1103/PhysRevA.96.043616} {\bibfield  {journal} {\bibinfo  {journal} {Phys. Rev. A}\ }\textbf {\bibinfo {volume} {96}},\ \bibinfo {pages} {043616} (\bibinfo {year} {2017})}\BibitemShut {NoStop}%
\bibitem [{\citenamefont {Haine}\ \emph {et~al.}(2014)\citenamefont {Haine}, \citenamefont {Lau}, \citenamefont {Anderson},\ and\ \citenamefont {Johnsson}}]{Haine:2014}%
  \BibitemOpen
  \bibfield  {author} {\bibinfo {author} {\bibfnamefont {S.~A.}\ \bibnamefont {Haine}}, \bibinfo {author} {\bibfnamefont {J.}~\bibnamefont {Lau}}, \bibinfo {author} {\bibfnamefont {R.~P.}\ \bibnamefont {Anderson}},\ and\ \bibinfo {author} {\bibfnamefont {M.~T.}\ \bibnamefont {Johnsson}},\ }\bibfield  {title} {\bibinfo {title} {Self-induced spatial dynamics to enhance spin squeezing via one-axis twisting in a two-component {Bose-Einstein} condensate},\ }\href {https://doi.org/10.1103/PhysRevA.90.023613} {\bibfield  {journal} {\bibinfo  {journal} {Phys. Rev. A}\ }\textbf {\bibinfo {volume} {90}},\ \bibinfo {pages} {023613} (\bibinfo {year} {2014})}\BibitemShut {NoStop}%
\bibitem [{\citenamefont {Haine}\ and\ \citenamefont {Lau}(2016)}]{Haine:2016}%
  \BibitemOpen
  \bibfield  {author} {\bibinfo {author} {\bibfnamefont {S.~A.}\ \bibnamefont {Haine}}\ and\ \bibinfo {author} {\bibfnamefont {W.~Y.~S.}\ \bibnamefont {Lau}},\ }\bibfield  {title} {\bibinfo {title} {Generation of atom-light entanglement in an optical cavity for quantum enhanced atom interferometry},\ }\href {https://doi.org/10.1103/PhysRevA.93.023607} {\bibfield  {journal} {\bibinfo  {journal} {Phys. Rev. A}\ }\textbf {\bibinfo {volume} {93}},\ \bibinfo {pages} {023607} (\bibinfo {year} {2016})}\BibitemShut {NoStop}%
\bibitem [{\citenamefont {Szigeti}\ \emph {et~al.}(2017)\citenamefont {Szigeti}, \citenamefont {Lewis-Swan},\ and\ \citenamefont {Haine}}]{Szigeti:2017}%
  \BibitemOpen
  \bibfield  {author} {\bibinfo {author} {\bibfnamefont {S.~S.}\ \bibnamefont {Szigeti}}, \bibinfo {author} {\bibfnamefont {R.~J.}\ \bibnamefont {Lewis-Swan}},\ and\ \bibinfo {author} {\bibfnamefont {S.~A.}\ \bibnamefont {Haine}},\ }\bibfield  {title} {\bibinfo {title} {Pumped-up {SU}(1,1) interferometry},\ }\href {https://doi.org/10.1103/PhysRevLett.118.150401} {\bibfield  {journal} {\bibinfo  {journal} {Phys. Rev. Lett.}\ }\textbf {\bibinfo {volume} {118}},\ \bibinfo {pages} {150401} (\bibinfo {year} {2017})}\BibitemShut {NoStop}%
\bibitem [{\citenamefont {Drummond}\ and\ \citenamefont {Hardman}(1993)}]{Drummond:1993}%
  \BibitemOpen
  \bibfield  {author} {\bibinfo {author} {\bibfnamefont {P.~D.}\ \bibnamefont {Drummond}}\ and\ \bibinfo {author} {\bibfnamefont {A.~D.}\ \bibnamefont {Hardman}},\ }\bibfield  {title} {\bibinfo {title} {Simulation of quantum effects in {Raman}-active waveguides},\ }\href {http://stacks.iop.org/0295-5075/21/i=3/a=005} {\bibfield  {journal} {\bibinfo  {journal} {EPL (Europhysics Letters)}\ }\textbf {\bibinfo {volume} {21}},\ \bibinfo {pages} {279} (\bibinfo {year} {1993})}\BibitemShut {NoStop}%
\bibitem [{\citenamefont {Gardiner}\ and\ \citenamefont {Zoller}(2004)}]{Gardiner:2004b}%
  \BibitemOpen
  \bibfield  {author} {\bibinfo {author} {\bibfnamefont {C.~W.}\ \bibnamefont {Gardiner}}\ and\ \bibinfo {author} {\bibfnamefont {P.}~\bibnamefont {Zoller}},\ }\href@noop {} {\emph {\bibinfo {title} {Quantum Noise: A Handbook of {Markovian} and Non-{Markovian} Quantum Stochastic Methods with Applications to Quantum Optics}}},\ \bibinfo {edition} {3rd}\ ed.\ (\bibinfo  {publisher} {Springer},\ \bibinfo {address} {Berlin and Heidelberg},\ \bibinfo {year} {2004})\BibitemShut {NoStop}%
\bibitem [{\citenamefont {Olsen}\ and\ \citenamefont {Bradley}(2009)}]{Olsen:2009}%
  \BibitemOpen
  \bibfield  {author} {\bibinfo {author} {\bibfnamefont {M.}~\bibnamefont {Olsen}}\ and\ \bibinfo {author} {\bibfnamefont {A.}~\bibnamefont {Bradley}},\ }\bibfield  {title} {\bibinfo {title} {Numerical representation of quantum states in the positive-{P} and {Wigner} representations},\ }\href {https://doi.org/10.1016/j.optcom.2009.06.033} {\bibfield  {journal} {\bibinfo  {journal} {Optics Communications}\ }\textbf {\bibinfo {volume} {282}},\ \bibinfo {pages} {3924 } (\bibinfo {year} {2009})}\BibitemShut {NoStop}%
\bibitem [{\citenamefont {Mehdi}\ \emph {et~al.}(2021)\citenamefont {Mehdi}, \citenamefont {Bradley}, \citenamefont {Hope},\ and\ \citenamefont {Szigeti}}]{Mehdi:2021}%
  \BibitemOpen
  \bibfield  {author} {\bibinfo {author} {\bibfnamefont {Z.}~\bibnamefont {Mehdi}}, \bibinfo {author} {\bibfnamefont {A.~S.}\ \bibnamefont {Bradley}}, \bibinfo {author} {\bibfnamefont {J.~J.}\ \bibnamefont {Hope}},\ and\ \bibinfo {author} {\bibfnamefont {S.~S.}\ \bibnamefont {Szigeti}},\ }\bibfield  {title} {\bibinfo {title} {{Superflow decay in a toroidal Bose gas: The effect of quantum and thermal fluctuations}},\ }\href {https://doi.org/10.21468/SciPostPhys.11.4.080} {\bibfield  {journal} {\bibinfo  {journal} {SciPost Phys.}\ }\textbf {\bibinfo {volume} {11}},\ \bibinfo {pages} {080} (\bibinfo {year} {2021})}\BibitemShut {NoStop}%
\bibitem [{\citenamefont {Prikhodko}\ and\ \citenamefont {Bidasyuk}(2021)}]{Prikhodko:2021}%
  \BibitemOpen
  \bibfield  {author} {\bibinfo {author} {\bibfnamefont {O.}~\bibnamefont {Prikhodko}}\ and\ \bibinfo {author} {\bibfnamefont {Y.}~\bibnamefont {Bidasyuk}},\ }\bibfield  {title} {\bibinfo {title} {Projected gross--pitaevskii equation for ring-shaped bose--einstein condensates},\ }\href {https://doi.org/10.15407/ujpe66.3.198} {\bibfield  {journal} {\bibinfo  {journal} {Ukrainian Journal of Physics}\ }\textbf {\bibinfo {volume} {66}},\ \bibinfo {pages} {198} (\bibinfo {year} {2021})}\BibitemShut {NoStop}%
\bibitem [{\citenamefont {Kitagawa}\ and\ \citenamefont {Ueda}(1993)}]{Kitagawa1993}%
  \BibitemOpen
  \bibfield  {author} {\bibinfo {author} {\bibfnamefont {M.}~\bibnamefont {Kitagawa}}\ and\ \bibinfo {author} {\bibfnamefont {M.}~\bibnamefont {Ueda}},\ }\bibfield  {title} {\bibinfo {title} {Squeezed spin states},\ }\href {https://doi.org/10.1103/PhysRevA.47.5138} {\bibfield  {journal} {\bibinfo  {journal} {Phys. Rev. A}\ }\textbf {\bibinfo {volume} {47}},\ \bibinfo {pages} {5138} (\bibinfo {year} {1993})}\BibitemShut {NoStop}%
\bibitem [{\citenamefont {Mertes}\ \emph {et~al.}(2007)\citenamefont {Mertes}, \citenamefont {Merrill}, \citenamefont {Carretero-Gonz\'alez}, \citenamefont {Frantzeskakis}, \citenamefont {Kevrekidis},\ and\ \citenamefont {Hall}}]{Mertes2007}%
  \BibitemOpen
  \bibfield  {author} {\bibinfo {author} {\bibfnamefont {K.~M.}\ \bibnamefont {Mertes}}, \bibinfo {author} {\bibfnamefont {J.~W.}\ \bibnamefont {Merrill}}, \bibinfo {author} {\bibfnamefont {R.}~\bibnamefont {Carretero-Gonz\'alez}}, \bibinfo {author} {\bibfnamefont {D.~J.}\ \bibnamefont {Frantzeskakis}}, \bibinfo {author} {\bibfnamefont {P.~G.}\ \bibnamefont {Kevrekidis}},\ and\ \bibinfo {author} {\bibfnamefont {D.~S.}\ \bibnamefont {Hall}},\ }\bibfield  {title} {\bibinfo {title} {Nonequilibrium dynamics and superfluid ring excitations in binary bose-einstein condensates},\ }\href {https://doi.org/10.1103/PhysRevLett.99.190402} {\bibfield  {journal} {\bibinfo  {journal} {Phys. Rev. Lett.}\ }\textbf {\bibinfo {volume} {99}},\ \bibinfo {pages} {190402} (\bibinfo {year} {2007})}\BibitemShut {NoStop}%
\bibitem [{\citenamefont {Hanna}\ \emph {et~al.}(2010)\citenamefont {Hanna}, \citenamefont {Tiesinga},\ and\ \citenamefont {Julienne}}]{Hanna2010}%
  \BibitemOpen
  \bibfield  {author} {\bibinfo {author} {\bibfnamefont {T.~M.}\ \bibnamefont {Hanna}}, \bibinfo {author} {\bibfnamefont {E.}~\bibnamefont {Tiesinga}},\ and\ \bibinfo {author} {\bibfnamefont {P.~S.}\ \bibnamefont {Julienne}},\ }\bibfield  {title} {\bibinfo {title} {Creation and manipulation of feshbach resonances with radiofrequency radiation},\ }\href {https://doi.org/10.1088/1367-2630/12/8/083031} {\bibfield  {journal} {\bibinfo  {journal} {New Journal of Physics}\ }\textbf {\bibinfo {volume} {12}},\ \bibinfo {pages} {083031} (\bibinfo {year} {2010})}\BibitemShut {NoStop}%
\bibitem [{\citenamefont {Ymai}\ \emph {et~al.}(2026)\citenamefont {Ymai}, \citenamefont {Wilsmann}, \citenamefont {Neves}, \citenamefont {Tonel}, \citenamefont {Links},\ and\ \citenamefont {Foerster}}]{Ymai:2026}%
  \BibitemOpen
  \bibfield  {author} {\bibinfo {author} {\bibfnamefont {L.~H.}\ \bibnamefont {Ymai}}, \bibinfo {author} {\bibfnamefont {K.~W.}\ \bibnamefont {Wilsmann}}, \bibinfo {author} {\bibfnamefont {J.~B.}\ \bibnamefont {Neves}}, \bibinfo {author} {\bibfnamefont {A.~P.}\ \bibnamefont {Tonel}}, \bibinfo {author} {\bibfnamefont {J.}~\bibnamefont {Links}},\ and\ \bibinfo {author} {\bibfnamefont {A.}~\bibnamefont {Foerster}},\ }\href {https://arxiv.org/abs/2605.09709} {\bibinfo {title} {Supersensitive rotation sensor from superintegrability}} (\bibinfo {year} {2026}),\ \Eprint {https://arxiv.org/abs/2605.09709} {arXiv:2605.09709 [quant-ph]} \BibitemShut {NoStop}%
\bibitem [{\citenamefont {Pethick}\ and\ \citenamefont {Smith}(2008)}]{Pethick2008}%
  \BibitemOpen
  \bibfield  {author} {\bibinfo {author} {\bibfnamefont {C.~J.}\ \bibnamefont {Pethick}}\ and\ \bibinfo {author} {\bibfnamefont {H.}~\bibnamefont {Smith}},\ }\href {https://doi.org/10.1017/cbo9780511802850} {\emph {\bibinfo {title} {Bose--{{Einstein}} Condensation in Dilute Gases}}}\ (\bibinfo  {publisher} {Cambridge University Press},\ \bibinfo {year} {2008})\BibitemShut {NoStop}%
\end{thebibliography}%

\begin{widetext}
\section{Appendix}

\subsection{Gaussian Ansatz for an effective 2D model}
\label{appendix:Gauss}
We reduce the dimension of our problem from a full 3D model to an effective 2D model by using a Gaussian ansatz (assuming a weakly interacting regime) and the variational method, similar to that done in Chapter 6.2.1 of Pethick and Smith \cite{Pethick2008}. The truncated Wigner simulations use the effective 2D interaction strength for all multimode simulations. 

To begin, a tight confinement is assumed in the $z$ direction, giving a frozen Gaussian for the ground state ansatz
\begin{equation}
    \psi({\bf r},t) = \psi(r_\perp ,t) \phi(z) ,
\end{equation}
where 
\begin{equation}
    \phi(z) = \frac{1}{\pi^{1/4} R_z^{1/2}} \exp(-z^2/2R_z^2).
\end{equation}
As of yet, we do not know the length scale $R_z$ to use here. We will use the variational method to determine the appropriate length scale.

The energy in a ring trap system is given by
\begin{equation}
    E(\psi) = \int d{\bf r} \psi^*({\bf r}) \hat H \psi({\bf r}),
\end{equation}
which is 
\begin{equation}
    E(\psi) = \int d{\bf r} \left[ \frac{-\hbar^2}{2m} \vert \nabla \psi({\bf r})\vert^2 + V({\bf r})\vert \psi({\bf r})\vert^2 + \frac{U_0}{2} \vert \psi({\bf r})\vert^4 \right].
\end{equation}
The potential for the ring trap is given by
\begin{equation}
    V({\bf r}) = \frac{1}{2} m \omega_r^2 (r_\perp-R)^2 + \frac{1}{2} m\omega_z^2 z^2, 
\end{equation}
where $r_\perp = \sqrt{x^2+y^2}$, and $\omega_r$ and $\omega_z$ are the trapping frequencies in the $r_\perp$ and $z$ directions respectively.
The ansatz for $\phi(z)$ is substituted into the expression for the energy to obtain
\begin{equation}
    E_{kin} = \frac{-\hbar^2}{2m}\int dr_\perp \int dz  \left(\vert \phi(z)\vert^2 \vert \nabla_\perp \psi(r_\perp)\vert^2 + \vert \psi(r_\perp)\vert^2 \frac{z^2}{R_z^4} \vert \phi(z)\vert^2  \right), 
\end{equation}
and
\begin{equation}
    E_{pot} = \int dr_\perp \int dz\left(\frac{1}{2} m \omega^2_r (r_\perp-R)^2 + \frac{1}{2} m \omega^2_z z^2 \right) \vert \psi(r_\perp) \vert^2 \vert \phi(z)\vert^2, 
\end{equation}
and
\begin{equation}
    E_{int} = \int dr_\perp \int dz \frac{U_0}{2} \vert \phi(z)\vert^4 \vert \psi(r_\perp)\vert^4. 
\end{equation}

We make use of the normalization of the state ($\int dr_\perp \int dz \vert \psi(r_\perp) \vert^2 \vert \phi(z) \vert^2 = N$) and integrate over $z$ to obtain
\begin{equation}
     E_{pot}+E_{kin} = \frac{\hbar^2 N}{2m} \left[-\int dr_\perp \vert \nabla_\perp \psi(r_\perp)\vert^2+   \int dr_\perp \frac{(r-R)^2}{a_{r_\perp}^4} \vert \psi(r_\perp)\vert^2  +  \left( \frac{R_z^2}{2 a_z^4} - \frac{1}{2 R_z^2} \right)\right],
\end{equation}
and
\begin{equation}
     E_{int} = \frac{U_0 N^2}{2(2\pi)^{1/2} R_z} \int dr_\perp \vert \psi(r_\perp)\vert^4
\end{equation}
where $a_i^2 = \hbar/m\omega_i$ is the length scale for the harmonic oscillator, so that $m\omega_i^2/2 = \hbar^2/2m a_i^4$.
To go beyond this point we will make an ansatz for the perpendicular direction. For this we choose a Gaussian centered at $r - R$. In cylindrical polar coordinates, the Gaussian takes the form
\begin{equation}
     \psi(r) = \frac{1}{(2\pi)^{1/2}\pi^{1/4}(R R_\perp)^{1/2}} \exp(-(r-R)^2/2R_\perp^2),
\end{equation}
where $R_\perp$ is the length scale for the perpendicular 2D direction. Now we see that we have two variational parameters $R_z, R_\perp$ with which to play with and which we need to find analytic values for in order to appropriately choose our length scales for our ring trap in the numerical simulations. All other parameters are fixed in our expression for the energy.
The Laplacian in polar coordinates in the 2D plane is given by
\begin{equation}
     \nabla_\perp =\left(\frac{\partial^2}{\partial r^2} +\frac{1}{r}\frac{\partial}{\partial r} + \frac{1}{r^2} \frac{\partial^2}{\partial \theta^2} \right) .
\end{equation}
The $\theta$ derivative will disappear when acting upon the ground state with no initial angular momentum, so that we have
\begin{equation}
-\frac{\hbar^2 N}{2m}\int dr_\perp \vert \nabla_\perp \psi(r_\perp)\vert^2= \frac{-\hbar^2 N}{2m} 2\pi  \int dr ~r \left[\frac{(r-R)^2}{R_\perp^4} - \frac{1}{R_\perp^2} + \frac{1}{r} \frac{R}{R_\perp^2} - \frac{1}{R_\perp^2} \right] \vert \psi(r)\vert^2.
\end{equation}
Integrating over $r$ for the kinetic and potential terms we have
\begin{equation}
     E_{pot}+E_{kin} = \frac{\hbar^2 N}{2m} \left[  \left( \frac{R_\perp^2}{2a_{r_\perp}^4} - \frac{1}{R_\perp^2} \right)  +  \left( \frac{R_z^2}{2 a_z^4} - \frac{1}{2 R_z^2} \right)\right].
\end{equation}
Likewise for the interaction term we have
\begin{equation}
     E_{int} = \frac{U_0 N^2}{8 \pi^2 R R_\perp R_z}.
\end{equation}

\subsubsection{Minimising the energy}
Let us keep the dimensions in our expression as we look to minimize the energy. 
We will minimize the energy with respect to the variational parameters. To minimize we take the derivative with respect to the parameters
\begin{equation}
\frac{\partial E(R_\perp,R_z)}{\partial R_\perp} = \hbar N \omega_r \left(\frac{R_\perp}{2 a_{r_\perp}^2} - \frac{a_{r_\perp}^2}{2R_\perp^3}  \right)- \frac{U_0 N^2}{8 \pi^2 R R_\perp^2 R_z},
\end{equation}
\begin{equation}
\frac{\partial E(R_\perp,R_z)}{\partial R_z} = \hbar N \omega_z \left(\frac{R_z}{2 a_{r_z}^2} - \frac{a_{r_z}^2}{2R_z^3}  \right)- \frac{U_0 N^2}{8 \pi^2 R R_\perp R_z^2}.
\end{equation}
We have the two equations to solve that are given by
\begin{equation}
\hbar N \omega_r \left(\frac{R_\perp^2}{2 a_{r_\perp}^2} - \frac{a_{r_\perp}^2}{2R_\perp^2}  \right)- \frac{U_0 N^2}{8 \pi^2 R R_\perp R_z} = 0,
\end{equation}
and
\begin{equation}
\hbar N \omega_z \left(\frac{R_z^2}{2 a_{r_z}^2} - \frac{a_{r_z}^2}{2R_z^2}  \right)- \frac{U_0 N^2}{8 \pi^2 R R_\perp R_z}=0,
\end{equation}
which is easily done numerically to obtain the parameters $R_z$ and $R_\perp$.

\subsubsection{Effective 2D interaction strength for our simulations}

We are now in a position to find the effective 2D interaction strength for the simulations based on the analytics obtained from the Gaussian ansatz and the variational principle. We will use the resulting length scale found for z to calculate $\int dz \vert \phi(z)\vert^4$, as 
\begin{equation}
    \int dz \vert \phi(z)\vert^4 = \frac{1}{R_z} \sqrt{\frac{1}{2\pi}}.
\end{equation}
Integrating out the z direction based on the Gaussian ansatz calculation, 
\begin{multline}
i \hbar \frac{\partial}{\partial t} \psi(r_\perp,t)  = \left(\frac{-\hbar^2}{2m} \nabla_\perp^2 + V(r_\perp) \right) \psi(r_\perp,t)  + \int dz \phi^*(z)\left(\frac{-\hbar^2}{2m} \frac{\partial^2}{\partial z^2} + V(z) \right) \psi(r_\perp,t) \phi(z) \\+  g_{3D} \left( \int dz \vert \phi(z)\vert^4 \right) \vert \psi(r_\perp)\vert^2\psi(r_\perp,t), 
\end{multline}
where we have the following terms from the kinetic and potential energy in $z$
\begin{equation}
 \int dz \phi^*(z)\left(\frac{-\hbar^2}{2m} \frac{\partial^2}{\partial z^2} + V(z) \right) \psi(r_\perp,t) \phi(z) = \psi(r_\perp,t) \left[\frac{\hbar^2}{4m R_z^2} + \frac{m\omega_z^2 R_z^2}{4} \right].
\end{equation}
This term is just a constant offset so we can ignore it in our simulations, but we will keep it here for posterity. Then the effective 2D GPE will be given (for the Gaussian ansatz) as:
\begin{equation}
i \hbar \frac{\partial}{\partial t} \psi(r_\perp,t)  = \left(\frac{-\hbar^2}{2m} \nabla_\perp^2 + V(r_\perp) \right) \psi(r_\perp,t)  +  \left[\frac{\hbar^2}{4m R_z^2} + \frac{m\omega_z^2 R_z^2}{4} \right] \psi(r_\perp,t) +  g_{3D} \frac{1}{R_z} \sqrt{\frac{1}{2\pi}} \vert \psi(r_\perp)\vert^2\psi(r_\perp,t). 
\end{equation}
\subsubsection{Interaction term for the few mode TWEs}
The above effective interaction strength and the length scale are used in order to determine the value in our numerical simulations for the few mode case.  The effective 2D interaction strength is given by
\begin{equation}
g_{2D} = g_{3D} \frac{1}{R_z} \sqrt{\frac{1}{2\pi}},
\end{equation}
where $R_z$ is found by numerically minimizing the energy for a given set of parameters. Then there is simply the matter of choosing the value for $\chi_{ij} (t) = \int d{r} \tfrac{U_{ij}}{2\hbar} \vert \phi_i ({\bf r})\vert^2 \vert \phi_j ({\bf r})\vert^2$ in the few mode simulations which we assume to be fixed in time, and we base this on the gaussian ansatz for the state. Here, we have that $U_{ij} = g_{2D}$. The form this takes is given by
\begin{equation}
\chi = \int d\theta\int r dr \frac{g_{3D}}{2\hbar}\frac{1}{R_z} \sqrt{\frac{1}{2\pi}}\vert \psi(r)\vert^4,
\end{equation}
which gives
\begin{equation}
\chi = \frac{g_{3D}}{8 \hbar \pi^2 R R_\perp R_z}.
\end{equation}

\subsection{Table of parameters for numerical simulations}
\label{appendix:parametertable}
These parameters are calculated from the Gaussian Ansatz for an effective 2D model, using a variational method.


 \begin{tabular}{|c|c||c|c|c|c|c|} \hline \multicolumn{7}{|c|}{Parameters for numerical simulations} \\ \hline Trap Radius ($\mu\mathrm{m}$) & Radial frequency (Hz/$2\pi$) & Calculated $R_z$ ($\mu\mathrm{m}$) & Calculated $R_r$ ($\mu\mathrm{m}$) & $\chi_{aa}$ & $\chi_{bb}$ & $\chi_{ab}$ \\ \hline 40 & 30 & 1.02747 & 2.32921 & 0.00645 & 0.00611 & 0.00628 \\ 70 & 30 & 1.01086 & 2.18719 & 0.00399 & 0.00378 & 0.00388 \\ \hline 40 & 50 & 1.04148 & 1.74346 & 0.00851 & 0.00805 & 0.00828 \\ 70 & 50 & 1.01916 & 1.65682 & 0.00523 & 0.00495 & 0.00508 \\ \hline 200 & 10 & 9.8855 & & & & \\ \hline \end{tabular}

\subsection{Truncated Wigner Equations for a four-mode model}
\label{appendix:4mTWE}

The truncated Wigner equations for the four-mode model are given by the following
\begin{equation}
    i \hbar \frac{d \alpha_{l}}{dt} = \frac{\hbar^2 {l}^2}{2m R^2}  \alpha_l +  4\hbar \chi_{aa}(t) \alpha_l\left(\frac{1}{2}\vert \alpha_l\vert^2 + \vert \alpha_{\kappa}\vert^2 -1 \right)+\hbar \chi_{ab}(t) \alpha_l\left(\vert \beta_{-l}\vert^2 +\vert \beta_{-\kappa}\vert^2 -1\right)+\hbar \chi_{ab}(t) \left( \alpha_{\kappa} \beta_{-\kappa} \beta_{-l}^* \right),
\end{equation}
\begin{equation}
    i \hbar \frac{d \alpha_{\kappa}}{dt} = \frac{\hbar^2 {(\kappa)}^2}{2m R^2}  \alpha_{\kappa} +  4\hbar \chi_{aa}(t) \alpha_{\kappa}\left(\frac{1}{2}\vert \alpha_{\kappa}\vert^2 + \vert \alpha_{l}\vert^2 -1 \right)+\hbar \chi_{ab}(t) \alpha_{\kappa}\left(\vert \beta_{-l}\vert^2 +\vert \beta_{-\kappa}\vert^2 -1\right)+\hbar \chi_{ab}(t)\left( \alpha_{l} \beta_{-l} \beta_{-\kappa}^*  \right)
\end{equation}
\begin{equation}
    i \hbar \frac{d \beta_{-l}}{dt} = \frac{\hbar^2 ({-l})^2}{2m R^2}  \beta_{-l} +  4\hbar \chi_{aa}(t) \beta_{-l}\left(\frac{1}{2}\vert \beta_{-l}\vert^2 + \vert \beta_{\kappa}\vert^2 -1 \right)+\hbar \chi_{ab}(t) \beta_{-l}\left(\vert \alpha_{l}\vert^2 +\vert \alpha_{-\kappa}\vert^2 -1\right)+\hbar \chi_{ab}(t)\left(\beta_{\kappa} \alpha_{-\kappa} \alpha_{l}^*  \right)
\end{equation}
\begin{equation}
    i \hbar \frac{d \beta_{-\kappa}}{dt} = \frac{\hbar^2 ({-\kappa})^2}{2m R^2}  \beta_{-\kappa} +  4\hbar \chi_{aa}(t) \beta_{-\kappa}\left(\frac{1}{2}\vert \beta_{-\kappa}\vert^2 + \vert \beta_{-l}\vert^2 -1 \right)+\hbar \chi_{ab}(t) \beta_{-\kappa}\left(\vert \alpha_{l}\vert^2 +\vert \alpha_{\kappa}\vert^2 -1\right)+\hbar \chi_{ab}(t)\left(\beta_{-l} \alpha_{l} \alpha_{\kappa}^*  \right).
\end{equation}
\end{widetext}

\begin{figure*}
\centering
  \includegraphics[width=.5\linewidth]{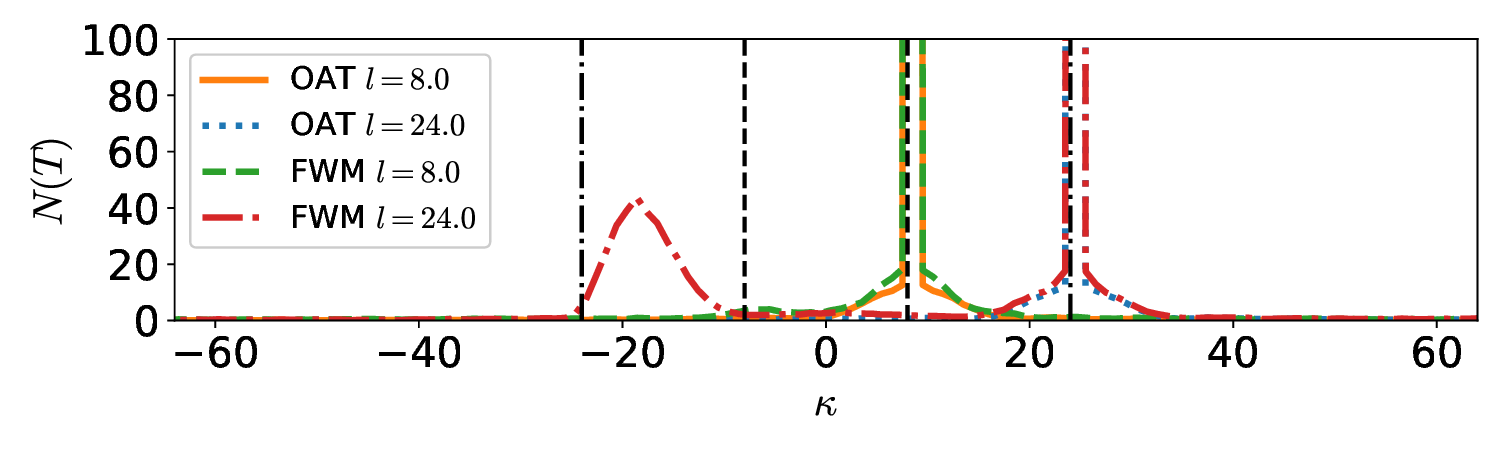}
\caption{Mode populations (from MTW) at $T=0.05$s for both OAT and FWM with charge $l=8,24$. The vertical lines give $\pm l$.
}
\label{fig:oatAMmode1}
\end{figure*}
\begin{figure*}
\centering
  \includegraphics[width=.5\linewidth]{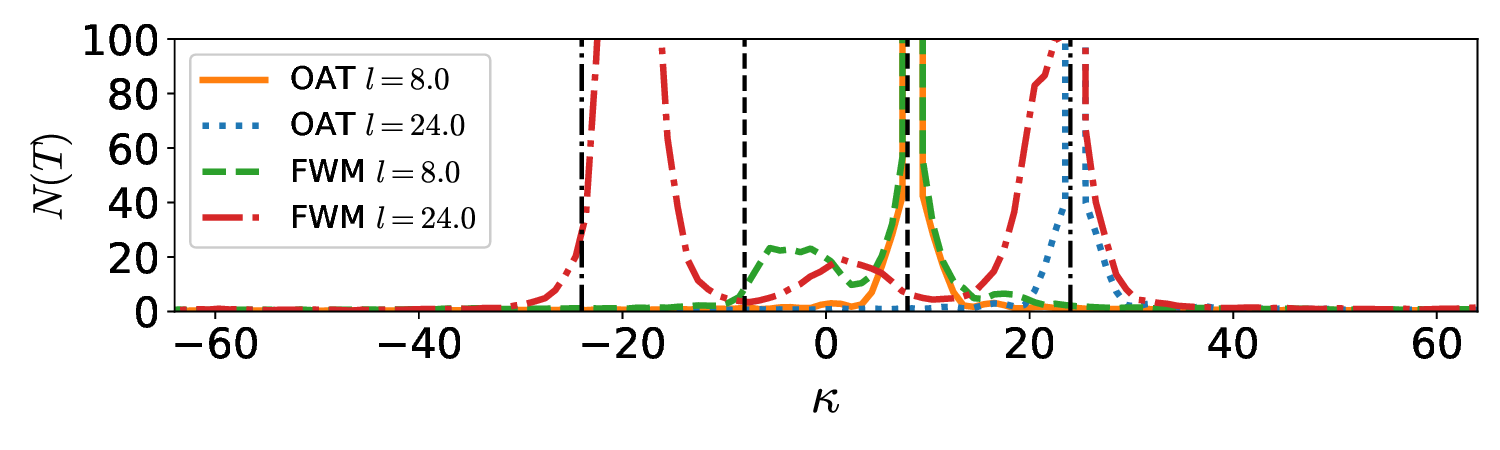}
\caption{Mode populations (from MTW) at $T=0.1$s for both OAT and FWM with charge $l=8,24$. The vertical lines give $\pm l$.
}
\label{fig:oatAMmode2}
\end{figure*}

\subsection{AM Mode population in OAT and FWM}
\label{appendix:OATMode}
In Figs. \ref{fig:oatAMmode1} and \ref{fig:oatAMmode2}, we demonstrate the similar AM mode population of both OAT and FWM that results from the radial coupling to the self interaction terms, resulting in four wave mixing. We show this for two different times ($T=0.05$s and $0.1$s) and for two different charges ($l=8,24$).

\subsection{Evolution of the non-interacting guided Sagnac}
\begin{figure*}
\centering
  \includegraphics[width=.80\linewidth]{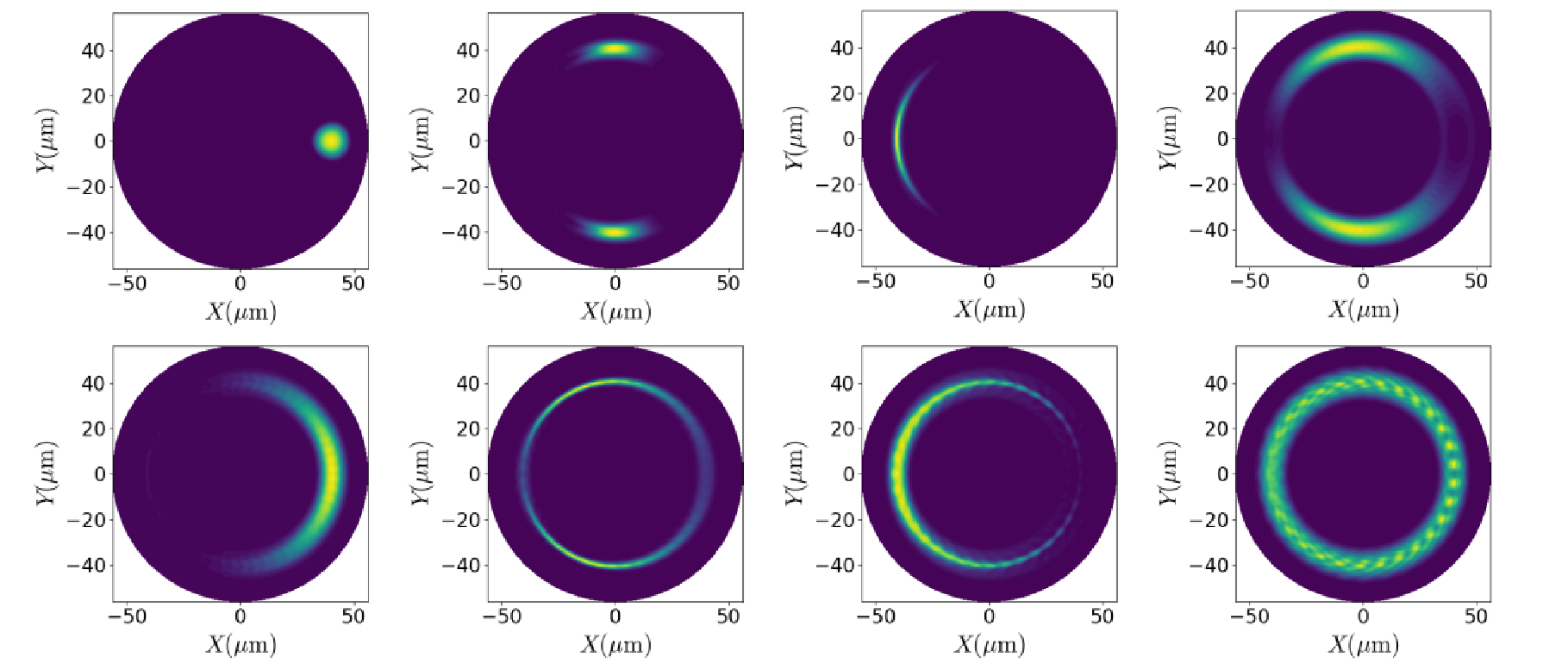}
  
  \label{fig:sub1}


\caption{Evolution of two components of non-interacting atoms in the guided Sagnac for a charge $l=50$. For the set of trapping parameters ($R=40 \mu m$, $\omega_r/2\pi = 30 Hz$). The snap shots are taken between $t=0$ and $t=0.49s$. }
\label{fig:BromDenstime}
\end{figure*}

In figure \ref{fig:BromDenstime} we give the snap shots of evolution in the non-interacting version of the guided Sagnac interferometer.

\end{document}